\documentclass[apj]{emulateapj}
\usepackage[colorlinks,linkcolor={blue},citecolor={blue},urlcolor={red}]{hyperref}
\usepackage{epsfig,graphicx,natbib,amsmath,amsfonts,amssymb}%,xfrac}
\usepackage{color}
\usepackage{multirow}
\usepackage{booktabs}
\usepackage{amssymb}
\usepackage{threeparttablex} % for "ThreePartTable" environment
\usepackage{booktabs} 

\usepackage{longtable}
\usepackage{array} % for extrarowheight
\usepackage[dvipsnames]{xcolor}  

\newcommand{\re}{\Reff}
\newcommand{\msolar}{${\rm M}_\odot$}
\newcommand{\mstar}{$M_\ast$($<$\re)}
\newcommand{\lgmstar}{$\log_{10}$($M_\ast(<$Re$)/$\msolar)}

\newcommand{\dindex}{{\rm D}$_n$(4000)}

\newcommand{\RSFRL}{${\rm SFR}_{\rm 5Myr}/{\rm SFR}_{\rm 800Myr}$}
\newcommand{\RSFR}{SFR79}
\newcommand{\SFRNI}{$\log_{\rm 10}{\rm SFR}_{\rm 800Myr}$}

\newcommand{\hd}{{\rm H}$\delta$}
\newcommand{\hda}{\hd$_{\rm A}$}
\newcommand{\ewhda}{{\rm EW}(\hda)}
\newcommand{\ha}{{\rm H}$\alpha$}
\newcommand{\hae}{\ha}
\newcommand{\ewhae}{{\rm EW}(\hae)}
\newcommand{\lgewhae}{$\log_{10}$\ewhae}

\newcommand{\Reff}{{$R_{\rm e}$}}

\newcommand{\myemail}{\email{enci.wang@phys.ethz.ch}}

\defcitealias{Wang-19}{W19}

\shorttitle{The time variability of star-formation in galaxies}
\shortauthors{Wang \& Lilly}

\graphicspath{{fig1/}{fig1/fit/}{fig1/fit_cor/}}
\begin{document}

\title{The variability of the star formation rate in galaxies: I. Star formation histories traced by EW(H$\alpha$) and EW(H$\delta_{\rm A}$)} 
\author{
Enci Wang\altaffilmark{1},
Simon J. Lilly\altaffilmark{1}
} \myemail

\altaffiltext{1}{Department of Physics, ETH Zurich, Wolfgang-Pauli-Strasse 27, CH-8093 Zurich, Switzerland}

\begin{abstract}  

To investigate the variability of the star formation rate (SFR) of galaxies, 
we define a star formation change parameter, \RSFRL\, which is the ratio of the SFR averaged within the last 5 Myr to the SFR 
averaged within the last 800 Myr. We show that this parameter can be determined from a combination of 
H$\alpha$ emission and H$\delta$ absorption, plus the 4000 \AA\ break, with an uncertainty of $\sim$0.07 dex for 
star-forming galaxies. We then apply this estimator 
to MaNGA galaxies, both globally within $R_e$ and
within radial annuli.  We find that the global \RSFRL, which indicates by how much a galaxy has changed its specific SFR (sSFR) is nearly 
independent of its sSFR, i.e. of its the position relative to the star formation main sequence (SFMS) as defined by SFR$_{\rm 800 Myr}$.  Also, 
at any sSFR, there are as many galaxies increasing their sSFR as decreasing it, as required if the dispersion in the SFMS is to stay the same.  The \RSFRL\ of the overall galaxy population is very close to that expected for the evolving Main Sequence. 
Both of these provide a reassuring check on the validity of our calibration of the estimator.  
We find that galaxies with higher global \RSFRL\ appear to have higher \RSFRL\ at all galactic radii, i.e. that 
galaxies with a recent temporal enhancement in overall SFR have enhanced star formation at all galactic radii.   
The dispersion of the \RSFRL\ at a given relative galactic radius and a given stellar mass decreases with the (indirectly inferred) gas depletion time: 
locations with short gas depletion time appear to undergo bigger variations in 
%the change of 
their star-formation rates on Gyr or less timescales.  In \cite{Wang-19} we showed that the dispersion in star-formation rate surface densities $\Sigma_{\rm SFR}$ %(averaged over the last $10^7$ yr) 
in the galaxy population appears to be inversely correlated with the inferred gas depletion timescale and interpreted this in terms of the dynamical response of a gas-regulator system to changes in the gas inflow rate.  In this paper, 
%we not only produce the analogous and largely independent result for the star-formation surface density on much longer $10^9$ yr timescales but, more importantly, 
we can now prove directly with \RSFRL\ that these effects are indeed due to genuine {\it temporal variations} in the SFR of individual galaxies on timescales between $10^7$ and $10^9$ years rather than possibly reflecting intrinsic, non-temporal, differences between different galaxies.
%We argue that \RSFRL\ contains valuable information of time-varying SFH of galaxies, which can shed
%light on the physical processes that govern changes in the star formation rate of galaxies. 

\end{abstract}

\keywords{galaxies: general -- methods: observational}

\section{Introduction}
\label{sec:introduction}
 
Studying the star formation histories (SFH) of galaxies is a major tool to study the formation
and evolution of galaxies.  Thanks to a series of deep surveys of galaxies, 
the evolution of the star formation rate (SFR) for the global galaxy population 
also known as the cosmic evolution of the star formation rate density (SFRD), is well established
up to redshift of $\sim$9 \citep[e.g.][]{Lilly-96, Schiminovich-05, Bouwens-11, Bouwens-14, Madau-14, 
Hagen-15, Alavi-16, Goto-19}.  The evolution of the SFRD is measured based on the galaxy population at 
different redshifts.
%, i.e. different snapshots of the real universe. 

Observationally, most star-forming (SF) galaxies form a narrow sequence
%with the slope of $\sim1$ 
on the stellar mass-SFR diagram 
up to at least redshift of 3 \citep[e.g.][]{Brinchmann-04, Daddi-07, Elbaz-07, Noeske-07, Elbaz-11}. 
This sequence is known as the star formation main sequence (SFMS). 
The relation is approximately linear, i.e. there is a characteristic specific SFR (sSFR).  The sSFR-normalization of the SFMS 
has been found to evolve with lookback time as  $(1+z)^{2.2}$ \citep{Pannella-09, Stark-13, 
Schreiber-15, Boogaard-18}. 
This increase with look-back time of the sSFR of typical galaxies at a given mass is due to the higher rate of accretion 
of cold gas by galaxies at high redshift.  The scatter of the SFMS is rather small 
at any given redshift, 0.2-0.4 dex,  depending on the exact definition of the SF populations 
and the method to obtain the stellar masses and SFRs.  
The origin of the SFMS and the small scatter of the SFMS is not well understood, but likely reflects 
the effect of a long-term quasi-steady state between gas accretion, 
star formation and gas outflow driven by feedback processes \citep[e.g.][]{Schaye-10, Bouche-10, 
Dave-11, Lilly-13, Tacchella-16, Rodriguez-Puebla-16, Wang-19}. 

It is clear that the scatter of the SFMS is relevant to the variability of SFHs of 
individual SF galaxies, and accurate measurements of SFHs could uncover the 
contributions of the scatter of SFMS by the variation of SFHs at different timescales.  However, the accurate star formation histories of individual galaxies 
are still poorly determined from observations, especially on short timescales ($<100$ Myr). 

Many physical processes have been proposed to account for the variability of the SFHs for individual galaxies.
These processes are generally separated into two types: internal processes and those driven by the external environment. Basically, these processes enhance or suppress (or even quench) the star formation 
by producing changes in the cold gas content and/or a change in the star formation efficiency (SFE), 
defined as the SFR per unit mass of cold gas. For instance, 
disk instabilities \citep[e.g.][]{Dekel-14, Zolotov-15, Tacchella-16} and 
bar-induced gas inflows \citep[e.g.][]{Wang-12, Lin-17, Chown-19} may 
enhance the star formation via an increase of star formation efficiency, 
while outflows driven by stellar feedback \citep[e.g.][]{Ceverino-09, Muratov-15, El-Badry-16} 
or tidal/ram-pressure stripping in massive halos \citep[e.g.][]{Gunn-72, Moore-96, Abadi-99, Poggianti-17}
may suppress the star formation in galaxies by removing cold gas.  

The variability of SFHs on short and long timescales is likely governed 
by physical processes operating on different timescales \citep{Sparre-15, Sparre-17, Broussard-19, Wang-19}. 
For instance, variations in gas accretion may drive the variation of SFR on relatively 
long timescales \citep{Sparre-15, Wang-19}, while feedback from supernovae or active 
galactic nuclei (AGN) may produce changes in the SFR on relatively short timescales \citep{Sparre-17}. 
Having more extensive information of how individual galaxies change their SFRs over time could reveal
which physical processes enhance or suppress the star formation during the lifetime of
galaxies, and which processes govern the variation of SFR over long and short timescales. 

Hydro-dynamical simulations can, in principle, produce accurate SFHs of simulated galaxies, 
which may be helpful to understand the origin of the scatter of the SFMS, regardless of
the poorly understood sub-grid physics.  Indeed, based on cosmological zoom-in simulations
of 26 moderately massive galaxies, \cite{Tacchella-16} found that SF galaxies oscillate about
the SFMS ridge on time-scales of $\sim$0.4 Hubble time ($t_{\rm Hubble}$) at $1<z<4$. 
The oscillation is the 
result of an interplay between gas compaction, gas depletion (including star formation and outflow), 
and accretion. Based on the EAGLE simulations, \cite{Matthee-19} investigated the 
evolution and origin of the scatter of the SFMS. They found that the scatter in sSFR in the local Universe 
originates in their simulation from a combination of fluctuations on short time-scales (0.2-2 Gyr), likely
associated with self-regulation of cooling, star formation and outflows, and variations on 
long time-scale ($\sim10$ Gyr) associated to different halo formation times. They found that
the long time-scale variations dominate the scatter of the SFMS in the local Universe. 
\cite{Rodriguez-Puebla-16} found that the scatter of the halo mass accretion rate ($\sim$0.3 dex) in the Bolshoi-Planck simulation \citep{Klypin-16}
is comparable to the observed scatter of SFMS. 
However, it should be noted that the halo mass accretion rate is averaged over 0.2$t_{\rm Hubble}$, 
which is much larger than the timescale of most SFR indicators in observations, 
such as H$\alpha$ and ultraviolet/infrared luminosity.  The scatter of halo mass accretion rate could be 
larger than 0.3 dex if it was averaged within a shorter timescale. 

Although the integrated spectral energy distribution (SED) of galaxies records the information of 
star formation at different lookback times, it is very challenging to obtain 
accurate SFHs by SED modelling \citep[e.g.][]{Papovich-01, Shapley-01, Muzzin-09,
Conroy-13, Carnall-19, Leja-19}. As shown in
the SED modelling test of \cite{Ge-18}, the SED fitting code is able to 
reproduce the input stellar population age, metallicity and mass-to-light ratio
with reasonable accuracy in dust-poor cases, while large discrepancies can occur in dust-rich cases. 
In addition, although SED modelling
can reproduce the overall shape of input SFHs in most cases, the 
short-time variations ($\sim100$ Myr) in SFHs are usually not well recovered 
\citep[e.g.][]{Ocvirk-06, Gallazzi-09, Zibetti-09, Leja-19}. 
Considering the possible variation of the initial mass function (IMF), and possibly different 
dust attenuations in young and old stellar populations \citep{Calzetti-00, Moustakas-06, Wild-11,
Hemmati-15, Reddy-15}, we are still quite a long way from obtaining accurate SFHs from SED modelling. 
An alternative way of obtaining individual SFHs is to analyze images of galaxies that are resolved 
down to individual stars \citep[e.g.][]{Tolstoy-09, Cignoni-15, Sacchi-19}, 
but this approach is only practical for the closest galaxies.

Given the difficulties of obtaining accurate SFHs of galaxies from observations, 
a number of measures comparing a longer-timescale SFR to a shorter-timescale SFR have been proposed in the literature \citep{Sullivan-00, Boselli-09, Wuyts-11, Guo-16, 
Sparre-17, Broussard-19, Emami-19, Faisst-19}.  These are often called ``burstiness" parameters, but we ourselves find this restrictive, as will be dismissed below.
It is clear that such burstiness contains information on the 
variability of recent SFHs. \cite{Weisz-12} found that the distribution of 
H$\alpha$-to-far-UV (FUV) flux ratios of a sample of 185 nearby galaxies can be 
well matched with simple, periodic SFH models, but can not be matched by
the constant SFHs with varying IMF.  
%\cite{Guo-16} investigated the burstiness of SFHs for a sample of 164 galaxies selected from CANDLES GOODS-N region, using the ratio of SFRs measured from H$\beta$ and FUV.  
\cite{Guo-16}  found a decrease in
H$\beta$-to-FUV ratio with decreasing galaxy mass, which can be explained by 
a bursty SFH on a timescale of a few tens of Myrs on galactic scales. 
More recently, \cite{Broussard-19} found that the dispersion of burstiness characterizes 
the stochasticity of a galaxy population's recent star formation, rather than the average 
value of burstiness.  Consistent with this, \cite{Caplar-19} have tried to use 
the scatter of the SFMS based on different SFR indicators to model the stochasticity of SFHs. 

A common method of quantifying burstiness is to use the 
average H$\alpha$-to-UV flux ratio. The H$\alpha$ emission is produced by the
recombination of gas ionized by photons from massive stars ($>15$\msolar), 
and is expected to be observed only within the typical lifetimes of these 
massive stars ($<5$ Myr). The UV continuum comes from non-ionizing photospheric 
emission from stars with mass greater than 3\msolar\ \citep[][and references therein]{Kennicutt-98}, 
which have lifetimes of $<$300 Myr. However, as pointed out by \cite{Caplar-19}, the
commonly-used SFR indicators, such as H$\alpha$, NUV, FUV, u-band and UV+IR luminosities,
do not exactly follow the recent SFHs within a specified timescale. Instead, 
the SFRs traced by any of these indicators can be considered 
to be a convolution of the  SFH with the luminosity evolution 
of these indicators for a single stellar population. 
This increases the complexity of studying the stochasticity of SFHs of a galaxy 
population. In addition, the measured burstiness based on
the above indicators strongly depends on
the dust attenuation correction, which may significantly broaden the scatter. 

In this work, we develop a new parameter to characterize the change of 
star formation, SFR$_{\rm 5Myr}$/SFR$_{\rm 800Myr}$. This is the ratio of the SFR averaged within the most recent 5 Myr to the SFR averaged
within the last 800 Myr.  The definition of this parameter is similar to 
the definition of the burstiness in the literature, but
in this work we prefer to call it a ``star formation change parameter'' 
(or simply the ``change parameter''), because galaxies can either enhance 
or suppress their star formation in the 
recent 5 Myr with respect to the average star formation within the last 800 Myr.  In other words, the parameter can take values
above or below unity, and indeed should average to unity over a large enough population.

Our change parameter \RSFR\ is calibrated using three diagnostic 
observational parameters: the equivalent width of H$\alpha$ emission (EW(H$\alpha$)), the 
Lick index of H$\delta$ absorption (EW(H$\delta$)$_{\rm A}$), and the size of 4000 \AA\ break (\dindex). 
The H$\alpha$ emission is a good tracer of the SFR averaged within the last 5 Myr, the \ewhda\ 
traces the star formation within the last roughly $\sim$1 Gyr, and the 4000 \AA\ break is sensitive 
to the light-weighted stellar age within 2 Gyr \citep[e.g.][]{Balogh-99, Kauffmann-03, Li-15, Wang-17, Wang-18}.
These three parameters can be directly measured from galaxy spectra, and each being measured at a single wavelength they are all, in principle,
insensitive to dust attenuation, although dust effects can still enter if different components of the system suffer different extinctions (this is explored in Section \ref{subsec:2.5} below). These observational parameters provide the basic means to measure the star formation change parameter \RSFRL. We then apply this estimator
to the MaNGA \citep[Mapping Nearby Galaxies at Apache Point Observatory;][]{Bundy-15} 
galaxies.  

We then use this change parameter to study the variations of SFR between and within galaxies on different timescales and
the recent change of star formation within and across galaxies. 
We establish a new observational result that strengthens the scenario proposed in 
\citet[][hereafter \citetalias{Wang-19}]{Wang-19} that the variation of 
SFR within and across galaxies is the result of the dynamic response of 
the gas-regulator system to the variation of the gas accretion.  

%In the second paper of this series, we will try to directly constrain the power spectrum distribution of sSFR histories of galaxies, based on the dispersion of the change parameter \RSFRL\ and the dispersion of the SFMS that are constructed in the present work. 

This paper is organized as follows. We develop the new change indicator \RSFRL\ of SFH 
in Section \ref{sec:2}. Specifically, in Section \ref{subsec:2.0}, we discuss 
the meaning of the star-formation change parameter. In Section \ref{subsec:2.1} and \ref{subsec:2.2},
we present the detailed calibration of the \RSFRL\ based on three the three observational diagnostic parameters,
and a wide suite of mock SFHs. 
We build the calibrator of \RSFRL\ and examine how good it is
in Section \ref{subsec:2.3}. We explore the dependence of the calibrator 
on different IMFs and different isochrones in Section \ref{subsec:2.4}.  
In Section \ref{subsec:2.5}, we present the recipes of the dust 
attenuation correction for \ewhae, \ewhda\ and \dindex\ when applying the 
calibrator to the observed spectra of galaxies.
In Section \ref{sec:3}, we apply the calibrator to a well-defined SF galaxy sample 
selected from MaNGA survey, and generate the maps and profiles of \RSFRL\ and 
the surface density of SFR$_{\rm 800Myr}$ for the sample galaxy. 
In Section \ref{sec:4}, we apply the calibrator to the integrated quantities of 
galaxies, and examine whether the calibrator can produce reasonable values of \RSFRL.
In Section \ref{sec:5}, we study the profiles of \RSFRL\ and 
SFR$_{\rm 800Myr}$, as well as the dispersion of \RSFRL\ and 
SFR$_{\rm 800Myr}$ within and across galaxies.  
We summarize this work in Section \ref{sec:6}. 

Throughout this paper, we assume a flat cold dark matter cosmology
with $\Omega_{\rm m}$=0.27, $\Omega_{\Lambda}$ = 0.73 and h=0.7 when computing 
distance-dependent parameters.  For convenience, the average star-formation over the last 5 Myr, SFR$_{\rm 5Myr}$, is denoted 
as SFR7, and that averaged over the last 800 Myr, SFR$_{\rm 800Myr}$, as SFR9\footnote{This is because 
the 5 Myr is close to $\sim$ 10$^7$yr, and 800 Myr is close to $\sim$10$^9$yr. }, and the ratio
SFR$_{\rm 5Myr}$/SFR$_{\rm 800Myr}$ is denoted as SFR79. 
 
\section{The construction and calibration of the change parameter of star formation}
\label{sec:2}

In this section, our task is to first construct and then calibrate our change parameter based on the 
diagnostic observational parameters, \ewhae, \ewhda, and \dindex.

% I SUGGEST WE HAVE A SUB-SECTION DEFINING SFR79 AND DISCUSSING ITS MEANING AND INTERPRETATION.

\subsection{The star-formation change parameter SFR79}
\label{subsec:2.0}

 The relatively long timescales for the formation of individual stars means that measurements of the rate of star-formation must necessarily represent averages over some even longer timescale, say $10^7$ years.  Ideally, we would like to have a change parameter that reflects the change of the star-formation rate, as measured within some fixed time interval, e.g. $10^7$ years, over some other, longer, time interval, say $10^9$ years.  Unfortunately this is not possible with current observational material, and it is in fact hard to see how it ever will be.  Practicalities therefore force us to instead compare the star-formation rates that are obtained by averaging over different periods of time prior to the epoch of observation, e.g. to compare the star-formation rate averaged over the previous 5 Myr with that averaged over the previous 800 Myr.  As noted above, we adopt a shorthand of SFR7 and SFR9 for these quantities, with the ratio denoted by SFR79.

The ratio SFR79 therefore mixes information both on short term ($10^7$ year) variations in the SFR, i.e. on the  ``burstiness" of star-formation, with longer-term drifts in the SFR of the galaxy taking place on longer timescales ($10^9$ year).   For this reason we prefer to think of the ratio SFR7/SFR9 as a star-formation ``change parameter" rather than simply as a measure of the ``burstiness" of the star-formation.  To think of ``bursts'' of star-formation implies values of SFR79 greater than unity. This may be appropriate for some subset of the galaxy population, but within the overall population, we would expect to find some values of SFR79 below unity, and indeed the average SFR79 should be roughly unity.  To be more precise on this point, we would expect the ratio of the average SFR7 divided by the average SFR9 (which will not be precisely the same as the average SFR79) to be unity, modulo any long term evolution of the SFR of galaxies with cosmic time.

A galaxy with a constant SFR will have an \RSFR\ of exactly unity. A galaxy with constant sSFR will have an \RSFR\ that is greater than unity by an amount that depends on that constant sSFR, because the increase in mass during the last Gyr will have produced an (exponentially) increasing SFR.  However, this effect is small if the mass-doubling timescale ${\rm sSFR}^{-1}$ is long compared with 1 Gyr, as will generally be the case for galaxies at the present epoch.  This effect will be discussed further below.

SFR79 will also give information on the movement of a galaxy in the SFR-mass plane.  If we neglect the changes in the stellar mass over the timescales of interest, i.e. if ${\rm sSFR}^{-1} >>$ 1 Gyr, then the SFR79 will tell us the present location of a galaxy on the SFR7-mass plane compared to the average position it has occupied over the last $10^9$ years.  In this sense, it tells 
us whether the individual galaxy is broadly moving up or down relative to its SFMS. 

\subsection{The diagnostic observational parameters of the recent SFHs}
\label{subsec:2.1}

The basis of the calibration is that the three chosen observational parameters contain 
information about the (specific) SFR averaged within 5 Myr and roughly 1 Gyr, and therefore the 
change parameter can be derived from a combination of these three diagnostic 
parameters. 

Here we briefly describe our overall approach to derive the change parameter.  The 
details will then be presented in the following subsections.
We first construct millions of mock SFHs 
of galaxies. These mock SFH span the whole of cosmic time and should cover
as much as possible the range of SFH encountered in the real Universe. 
We then generate synthetic spectra of these mock galaxies at the present epoch based on stellar 
population models for a range of different metallicities. We then measure the three diagnostic parameters 
of interest.  
We also compute the actual \RSFR\ from the mock SFHs.  Finally, we search for 
the solution of \RSFR\ in terms of the three diagnostic observational parameters. 

%In this section, we will present the details of the calibration, including the basis of
%this method, mock SFH construction, the formula of calibrator, exploration of the 
%stability of calibrator, and the correction of dust attenuation 
%of three diagnostic parameters in the application. 

\begin{figure*}
  \begin{center}
    \epsfig{figure=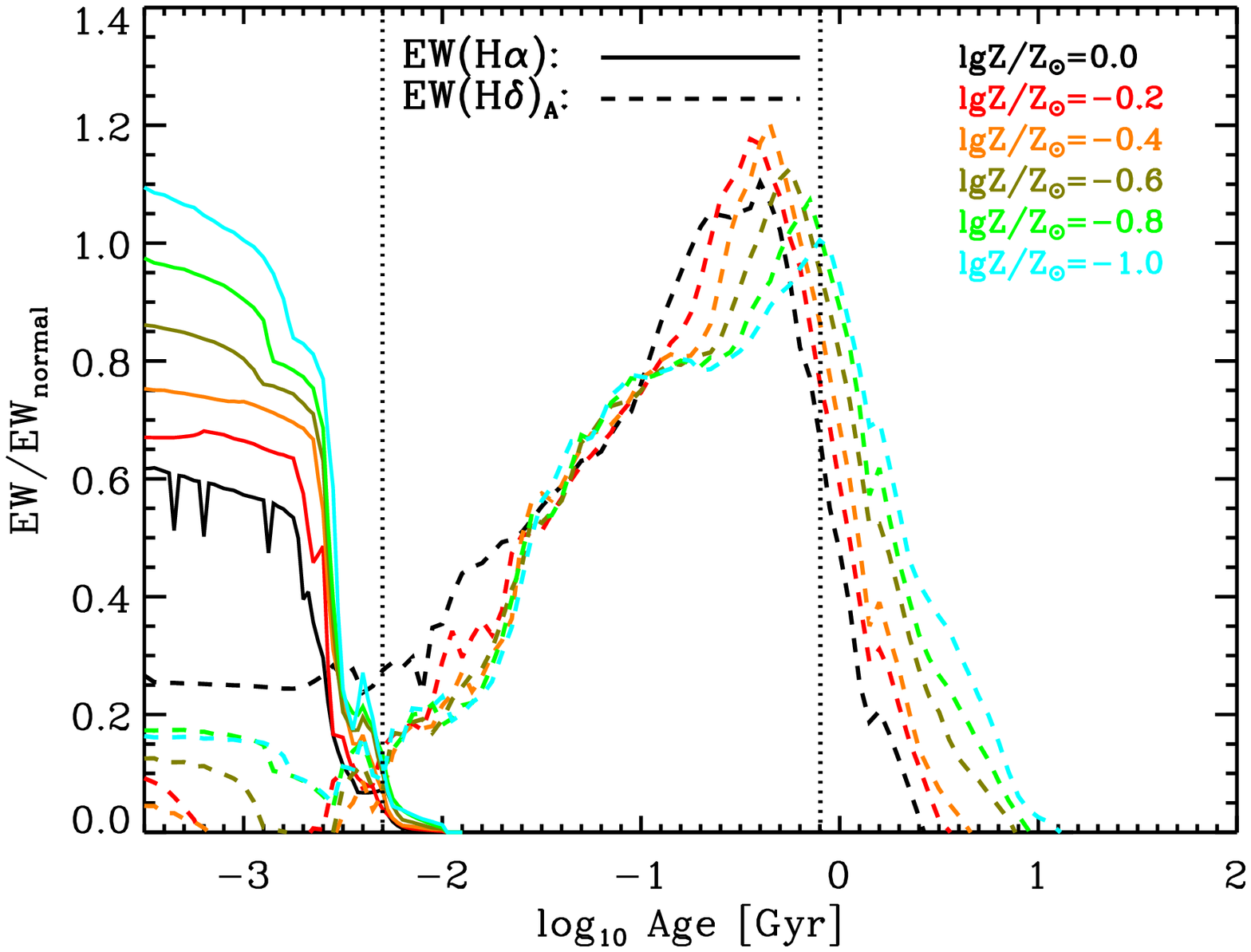,clip=true,width=0.48\textwidth} % size
    \epsfig{figure=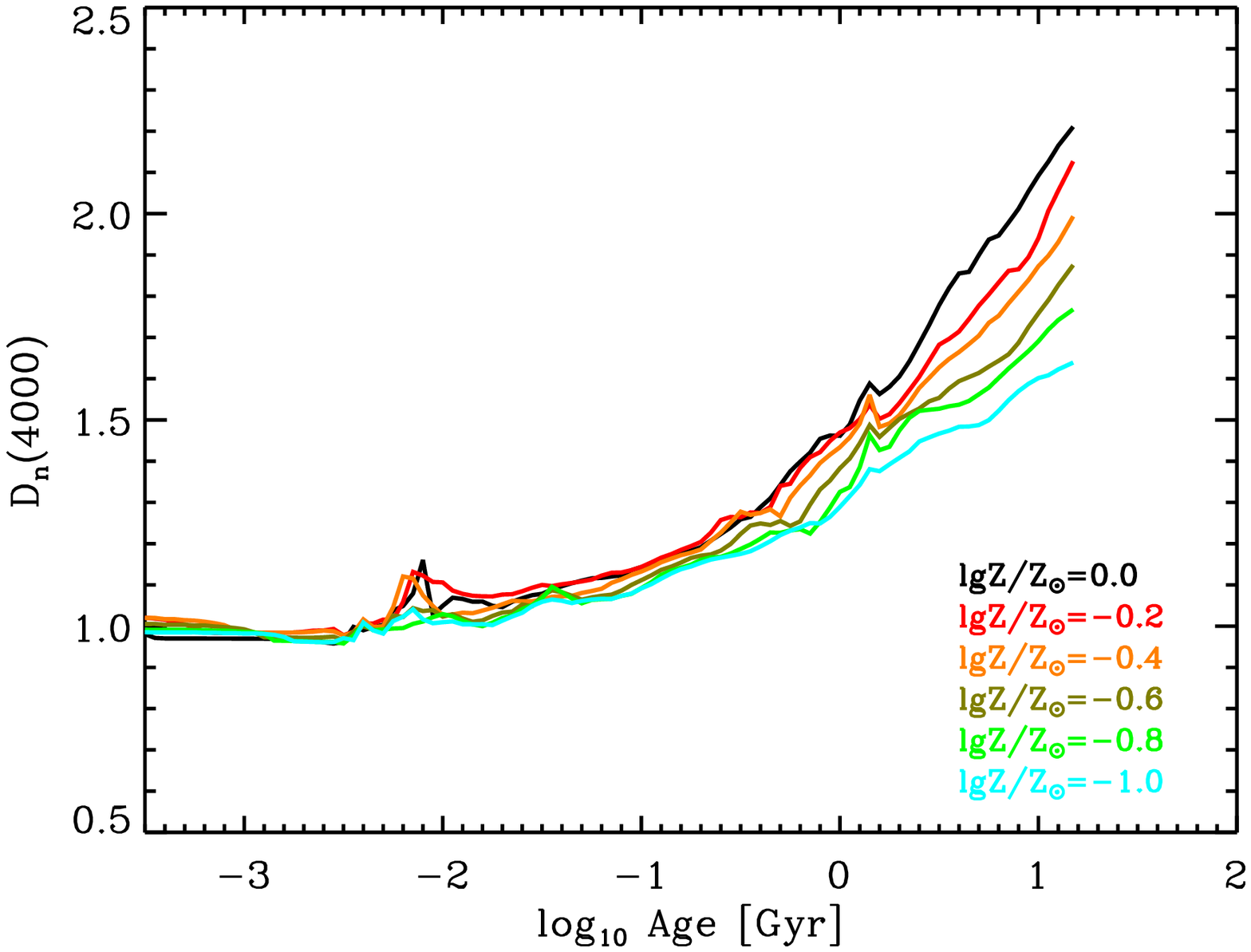,clip=true,width=0.48\textwidth} % bt 
    \end{center}
  \caption{Left panel: The scaled equivalent width of H$\alpha$
  emission (solid lines) and H$\delta$ absorption (dashed lines)
  lines as a function of the age of single stellar populations for six different metallicities. The \ewhae\ is scaled to 
10000 \AA, and the \ewhda\ is scaled to 10 \AA.
  The two vertical dotted lines represent the fiducial ages of 5 Myr and
  800 Myr, defining the windows of age traced by the H$\alpha$
  emission and H$\delta$ absorption lines. 
  Right panel: The 4000 \AA\ break as a function of the age of
  single stellar populations for the different metallicities. 
  %Different colors are for the SSPs
  %of different metallicities, as denoted in each panel.  
  }
  \label{fig:ha_hd}
\end{figure*}

The three diagnostic parameters have long been used to indicate the recent SFHs
on different timescales \citep[e.g.][]{Worthey-97, Balogh-99, Kauffmann-03, Li-15, Wang-17, Wang-18}.  
In SF galaxies, the H$\alpha$ emission mainly comes from the  recombination  of  
gas  ionized  by  photons from extremely
massive stars ($>$15\msolar), which is therefore expected to trace the SFHs within the lifetime 
of these massive stars ($\sim$5 Myr).  However, the \ewhda\ 
traces the recent star formation within a longer timescale of more like 1 Gyr.  The Balmer absorption lines 
arise from intermediate mass main-sequence stars with lifetimes of $\sim$1 Gyr.  They 
are relatively insensitive to the metal abundance because they depend mostly on the 
behavior of the main-sequence turn-off temperature rather than the behavior 
of the red giant branch temperature \citep{Worthey-94, Worthey-97}. 
Finally, the 4000 \AA\ break is determined by the SFHs on still longer timescales with respect to \ewhda,
and is found to be sensitive to the light-weighted stellar age \citep{Balogh-99}.

The evolution of these three diagnostic parameters for single 
stellar population (SSP) models of different metallicities can be seen 
by using the Flexible Stellar Population Synthesis code 
\citep[{\tt FSPS};][]{Conroy-09}. {\tt FSPS}\footnote{https://github.com/cconroy20/fsps} 
is a powerful code that can generate 
spectra and absolute magnitudes of arbitrary stellar populations, with a series of 
flexible settings, such as metallicity, choice of stellar library, different IMF, and different evolutionary isochrones. 
Throughout this work, we will adopt the {\tt MILES} stellar library 
\citep{Sanchez-Blazquez-06, Falcon-Barroso-11}, a 
\cite{Chabrier-03} IMF, and the {\tt Padova} isochrones \citep[e.g.][]{Bertelli-94, 
Bertelli-08}, unless specified otherwise. 

Nebular emission is produced by using the {\tt FSPS} implementation of the photoionization 
code, {\tt CLOUDY} \citep{Byler-17}.
By simulating physical conditions within a gas cloud, {\tt CLOUDY} predicts the thermal, ionization, and chemical structure of the cloud, and further produces the resultant spectrum of the diffuse emission \citep{Ferland-13}. In {\tt FSPS} model, the ionizing radiation is produced by a point source at the central of a spherical shell of cloud, with assuming a constant gas density of $n_{\rm H}=100 {\rm cm}^{-3}$. The fraction of the ionizing luminosity to escape from the HII region is assumed to be zero \citep{Byler-17}. 
%which assumes a constant-density spherical shell of gas surrounding the stellar population. 

For each of our mock SFH, we produce the current-epoch spectrum for six metallicities ($\log_{10}Z/Z_{\odot}=$0.0, $-$0.2, $-$0.4, 
$-$0.6, $-$0.8 and $-$1.0), without implementing any dust attenuation. This means that the 
diagnostic parameters in the calibration are assumed to be dust-free.  Being equivalent widths, the observed diagnostic parameters should, ideally, be independent of dust attenuation, but this will only be true if the nebular emission and stellar continuum have the same attenuation, which is unlikely to be the case.  There are likely be second-order effects if different stellar populations have different dust obscuration.
The correction of the observed diagnostic parameters for these second order dust effects
will be presented in Section \ref{subsec:2.5}.  Although we will only calibrate the \RSFR\ 
for the six discrete metallicities, we can obtain the \RSFR\ calibration for galaxies of other 
metallicities by linear interpolation in $\log_{10}Z/Z_{\odot}$ (see details in Section \ref{subsec:3.1}). 

Figure \ref{fig:ha_hd} shows the evolution of \ewhae, \ewhda\ and \dindex\ 
as a function of stellar age for SSP models at the six different metallicities.  
We present the evolution of \ewhae\ (solid lines) and \ewhda\ (dashed lines) 
on the left panel of Figure \ref{fig:ha_hd}.  For all the different metallicities, 
the \ewhae\ is scaled to 10000 \AA, and the \ewhda\ is scaled to 10 \AA.
The two vertical dotted lines represent the ages of 5 Myr and 800 Myr respectively.  
As expected, after a single burst of star formation, \ewhae\ is large at first, but
then quickly decays and becomes only 10\% of its maximum value after 5 Myr, for
all the six different metallicities.  In addition, \ewhae\ is higher at lower
metallicities. The dependence of \ewhae\ on metallicity comes from the nearly equal
contribution of the variation in stellar continuum and the variation in H$\alpha$ emission. 
%The evolution curves of \ewhae\ are quite similar for different metallicities of SSP models.

For an SSP model with given metallicity, \ewhda\ instead shows a peak at an age of 
a few hundred Myrs. The stellar population age corresponding to the peak \ewhda\ decreases with
metallicity.  Rather than having a different timescale for each metallicity, we choose instead a standard timescale of 800 Myr for all metallicities that enables the defined 
change parameter \RSFR\ to be reasonably well calibrated at all the six metallicities. This however makes the \RSFR\ calibration dependent on metallicity.
The \dindex\ increases with increasing stellar age at a given metallicity, 
and increases with increasing metallicity at given stellar age. 

Our definition of \RSFR\ is based on the SFR 
across two orders of magnitude in timescale, which is much larger than that of the 
widely-used ``burstiness" based on H$\alpha$-to-UV ratio. 

Further, as already noted the diagnostic parameters based on equivalent widths are insensitive 
to the dust attenuation and can be readily measured from the existing 
large body of optical spectra of galaxies, and are not strongly model-dependent.  
These conditions make the three diagnostic parameters to be an ideal choice
to study the variability of SFR in galaxies. 

\subsection{Construction of the mock SFHs}
\label{subsec:2.2}

\begin{figure*}
  \begin{center}
    \epsfig{figure=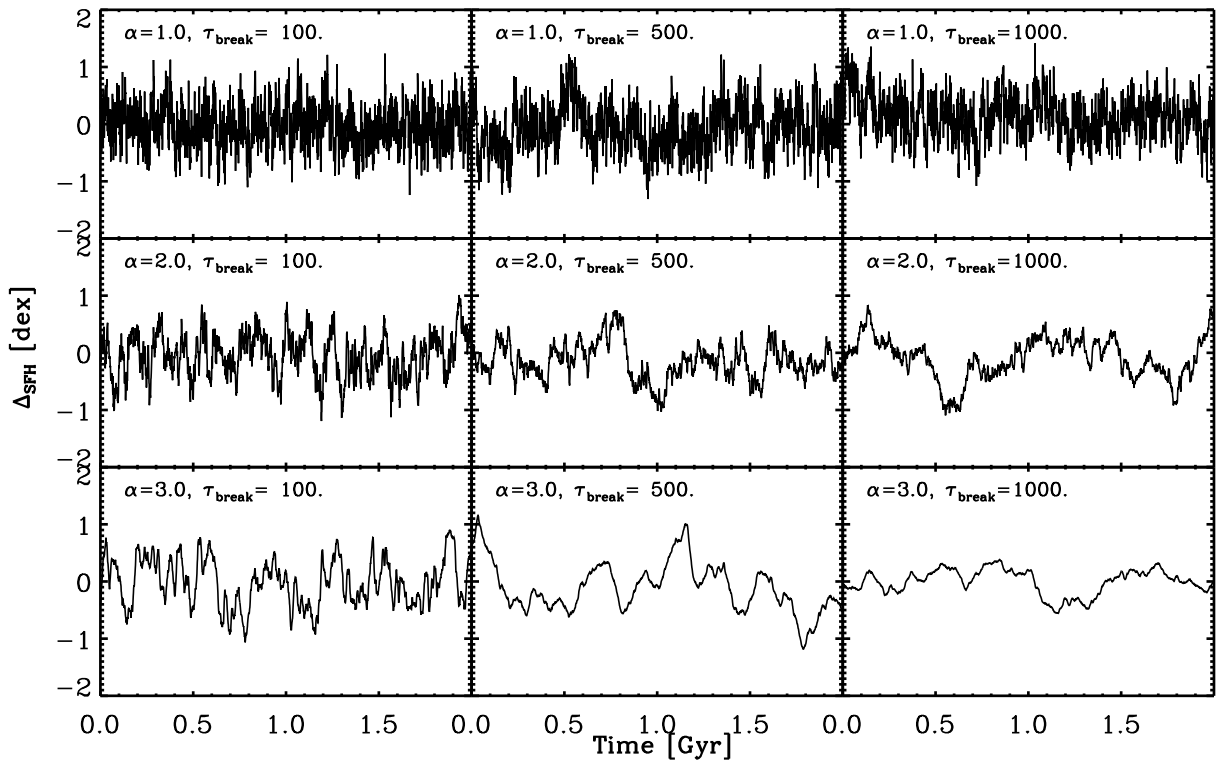,clip=true,width=0.99\textwidth} % size
    \end{center}
  \caption{Illustrating the range of variation in the SFHs 
  %with a series of different
  %stochastic processes. 
 that are used in the calibration of \RSFR.
 Each column of panels has the same
  $\tau_{\rm break}$, from left to right: 100 Myr,
  500 Myr and 1000 Myr.  Each row of panels has the same $\alpha$, from 
  top to bottom: 1.0, 2.0 and 3.0.}
  \label{fig:sfh_example}
\end{figure*}

In parametric SED modelling, strong priors are usually imposed on the SFHs.
One of the widely used ones is the exponentially declining SFH \citep[e.g.][]{
Bruzual-A.-83, Papovich-01, Shapley-05, Pozzetti-10, Carnall-19}, 
i.e. the SFR is assumed to decline exponentially with some e-fold timescale $\tau$: 
${\rm SFH}(t) \propto \exp(-t/\tau)$. However, it is clear in the real Universe, that the SFH of galaxies may be
much more complicated than any assumed analytic formula.  
Motivated by the fact that the global SFD is well fit by a log-normal in time, \cite{Gladders-13} 
proposed that the log-normal form might also characterize the SFHs of individual galaxies 
\citep{Dressler-13, Oemler-13, Abramson-15}.  
Using the Illustris simulation, \cite{Diemer-17} investigated the SFHs for individual galaxies, and 
found that the log-normal form fits the overall shape of the majority of SFHs very well: 85\% of 
cumulative SFHs are fitted to within a maximum error of 5\% of the total stellar mass formed. 
The log-normal works systematically better than the commonly used exponentially declining model, and 
appears to be a reasonably good description for the global shape of SFHs for individual galaxies. 
Therefore we adopt the log-normal fits of the SFHs for Illustris galaxies \citep{Diemer-17} as being representative 
of the global shape for the long-term variation of SFHs in the real Universe. 

On top of these smoothly varying underlying SFH must be added short-term stochastic variations. 
We describe the stochastic variations in SFR in the frequency (or time) domain using
the power spectrum distribution (PSD) of variations in the SFR.  
To construct the mock SFH, 
we use, for simplicity and following the work of \cite{Caplar-19}, a broken power-law PSD
to characterize the possible variations in SFR that are superposed on the broad underlying log-normal SFH. 
The PSD can be written as: 
\begin{equation}
{\rm PSD}(\nu) = \frac{\sigma^2}{1+(\tau_{\rm break}\nu)^\alpha}, 
\end{equation}
where $\nu$ is the frequency, $\sigma$ defines the amplitude of the PSD, $\alpha$ 
is the slope of PSD at the high frequency end, and $\tau_{\rm break}$ defines 
the break point where the PSD becomes flat towards lower frequency. We refer 
the reader to \cite{Caplar-19} for more details of the properties of this kind of PSDs. 

We then use a public IDL 
code\footnote{https://github.com/svdataman/IDL/tree/master/src} 
to generate the random time series of variation in SFR
with a given power spectrum distribution. 
Note that the variations of SFR are generated in logarithmic space. 
Figure \ref{fig:sfh_example} shows examples of the variations of stochastic 
components with different $\alpha$ and $\tau_{\rm break}$. Here we only
show the variations with a time range of 2 Gyr, while in practise we 
generate the variation in time series over the full lifetime of the Universe. 
Following the work of \cite{Caplar-19}, the 1$\sigma$ scatter of the variations 
are normalized to 0.4 dex, which is comparable to
the maximum scatter of SFMS in the observations  \citep[e.g.][]{Whitaker-12,Speagle-14,Schreiber-15,Davies-19}. The time resolution 
is set to be 1 Myr, which is much smaller than the SFH timescale 
traced by H$\alpha$, and smaller than the free-fall timescale of 
molecular clouds \citep{Murray-10, Hollyhead-15, Freeman-17}. 
As shown in Figure \ref{fig:sfh_example}, larger $\alpha$ and 
longer $\tau_{\rm break}$ result in slower oscillations and stronger
correlation in time domain.  

In total, there are 29,203 galaxies in the Illustris simulation 
with stellar mass greater than $10^9$\msolar\ \citep{Diemer-17}. 
We do not yet exclude the quenched galaxies, but will do so 
later in Section \ref{subsec:2.3} based on the \ewhae\ and \ewhda\
of the mock spectra. 
For each of these 29,203 galaxies, with a given underlying log-normal SFH, 
we then construct 100 stochastic variations of the SFH by varying both
$\alpha$ and $\tau_{\rm break}$ in logarithmic space, $\Delta_{\rm SFH}$ (see examples in Figure \ref{fig:sfh_example}). The mock SFHs are then constructed by multiplying the log-normal SFH from Illustris galaxies with a factor 10$^{\Delta_{\rm SFH}}$. We therefore have 2,920,300 
mock SFHs in total.  
In practise, we make a 10$\times$10 grid for $\alpha$ and 
$\tau_{\rm break}$, where the two parameters are evenly 
spaced with $\alpha$ in the range of 1 to 3, 
and $\tau_{\rm break}$ in the range of 100 Myr to 1000 Myr. 
The ranges of the two parameters are chosen according to the result of \cite{Caplar-19}. Further, we find that the constrained slope of PSD is within this range in the second paper of this series. 
For each point on this grid, we generate a stochastic variation of SFH
based on its $\alpha$ and $\tau_{\rm break}$ according to the 
approach above. We stress that our propose is not to try to model or reproduce the stochastic 
variation in SFHs in the real Universe, but is instead to simply generate a huge range 
of SFHs, which should cover the range of \RSFR\ that will be encountered in normal galaxies in the real Universe\footnote{We will attempt to constrain the 
PSD of specific SFHs in the second paper of this series. 
We find that the constrained slope of PSD is $\sim$1.5 with assuming 
no intrinsic scatter of SFMS. This indicates that the constructed mock SFHs 
covers the cases of the \RSFRL\ in the real Universe.}.   

\subsection{Calibration of SFR79}
\label{subsec:2.3}

For each of the 2.9 million mock SFHs constructed above, we calculate the change parameter \RSFR\ at the present epoch, i.e. the simple ratio of the SFR averaged over the last 5 Myr to that averaged over the last 800Myr. 
Using the mock SFH and the SSP models, we can obtain the mock spectrum of the composite stellar population produced by each mock SFH by convolving the time varying spectrum of the SSP (at a given metallicity) with the detailed age distribution of each mock SFH.  We can further compute the three diagnostic spectral parameters for each mock spectrum at the present epoch, for each of the six metallicities. 

In practise, we do not of course need to produce an entire high resolution composite spectrum but simply measure the relevant input fluxes (or flux deficits) of the SSP once as a function of age from its evolving spectrum and then produce the diagnostic parameters for each mock SFH at the present epoch through a straight convolution of these functions with the age distribution of the SFH in question.  For instance, the \dindex\ is defined as the ratio
of the flux density between the 4000 and 4100 \AA\ ($f_{\rm red}$) and that 
between 3850 and 3950 \AA\ ($f_{\rm blue}$) \citep{Balogh-99}.  
We first compute the evolution of $f_{\rm red}(t)$ and $f_{\rm blue}(t)$ with time
for a SSP at a given metallicity. Then the $f_{\rm red}$ (or $f_{\rm blue}$) for a given
mock SFH is the convolution of the corresponding age distribution $n(t)$ with the $f_{\rm red}(t)$
(or $f_{\rm blue}(t)$)
evolution curve. In the similar way, the \ewhae\ 
and \ewhda\ can also easily be obtained. 
The bandpasses for calculating \dindex\ and \ewhda\ are defined in \cite{Balogh-99}. Specifically, the blue and red bandpass of wavelength (in \AA) in calculating the \dindex\ are [3850,3950] and [4000,4100]. The three bandpasses for the index of H$\delta$ absorption are [4083.50, 4122.25], [4041.60, 4079.75], and [4128.50, 4161.00].
In calculating emission (or absorption) line flux of H$\alpha$ (or H$\delta$), the contamination of the absorption (or emission) in H$\alpha$ (or H$\delta$) is corrected. We note that the approach in calculating the three diagnostic parameters for the mock spectra is exactly the same as that used in analysis of the observations in Section \ref{subsec:3.1} \cite[also see][]{Wang-18}.   

\begin{figure*}
  \begin{center}
    \epsfig{figure=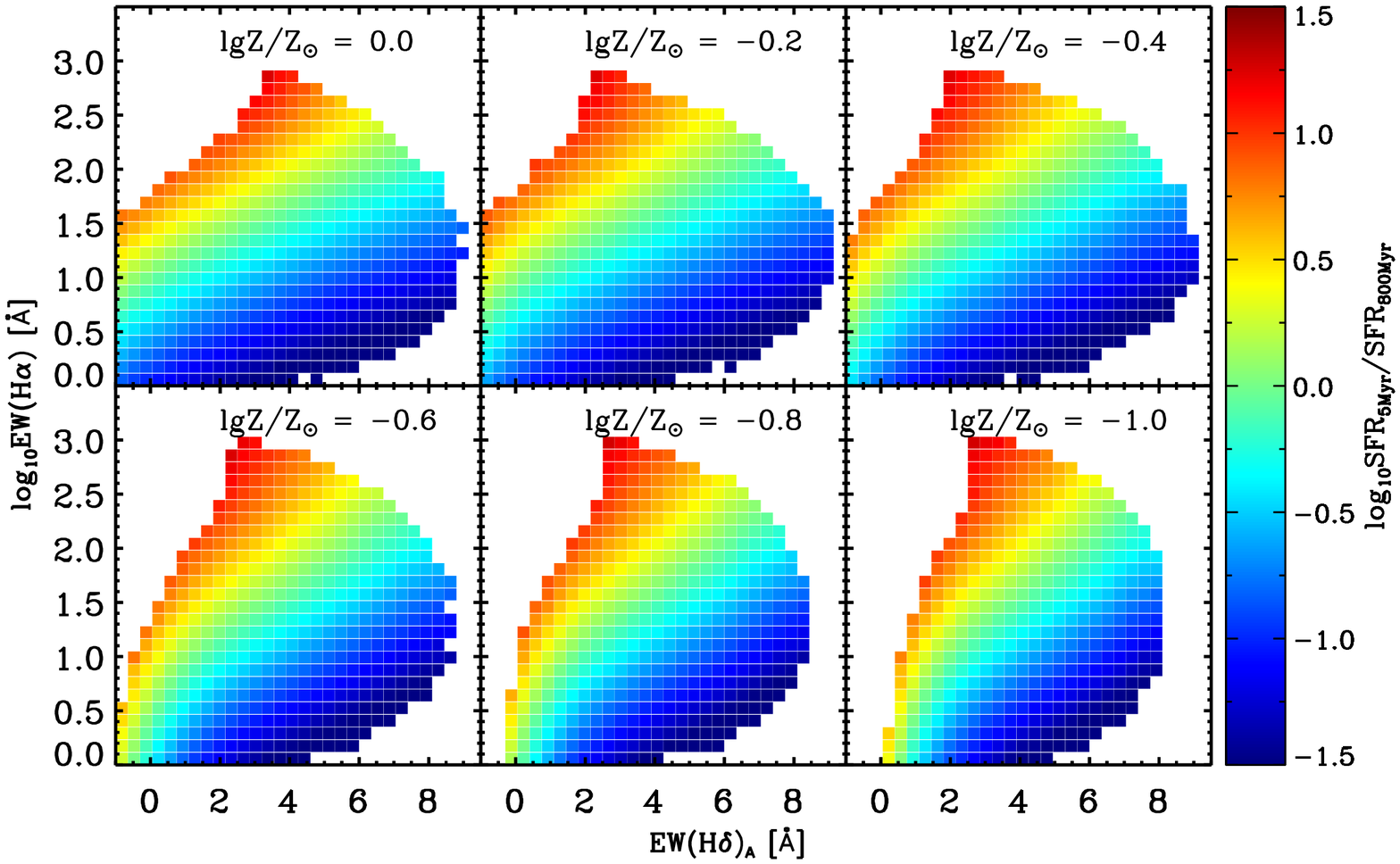,clip=true,width=0.88\textwidth} % size
    \end{center}
  \caption{The calibration of the \RSFR\ estimator in terms of \ewhae\ and \ewhda.  The panels show the mean \RSFR\ as a function of the observed
  \lgewhae\ vs. \ewhda\ for the six different metallicities.  Any effects of any differential reddening are not included in this calibration.  Each panel represents the outcome of 2.9 million mock SFHs, as described in the text. The actual calibration used in this paper (Equation \ref{eq:2} and Table \ref{tab:1}) includes also the \dindex. }
  \label{fig:rsfr_ha_hd}
\end{figure*}

\subsubsection{\RSFR\ as a function of $EW(H\alpha)$ and $EW(H\delta_A)$} 
\label{subsec:2.3.1}

\begin{table*}[ht]
\renewcommand\arraystretch{1.5}
\begin{center}
\caption{The fitting parameters of the calibrator at different metallicities in Figure \ref{fig:rsfr_fit} \label{tab:1}}
\begin{tabular}{@{}lrrrrrrrrrrrrr@{}}
\tableline
\tableline

$\log Z/Z_{\odot}$                 & a1    &  a2 &  a3 & {}  & b1  & b2 & b3 & {}  &  c1  &   d  &  scatter  & \ewhae &\ewhda \\
\tableline
    0.0                 &   1.082  &   $-$0.06909 & $-$0.01134 & {} &  $-$0.2922   &  0.01167  & $-$5.999e-05  & {} &   $-$1.268   &   1.126  & 0.063  & $>$1.0\AA & $>-$1.0\AA  \\
   $-$0.2               &  1.014  &   0.006220 &  $-$0.03168 & {} &    $-$0.2958  &   0.01855  & $-$7.570e-04 & {} &   $-$0.8483  &   0.5110 & 0.065 & $>$1.0\AA & $>-$1.0\AA   \\
   $-$0.4               &  1.006  &   0.04190 & $-$0.03518 & {} &    $-$0.2576  &  0.01338  & $-$5.561e-04 & {} &   $-$0.3489  &   $-$0.3272 & 0.062 & $>$1.0\AA &  $>$0.0\AA  \\
   $-$0.6               &  0.9310  &  0.1042  & $-$0.04182 & {} &   $-$0.3572  &  0.03650  & $-$0.002370 & {} &    0.1615    &  $-$0.8405 & 0.071  & $>$1.0\AA &   $>$0.5\AA  \\
   $-$0.8               &  0.9319  &  0.1067  & $-$0.03414 & {} &    $-$0.4048  &  0.04930  & $-$0.003331 & {} &   0.7171   &   $-$1.543 & 0.080 & $>$1.0\AA &  $>$1.0\AA  \\
   $-$1.0               &  0.8920 &   0.1293  & $-$0.03305 & {} &   $-$0.4694  &  0.06127   & $-$0.004017 & {} &    1.244    &  $-$2.091 & 0.090  & $>$1.0\AA &  $>$1.5\AA   \\
\tableline
\tableline
\end{tabular}
\end{center}
\end{table*}

\begin{figure*}
  \begin{center}
    \epsfig{figure= 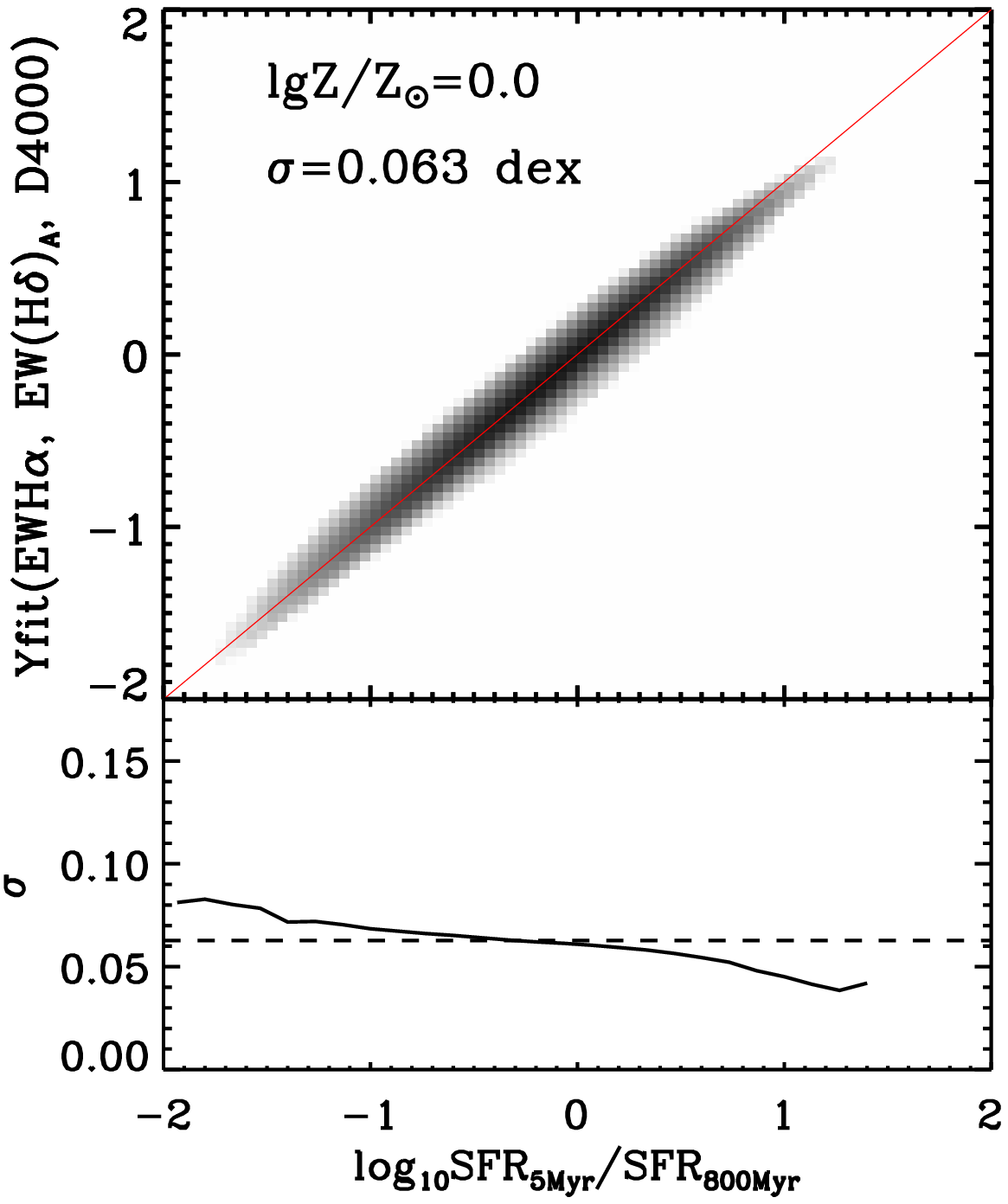,clip=true,width=0.28\textwidth} % size
    \epsfig{figure=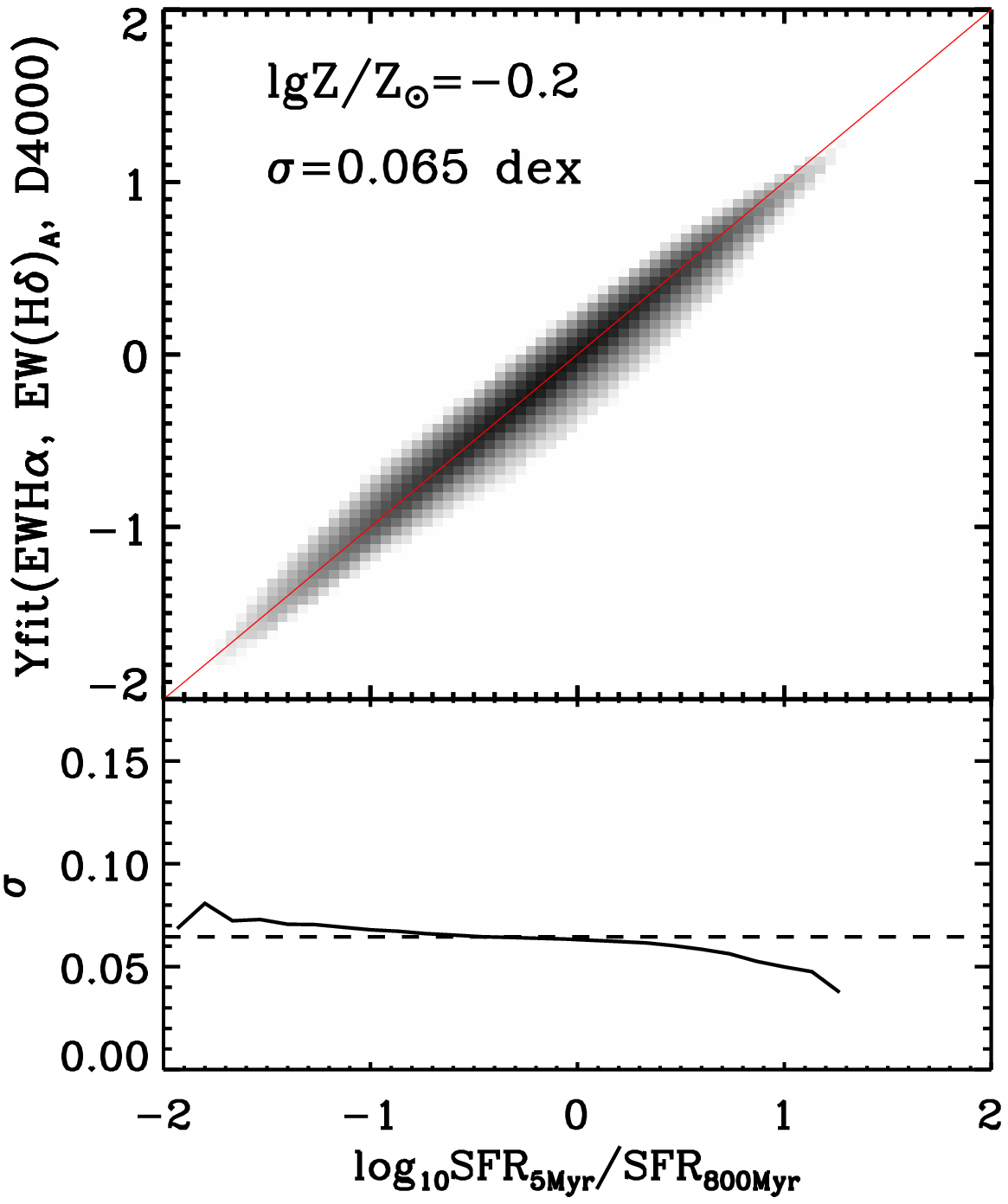,clip=true,width=0.28\textwidth} % size
    \epsfig{figure=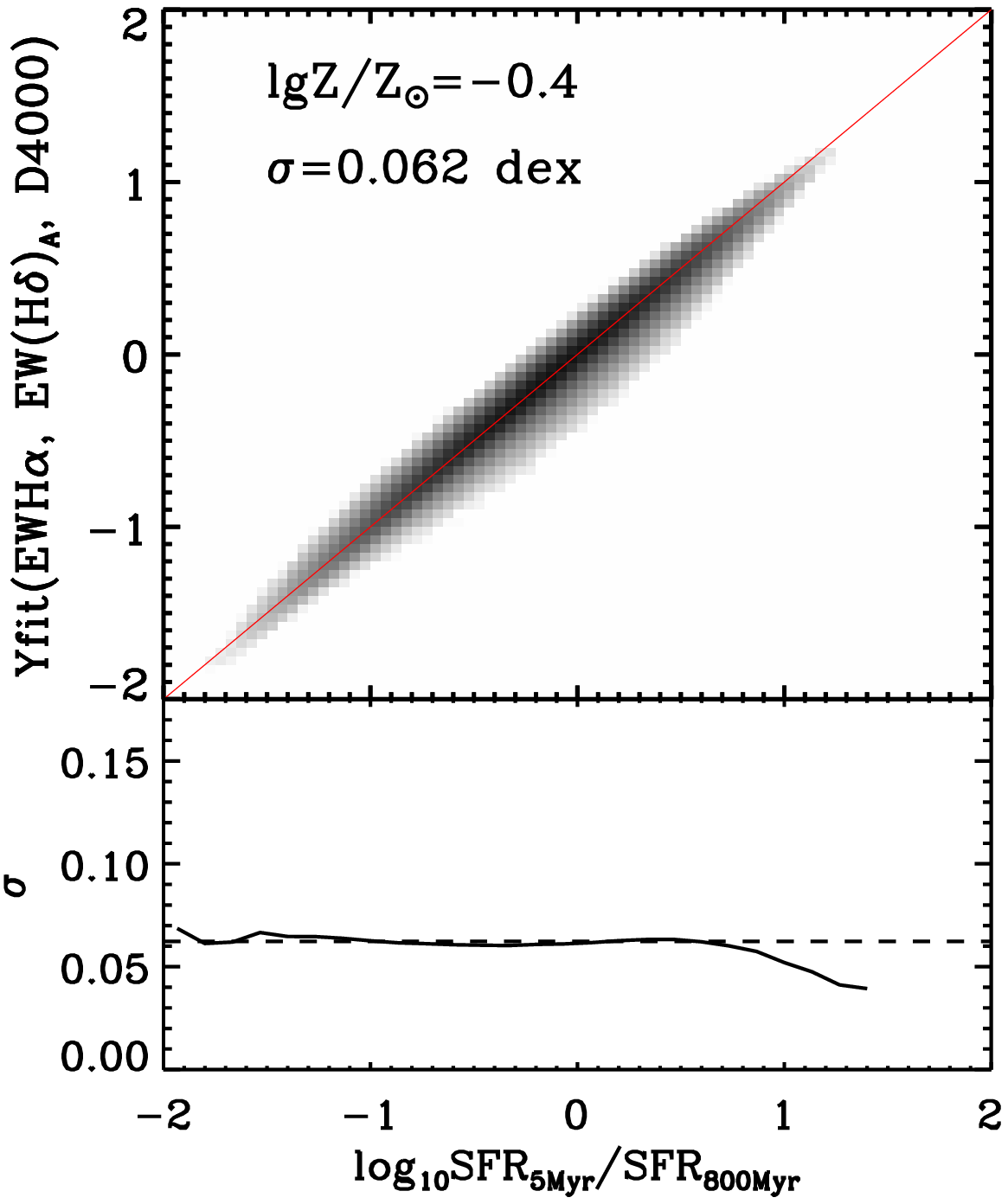,clip=true,width=0.28\textwidth} % size
    \epsfig{figure=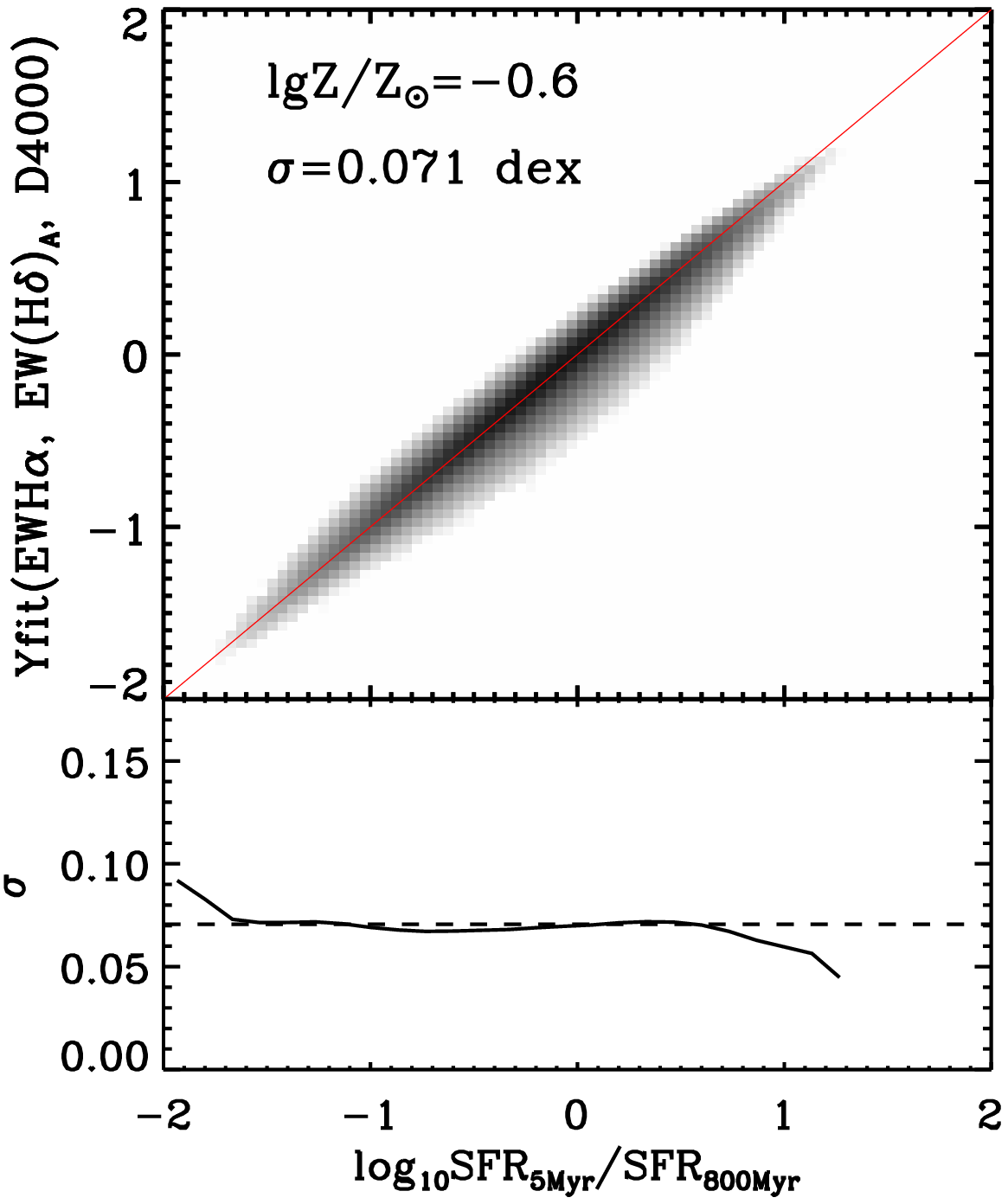,clip=true,width=0.28\textwidth} % size
    \epsfig{figure=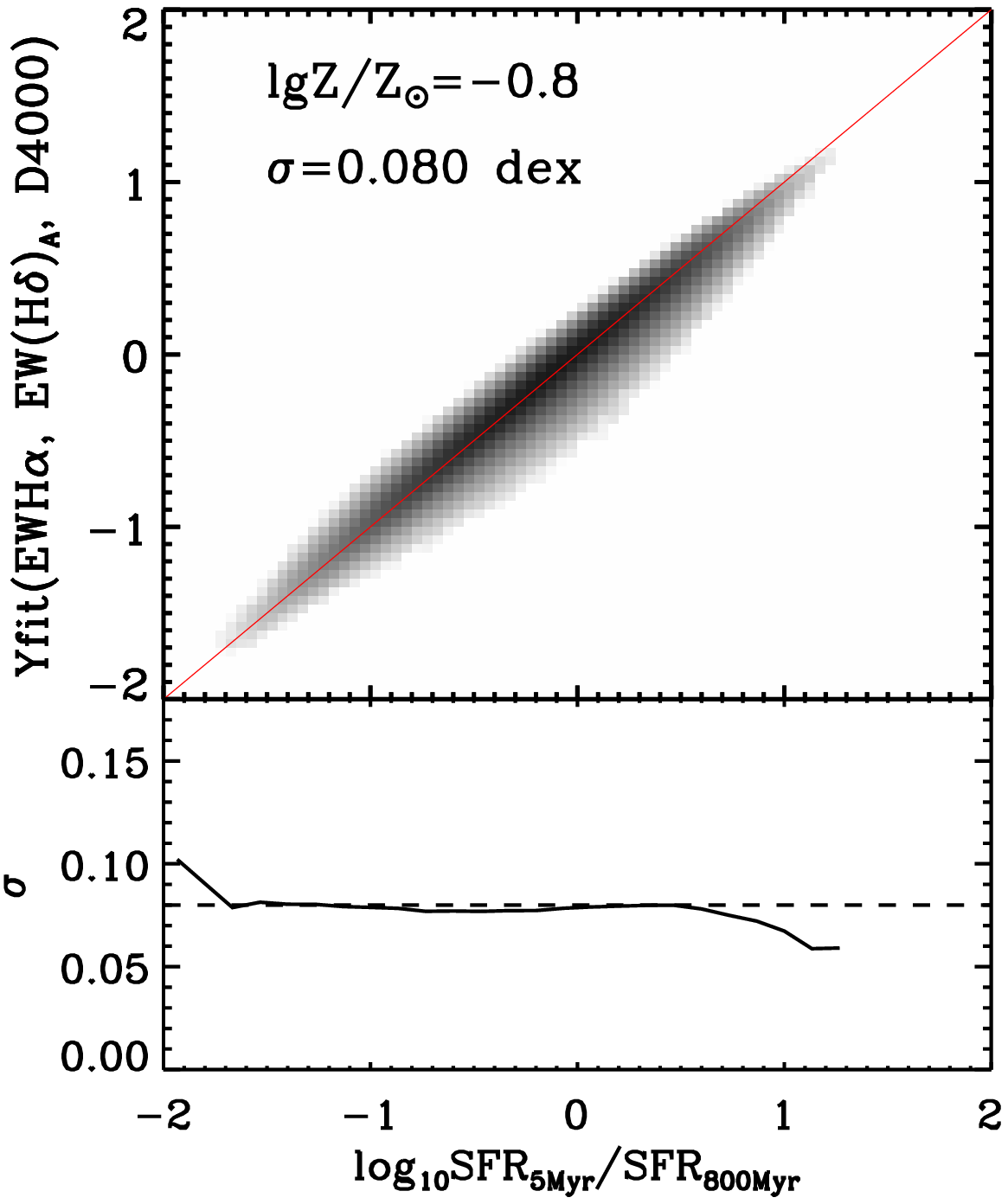,clip=true,width=0.28\textwidth} % size
    \epsfig{figure=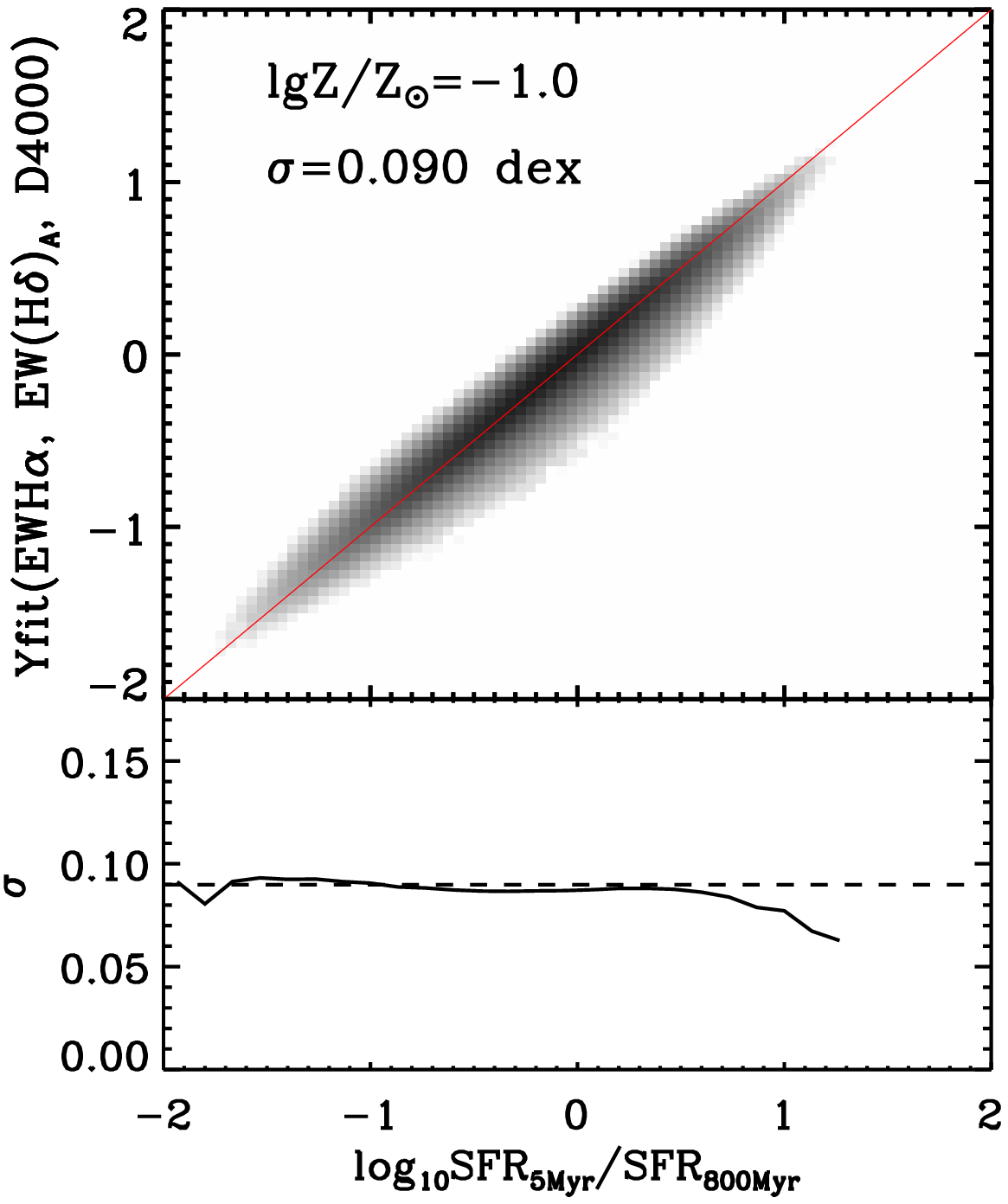,clip=true,width=0.28\textwidth} % size
    \end{center}
  \caption{ Comparison of the \RSFR\ that is derived from the combination of \lgewhae, \ewhda, 
  and \dindex\ with the actual \RSFR\ as directly calculated from the SFH for the 2.9 million mock galaxies, for six different metallicities. The lower panels in each figure, show the dispersion between the fitted \RSFR\ and the
  real \RSFR\ as a function of the real \RSFR.  This is largely independent of \RSFR, but increases somewhat to lower metallicities because of the variation of \ewhda(t) shown in Figure \ref{fig:ha_hd}.
  %In the upper sub-panel of each panel, the grayscale indicate the number 
  %density of mock galaxies. In the lower sub-panel of each panel, the solid line
  %show the scatter as a function of \RSFR, while the  horizontal
  %dashed lines show the median scatter, which are also denoted in the top 
  %of the upper sub-panel. As above, different panels represent the 
  %relation for different metallicities, as denoted in the bottom corner of the
  %upper sub-panel.
}
  \label{fig:rsfr_fit}
\end{figure*}

Now that we have the measurements of SFR79 as well as the diagnostic parameters for 
millions of mock SFHs, it is straightforward to search for the solution of \RSFR\ as a function of the three 
diagnostic parameters. 
Figure \ref{fig:rsfr_ha_hd} shows the \lgewhae\ vs. \ewhda\ diagram with the color-coding 
of \RSFR\ for the six different metallicities. The \RSFR\ shows clear 
gradients on this diagram, confirming that the \ewhda\ and \lgewhae\ indeed contain
information on the change parameter. 
For all metallicities, at fixed \ewhda, the \RSFR\ 
increases with increasing \lgewhae; at fixed \lgewhae, the \RSFR\ decreases with 
increasing \ewhda.  This is as expected, since the two parameters
indicate the strength of star formation at two different timescales.  
Another interesting feature is that the range of \ewhda\ becomes smaller towards
smaller metallicity. This is due to the different evolution curves of \ewhda\ for 
the SSP models at the different metallicities (see Figure \ref{fig:ha_hd}). For the lowest
metallicity ($\log_{10}Z/Z_{\odot}=-1.0$), the \ewhda\ of the SSP is greater than zero over 
the entire age range of 10 Gyr after a single starburst. This is the reason why there is 
no data points with \ewhda\ below zero at the lowest metallicity.  Note that \ewhda\ is here defined to be positive for absorption.

After exploring many kinds of combination of the three parameters, we find that a 
combination of polynomials to the third-order can well reproduce the values of \RSFR\, to within a scatter of 0.06-0.09 dex.  The form of the polynomials can 
be written as:
\begin{equation} \label{eq:2}
\begin{split}
\log_{10}{\rm SFR79} = &a1\times x+a2\times x^2 +a3\times x^3 \\
                      &+b1\times y + b2\times y^2 + b3\times y^3 \\
                      &+c1\times z+d, \\
\end{split}
\end{equation}
where $x =$ \lgewhae, $y =$ \ewhda, $z =$ \dindex, and $a1$, $a2$, $a3$, $b1$, $b2$,
$b3$, $c1$ and $d$ are parameters determined from the fittings. The fitting parameters
for different metallicities are listed in Table \ref{tab:1}.  
During the fittings, we exclude mock spectra with extremely 
low \ewhae\ and \ewhda, since these are not encountered in the star-forming galaxies that are of interest in this work. The exclusion thresholds of these two parameters at different metallicities 
are listed in the last two column of Table \ref{tab:1}. 
As shown in Table \ref{tab:1},  the threshold of \ewhae\ is 1\AA\ for all metallicities, while the threshold of \ewhda\ increases with decreasing metallicity, varying from $-$1\AA\ to 1.5\AA.
The exclusion of quenched galaxies with very small \ewhae\ and \ewhda\ is immaterial for the present purposes.  
We also note that very small values of these two parameters would anyway be associated
with relatively large observational uncertainties.  

Figure \ref{fig:rsfr_fit} compares the \RSFR\ as directly measured from the 
mock SFHs with the results obtained from Equation \ref{eq:2} using the 
three diagnostic parameters. As shown, even though we included a huge range 
of possibilities in the mock SFH construction, the \RSFR\ can be very well 
determined by the combination of these three diagnostic parameters with a 
scatter of 0.06-0.09 dex. The scatter shows 
very little dependence on \RSFR\ for almost all the metallicities examined. 
Another interesting feature is that the scatter becomes larger 
with decreasing metallicity. This is due to the fact that we use a fixed
timescale of 800 Myr to define the change parameter for all the metallicities.
In principle, we could have increased the averaging timescale to $\sim$1 Gyr to 
reduce the scatter in the calibrator at low metallicities. However, this variable timescale 
would introduce more complexity in analyzing the results in Section \ref{sec:4} and \ref{sec:5}.
We therefore decided to keep the timescale constant for the different
metallicities. The chosen timescale of 800 Myr was selected to minimize 
the scatter of calibrators at the three highest metallicities 
($\log_{10} Z/Z_{\odot}=$0.0, $-$0.2, and $-$0.4), 
in which most of our sample galaxies are in fact located (see Section \ref{subsec:2.3.2}). 

\subsubsection{The performance of the calibrator} 
\label{subsec:2.3.2}

\begin{figure*}
  \begin{center}
    \epsfig{figure=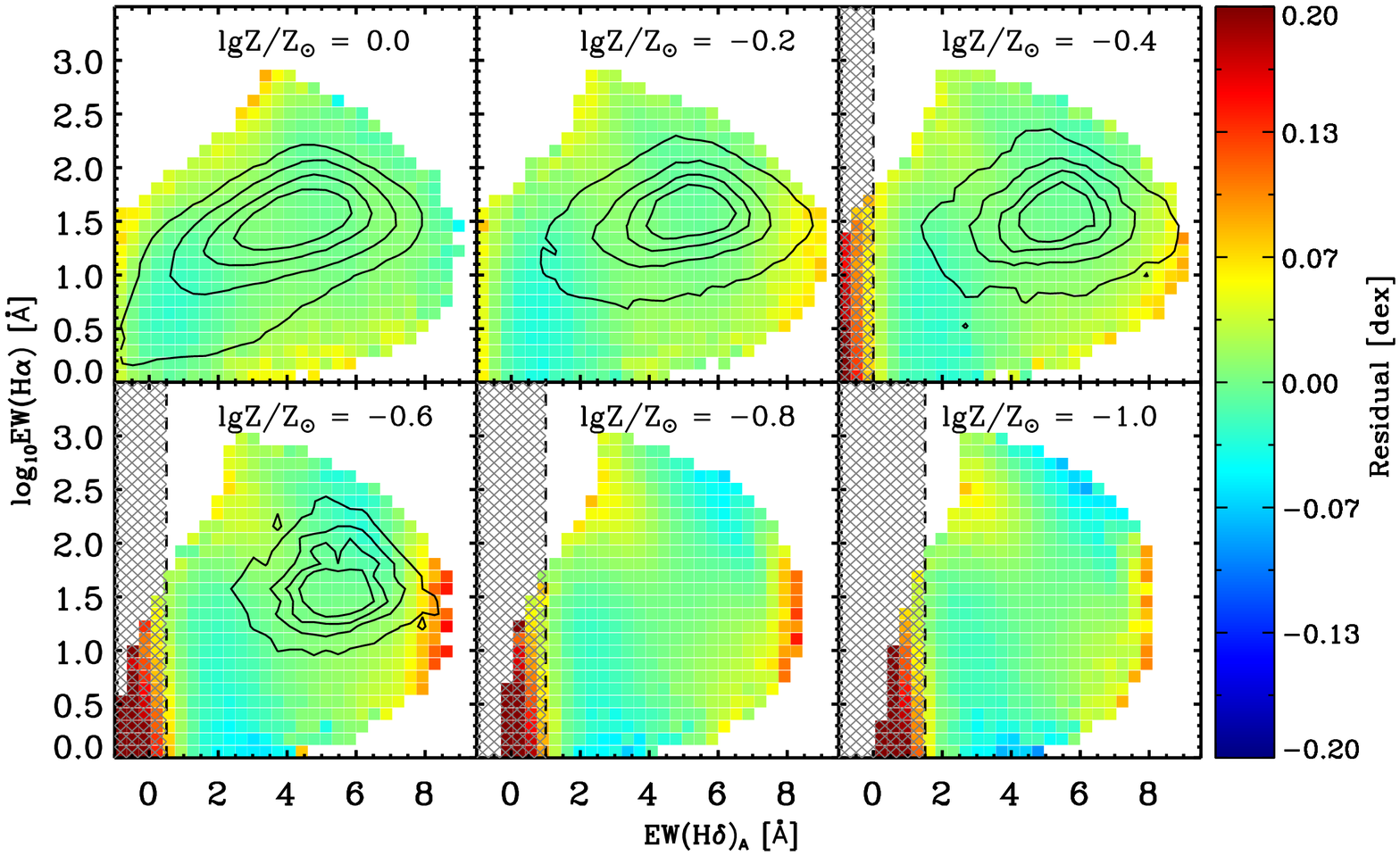,clip=true,width=0.88\textwidth} % size
    \epsfig{figure=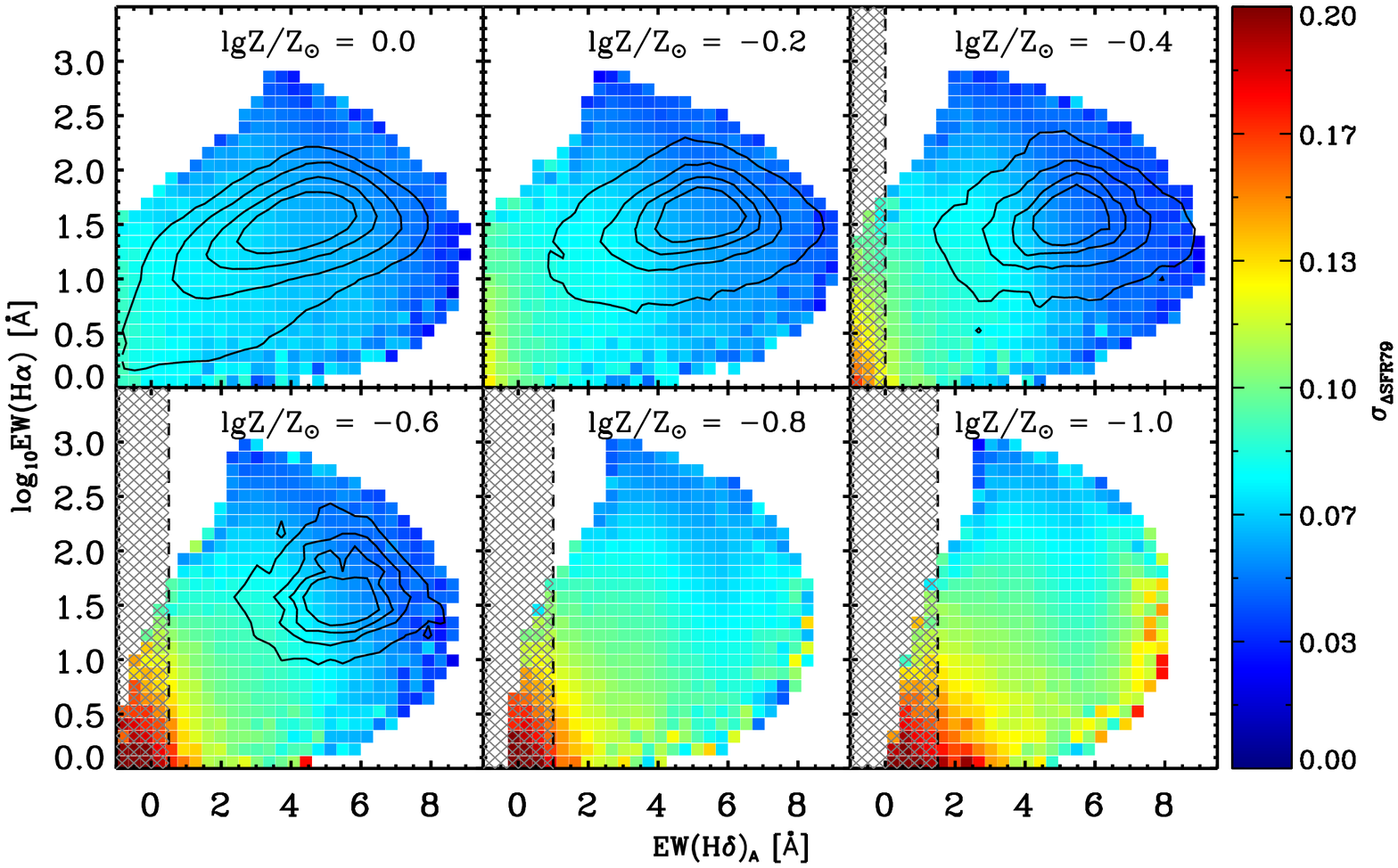,clip=true,width=0.88\textwidth} % size
    \end{center}
  \caption{The mean offset (upper panels) and dispersion (lower panels) in the recovered \RSFR\ compared with the true \RSFR\ across the \lgewhae\ vs. \ewhda\ diagram for the mock galaxies at the six metallicities. The recovered SFR79 is calculated based on Equation \ref{eq:2}. This means that the \dindex\ is also included in calculating SFR79, although here we only present the dependence of \ewhae\ and \ewhda. 
  %color-coding of the residuals (top group of panels) and scatters in $\Delta$\RSFR\ 
  %(bottom group of panels), i.e. the difference 
  %between the fitted \RSFR\ and the \RSFR\ measured from the mock SFHs. Different 
  %panels are for different metallicities, as denoted in the top of each panel.
  The shaded regions are considered to be beyond the valid regions of the calibrator %on 
  %\lgewhae\ vs. \ewhda\ diagram 
  (see also Table \ref{tab:1}). In each panel, the black 
  contours show the distribution of the spaxels from the MaNGA SF galaxies 
  of the corresponding metallicity. The observed \ewhae\ and \ewhda\ of MaNGA spaxels are corrected
  for the dust attenuation according to the approach described in Section \ref{subsec:2.5}. 
  The black contours enclose 30\%, 50\%, 70\% and 90\% 
  of spaxels from the inside outwards.}
  \label{fig:residual}
\end{figure*}

We showed in the above subsection that \RSFR\ could be calibrated with an
overall uncertainty of 0.06-0.09 dex.  In this subsection, we examine the performance 
of the calibrator in the \lgewhae\ vs. \ewhda\ diagram, to examine the performance of the \RSFR\
estimator in different regions of the diagram. 

We show the \lgewhae\ vs. \ewhda\ diagram for the mock SFHs with the color-coding
of the $\Delta$\RSFR\ in the top group of panels in Figure \ref{fig:residual}. 
The $\Delta$\RSFR\ is defined as the difference between the \RSFR\ from 
Equation \ref{eq:2} and the true \RSFR\ in the mock SFHs. 
At each metallicity, we also present the distribution of the spaxels from the 
MaNGA SF galaxies (of the corresponding metallicity, with bins of width of 0.2 dex) 
on the \lgewhae\ vs. \ewhda\ diagram, shown in black contours. 
The stellar metallicity of the sample galaxies is taken from the empirical mass-metallicity 
relation from \cite{Zahid-17} (see Equation \ref{eq:4} in Section \ref{subsec:3.1}). 
Most of the sample galaxies are in the three highest metallicity bins, and there are no
contours in the two lowest metallicity bins. This is because the number of spaxels 
in the two lowest metallicity bins are quite limited (only one galaxy in the sample has a metallicity in each of these two lowest bins). 

As shown, the calibration formula can indeed give an excellent estimation 
of \RSFR. However, we note that the estimator does not
work well when \ewhda\ and \ewhae\ lie beyond the threshold criteria of the 
calibrator (see the shaded regions of Figure \ref{fig:residual} and also Table \ref{tab:1}). 
In addition, although our mock SFHs cover a very wide range on the diagram 
of \lgewhae\ vs. \ewhda, some regions of parameter space are still not covered (the white regions
in Figure \ref{fig:residual}). The calibration polynomial is clearly 
not valid for the data points 
that are beyond the colored regions. 
This limitation does not affect our application to MaNGA galaxies, 
because almost all the spaxels of the SF galaxies in MaNGA are within the 
regions of validity for the estimator (see the black contours). 

At the three lowest metallicities, the estimator appears to systematically underestimate the 
\RSFR\ by up to $\sim$0.1 dex at the high \ewhda\ end 
(see yellow and red colors at high \ewhda\ in Figure \ref{fig:residual}).  
This may be due to the fact that at the edge of the parameter space, the number 
density of mock SFH is relative low, which contribute a small weight 
during the fitting.  
However, this systematic deviation in the calibrator 
is not a problem, since it can be easily corrected based on the position 
of the \lgewhae-\ewhda\ diagram in the application.  We perform this correction 
in applying the calibrator to MaNGA galaxies. We note that this correction
is very minor, since in the observation almost all the data points from MaNGA 
located in the regions that the calibrator operates rather well. 

In principle, we could of course simply establish a look-up table for SFR79 over the whole range of the 
three diagnostic parameters, so as to avoid the above correction. However, in this work,
we prefer to present a simple empirical formula of SFR79 which can be used by readers. 

We show the scatter in $\Delta$\RSFR\ on the \lgewhae\ vs. \ewhda\ diagram
in the bottom group of panels in Figure \ref{fig:residual}. As a whole, 
for all the metallicities explored, the scatter of the calibrator is small at the high 
end of both \lgewhae\ and \ewhda, and increases towards the lower end of 
\lgewhae\ and \ewhda.  This may be due to the fact that for high 
\lgewhae\ and \ewhda\ (corresponding to high recent SFRs),
the contribution from the older stellar populations in the 
measurement of these two parameters is correspondingly small. 
With increasing metallicity, the scatter decreases 
as a whole, which is likely due to the timescale of 800 Myr in definition 
of the change parameter for all the metallicities discussed above. 

Based on the bottom panels of Figure \ref{fig:residual}, we assign an
uncertainty in the \RSFR\ according to the location on the \lgewhae\ vs. \ewhda\ 
diagram in the application of the calibrator.  While the scatter of SFR79 in the mock SFHs may over-estimate the uncertainty of \RSFR\ for an individual galaxy because the mock SFHs may contain not found ones in the real Universe, we think that this is a more reasonable approach 
than simply assigning a constant uncertainty of \RSFR\ at a given metallicity 
based on Figure \ref{fig:rsfr_fit}. 
Since this uncertainty is due to the estimator itself, we refer to this uncertainty as ``model uncertainty''. This is distinct from the uncertainty due to measurement uncertainties of the three diagnostic parameters from the observations (see Section \ref{subsec:4.1}).

\subsection{The effect of changing the IMF and the evolutionary isochrones} \label{subsec:2.4}

\begin{figure}
  \begin{center}
    \epsfig{figure=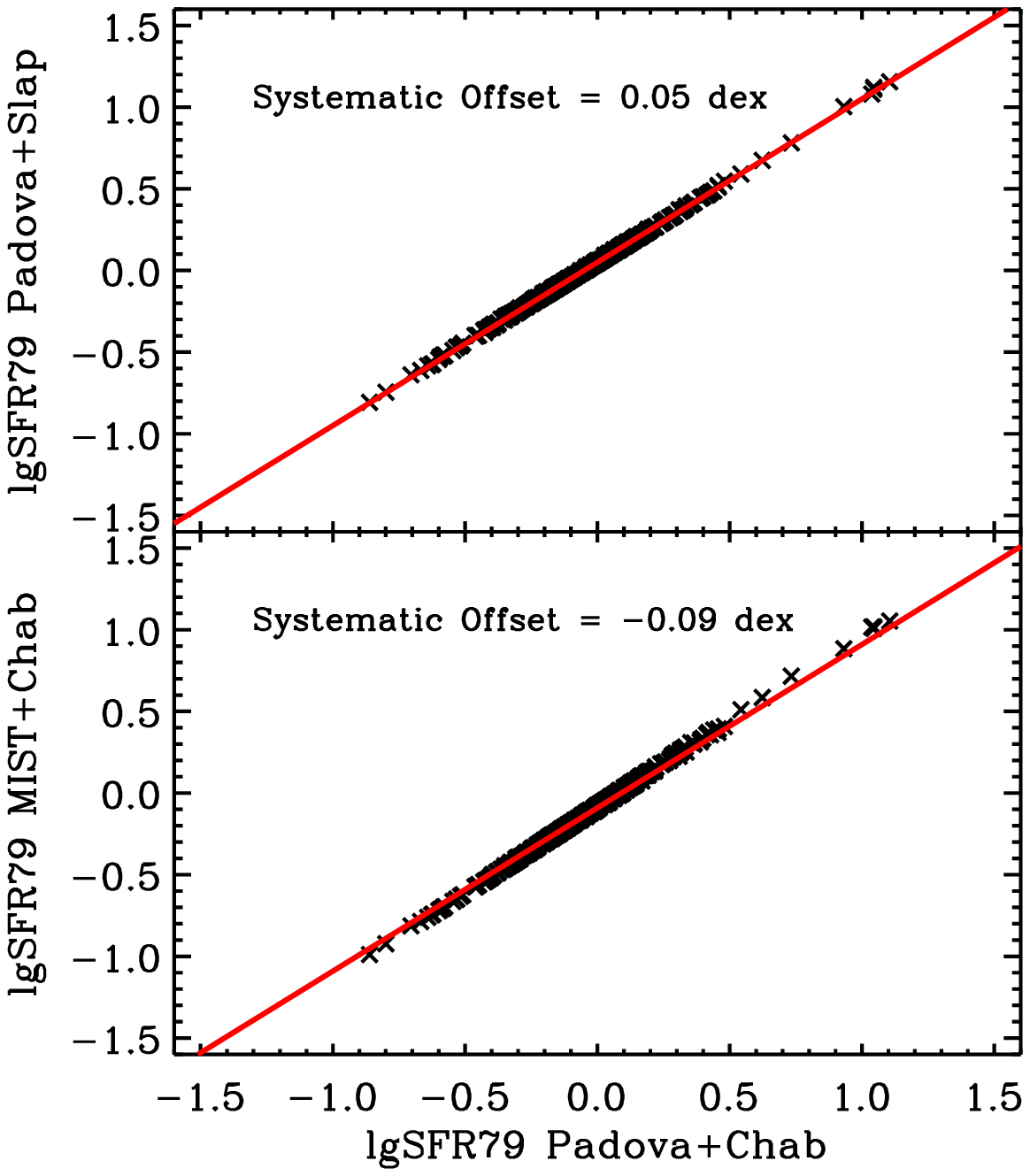,clip=true,width=0.40\textwidth} % size
    \end{center}
  \caption{The impact of changing the IMF and evolutionary isochrones on the calibration of \RSFR\ for MaNGA galaxies with stellar mass 
  greater than $10^{10}$\msolar. Upper panel:
  the \RSFR\ calibrated for Padova isochrones and the Chabrier IMF versus the 
  \RSFR\ calibrated with same isochrones and the Salpeter IMF.  
  Lower panels: the \RSFR\ calibrated with the Padova isochrones and the Chabrier IMF versus the 
  \RSFR\ calibrated with MIST isochrones and the same IMF. In the upper panel, 
  the red line is parallel to but 0.05 dex above the one-to-one line, while in the
  lower panel, the red line is parallel to but 0.09 dex below the one-to-one line, indicating that the effects of these two changes is to introduce a uniform offset in \RSFR.  
  }
  \label{fig:imf}
\end{figure}

In deriving the calibration for \RSFR\, we assumed for each mock galaxy the same, non-evolving, IMF of \cite{Chabrier-03}.  This assumption may not be the case in the real Universe. In principle, a time-varying 
IMF and a time-varying SFH are deeply degenerate, since we have no way of knowing when stars below the stellar Main Sequence turn-off stars produced.
%, and one could not break
%the degeneracy purely based on the SFR indicators, like H$\alpha$ and UV emission.  
We do not consider this issue further in the current work, but this should not be a problem unless the cosmic evolution 
of the IMF was significant in the last $\sim$1 Gyr (the timescale used in
definition of the change parameter), which we consider unlikely. 

However, the IMF may also be different from galaxy to galaxy, or even in different parts of the same galaxy.
Based on a very sensitive index of the IMF, $^{13}$CO/C$^{18}$O,  
\cite{Zhang-18} found evidence of a top-heavy stellar 
IMF (with respect to Chabrier IMF) in the 
dusty starburst galaxies at redshift $\sim$2-3. 

% XXXX I AM NOT SURE THIS IS RELEVANT... GIVEN THE NEXT PARAS
%This suggests that the calibrator 
%built in Section \ref{subsec:2.3} may not be valid for the extremely dusty SF galaxies, 
%which usually have extremely bright in infrared and lie significantly above the SFMS. 
%In the current work, we focus on SF main-sequence galaxies only, and do not consider 
%the variation of IMF across the galaxy population. 

In this subsection, we examine the stability of our calibrator with 
different IMFs, choosing to do this, for simplicity, only at a single (solar) metallicity. 
Based on the same approach in Section \ref{subsec:2.2} and \ref{subsec:2.3}, we construct
a new calibration of \RSFR\ using a Salpeter IMF \citep{Salpeter-55} without changing 
the other settings. Then we compare the two calibrators by applying them to 
a sample of MaNGA galaxies with stellar mass greater than $10^{10}$\msolar. The
three diagnostic parameters of MaNGA galaxies are calculated based on the spectra 
binned within the effective radius. Details of the binning scheme are in Section \ref{subsec:3.3}.
Note that the three diagnostic parameters are corrected for the dust attenuation,
according to the approach in Section \ref{subsec:2.5}. 
As shown in the top panel of Figure \ref{fig:imf}, the change of IMF gives the same result but with a systematic offset of about 0.05 dex.  
It is to be expected that a change in the IMF produces a systematic offset in \RSFR\ because both IMF and \RSFR\ will change the relative number of stars of different masses.  

In our work, the absolute value of \RSFR\ is of less interest than the dispersion in \RSFR. Therefore we argue that the choice of IMF is, at least within the range of plausible possibilities, not important. 

We next examine the stability of the \RSFR\ calibration with respect to the use of different stellar evolution 
model, i.e. the isochrones.  In the similar way, we generate a new calibrator by adopting the
{\tt MIST} isochrone  \citep[e.g.][]{Paxton-11, Choi-16, Dotter-16, Paxton-18}
with all other settings unchanged. The result is shown in the bottom panel of Figure
\ref{fig:imf}.  Again we see that the use of {\tt MIST} isochrones introduces a small systematic offset 
of $-$0.09 dex. 

While these effects introduce systematic offsets to \RSFR, this will not affect the 
investigation of the temporal variation of SFR in galaxy populations, 
because it is the {\it scatter} of \RSFR\ that characterizes this variation, rather
than the average value \citep{Broussard-19}.  A significant problem would occur only if the IMF or the appropriate isochrones varied significantly from galaxy to galaxy. While the former is possible, the latter is presumably not.  We note at this point that the variation with IMF (systematic offsets of 0.05 dex) is rather small compared with the observed dispersion in \RSFR\ within the population, which is 0.23 dex (see Section \ref{sec:4} below).  So, we can assume that any variations in IMF are probably a negligible contributor to this scatter.

\subsection{Correction for dust attenuation} \label{subsec:2.5}

\begin{figure*}
  \begin{center}
    \epsfig{figure=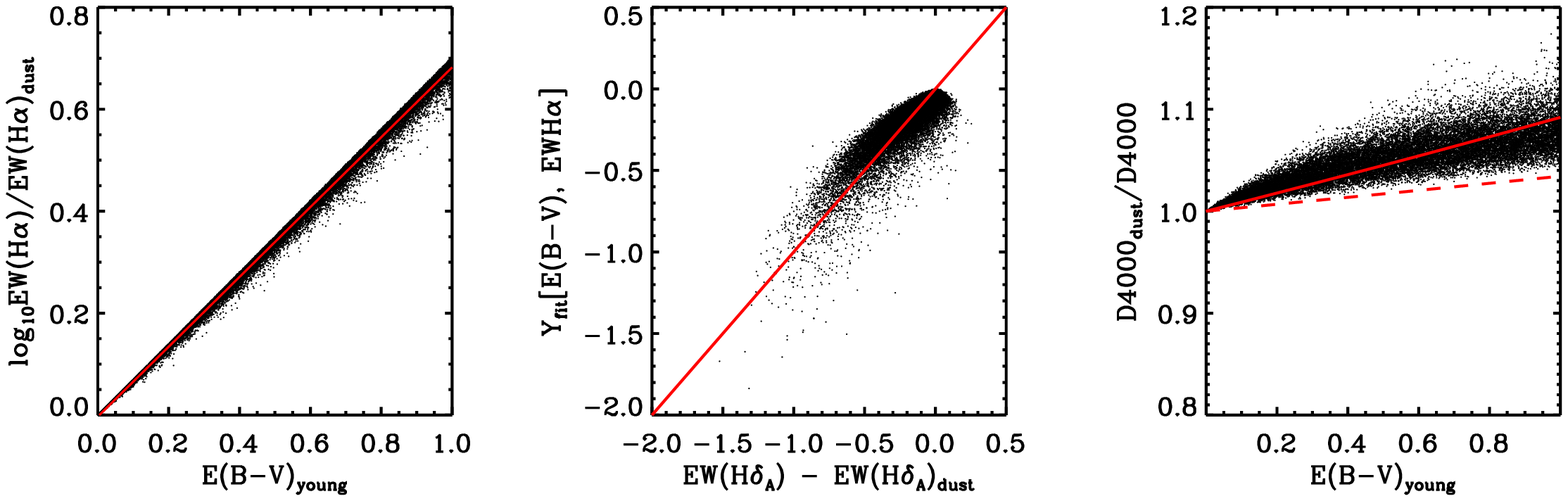,clip=true,width=0.99\textwidth} % size
    \end{center}
  \caption{Correction of the observed values of \lgewhae, \ewhda\ and \dindex\ for
dust attenuation as required before application of the calibrator to the observational data.
Left-hand panel: the change in $\log_{10} {\rm EW}({\rm H}\alpha)$ 
%/{\rm EW}(H\alpha)_{\rm dust}$
as a function of E(B$-$V)$_{\rm young}$. Middle panel: the correction of 
  \ewhda\ with E(B$-$V)$_{\rm young}$ and EW(H{$\alpha$}). 
  Right-hand panel: the change in \dindex$_{\rm dust}/$\dindex\ as a function 
  of E(B$-$V)$_{\rm young}$. The red lines are the fitting result of the 
  corrections (see Equation \ref{eq:3}). In the right-hand panel, the red dashed line 
  is the analytical relation computed with assuming that the value of E(B$-$V) 
  does not change with time. Here we only present the correction at solar metallicity
  for illustration. } 
  \label{fig:dust_cor}
\end{figure*}

%%%... add a table for the dust attenuation correction -------

\begin{table*}[ht]
\renewcommand\arraystretch{1.5}
\begin{center}
\caption{The fitting parameters of the dust attenuation for three diagnostic parameters in Figure \ref{fig:dust_cor} \label{tab:2}}
\begin{tabular}{@{}lrrrrrrrrrr@{}}
\tableline
\tableline

$\log_{\rm 10} Z/Z_{\odot}$                 & p1   &  p2 & {}  & p3  & p4 & {}  &  p5  &  \lgewhae$_{\rm ERR}$ & \ewhda$_{\rm ERR}$ &  \dindex$_{\rm ERR}$  \\
\tableline
    0.0                  &   0.685  &   $-$0.00582 &  {} &  0.580    &  $-$4.29  & {} &    0.0382    &   0.01 dex & 0.098 \AA\  & 0.019  \\
   $-$0.2                  &   0.687  &   $-$0.00691 & {} &   0.569  &   8.23  & {} &   0.0345  &    0.01 dex &  0.084 \AA\ & 0.016   \\
   $-$0.4                  &   0.689  &   $-$0.00595 & {} &   0.576  &   12.5  & {} &   0.0318  &    0.009 dex & 0.072 \AA\ & 0.014   \\
   $-$0.6                  &   0.691  &   $-$0.00496  & {} &  0.564  &   14.4  & {} &   0.0306  &   0.008 dex &  0.063 \AA\  & 0.014   \\
   $-$0.8                  &   0.693  &   $-$0.00407  & {} &  0.565  &   15.1  & {} &    0.0287   &   0.007 dex & 0.052 \AA\ &  0.012   \\
   $-$1.0                  &   0.692 &   $-$0.00388  & {} &  0.563  &    16.4   & {} &   0.0275   &  0.007 dex &  0.050 \AA\  & 0.011  \\
\tableline
\tableline
\end{tabular}
\end{center}
\end{table*}

As described above, we did not include the effects of dust attenuation in the calibration of \RSFR. 
This means that any correction for dust attenuation correction must be made to the 
three observational parameters before feeding them into the \RSFR\ calibration. As already noted,
the use of equivalent widths makes them relatively insensitive
to extinction.  However, differential extinction between stars of different ages, or between line and continuum emission is more of a problem. Newly formed stars ($<10$ Myr) may
have different extinction than older stars, 
since the extinctions of stellar continuum and of nebular emission are usually 
different \citep[e.g.][]{Calzetti-00, Moustakas-06, Wild-11, Hemmati-15}. 
\cite{Charlot-00} proposed a two-component dust model, where the optical 
depth for stellar populations older than 10 Myr is around one-third of the 
optical depth of stellar populations younger than 10 Myr. 
In this model, the regions of young stellar populations are more
dusty than the regions of older stellar populations, because
the newly formed stars are embedded in molecular clouds.  
The demarcation timescale of 10 Myr comes from the timescale of 
disruption of molecular clouds \citep{Blitz-80, Murray-10, Conroy-09}.  
In this work, we make the same assumption of this stellar age dependent 
dust model, and we adopt the Cardelli-Clayton-Mathis (CCM) dust 
attenuation curve \citep{Cardelli-89}.  

With this dust model, we correct the dust attenuation of the three diagnostic 
parameters in the following way. First, we generate as before mock SFHs based on the
SFHs of Illustris galaxies adding stochastic processes, but here adopt 
a broken power-law PSD of SFHs with $\alpha=$1.5 and $\tau_{\rm break}=$20 Gyr
(The large $\tau_{\rm break}$ make the PSD close to a single power-law PSD with $\alpha$=1.5). 
And again, the scatter of the variations are normalized to 0.4 dex.
%with a truncation timescale\footnote{The timescale is comparable  to the lifetime of galaxies, which 
%should be less than the age of the Universe.} of 10 Gyr. 
In contrast to the situation in Section \ref{subsec:2.2}, we here try to generate 
SFHs that resemble the observations, rather than a huge range of all possibible SFH. 
A power-law PSD with $\alpha=$1.5 is likely a good description of the stochasticity of SFHs 
without considering the intrinsic scatter of the SFMS, according to 
the analysis of our second paper of this series. 
Actually, in the second paper, we will find that, if we assume a single power-law form for the PSD of the specific SFH, a slope of 1.5 is the best to reproduce the distribution of galaxies on the $\Delta$sSFR7-$\Delta$sSFR9 plane \footnote{The $\Delta$sSFR7 and $\Delta$sSFR9 are defined as the offset of galaxies to the ``nominal'' SFMS based on SFR7 and SFR9 (see the definition in Section \ref{subsec:4.1} and Figure \ref{fig:global}).}. The slope of PSD adopted here is slightly shallower than the one assumed in \cite{Caplar-19}, who assumed a PSD index of 2, corresponding to a random walk process.
We refer the reader to the second paper of this series for details.

We then calculate two sets of diagnostic observational parameters based on the
mock SFHs with and without the dust attenuation.
For the mock spectra, we know the contribution of the stellar populations at different ages, from the mock SFHs and evolving spectra of the SSP models. This enables us to apply a stellar-age-dependent extinction model to obtain the reddened spectra and the values of the three diagnostic parameters.
The diagnostic parameters 
with dust reddening are denoted as \dindex$_{\rm dust}$, \ewhda$_{\rm dust}$ 
and \ewhda$_{\rm dust}$. In this process, we 
assign for each galaxy an E(B$-$V)$_{\rm young}$\footnote{The E(B$-$V) for old stellar 
populations (E(B$-$V)$_{\rm old}$) is 0.3E(B$-$V)$_{\rm young}$
according to the dust model we assumed \citep{Charlot-00}. }, i.e. the E(B$-$V) for 
the stellar population younger than 10 Myr, 
that is randomly distributed between 0.0 and 1.0.
At last, we compare the two set of observed parameters and define recipes for
the correction. 

After exploring many forms for the dust correction of the three observational parameters,
we found good expressions with rather small uncertainties. 
Figure  \ref{fig:dust_cor} presents the formulae to correct the dust 
attenuation (solid red lines) by comparing the two set of diagnostic 
parameters at solar metallicity.   The correction of \ewhae\ and \dindex\ 
are a function of E(B$-$V)$_{\rm young}$, while the correction of \ewhda\ is 
a function of both E(B$-$V)$_{\rm young}$ and \ewhae. In the observation, 
the E(B$-$V)$_{\rm young}$ can be determined from the flux ratio of H$\alpha$ 
and H$\beta$ using the Balmer decrement. The adopted expressions for the 
correction for the three parameters are as follows: 
\begin{equation} \label{eq:3}
\begin{split}
\log_{10}{\rm EW}({\rm H}_{\alpha})/{\rm EW}({\rm H}_{\alpha})_{\rm dust}  = & p1\times {\rm E(B-V)}_{\rm young} \\ 
{\rm EW}({\rm H}\delta_{\rm A})  - {\rm EW}({\rm H}\delta_{\rm A})_{\rm dust}  = & p2\times {\rm E(B-V)}_{\rm young}^{p3} \\
                            &  \times ({\rm EW}({\rm H}_{\alpha}) + p4)   \\
\log_{10}D_n(4000)_{\rm dust}/D_n(4000)  = & p5\times {\rm E(B-V)}_{\rm young},  
\end{split}
\end{equation}

%\begin{equation} \label{eq:3}
%\begin{split}
%\lgewhae/\ewhae$_{\rm dust}$  = & p1\times {\rm E(B-V)}$_{\rm young}$ \\ 
%{\rm EW}({\rm H}$\delta_{\rm A}$)  - \ewhda$_{\rm dust}$  = & p2\times {\rm E(B-V)}$_{\rm young}^{p3}$ \\
%                            &  \times (\ewhae\ + p4)   \\
%\log_{10}\dindex$_{\rm dust}$/\dindex  = & p5\times {\rm E(B-V)}$_{\rm young}$,  
%\end{split}
%\end{equation}

where $p1$, $p2$, $p3$, $p4$, and $p5$ are parameters determined by fitting.  Table \ref{tab:2} shows these fitting parameters to be used in Equation \ref{eq:3},
as well as the resulting uncertainties of the observational diagnostic parameters (listed in the 
last three columns in Table \ref{tab:2}) produced by the 
dust attenuation correction, at different metallicities. 
The uncertainties are determined by the scatters in the three panels of 
Figure \ref{fig:dust_cor}.  As can be seen in Table \ref{tab:2}, 
the uncertainties due to this correction 
are small. 

However, there may be additional uncertainties in the dust attenuation correction, due to possible variations in the f-factor \citep[e.g.][]{Kashino-13, Lin-19, Faisst-19}, defined as the E(B$-$V)$_{\rm star}$/E(B$-$V)$_{\rm nebular}$\footnote{The E(B$-$V)$_{\rm star}$ is close to, but not the same as the E(B$-$V) for the stellar population with the age greater than 10 Myr. }, from galaxy to galaxy or from regions to regions. 
This effect is not included in our dust model in the present work. Specifically, by using the MaNGA galaxies,  \cite{Lin-19} found that the f-factor decreases with increasing stellar mass, and slightly increases with 
increasing sSFR. For most of SF spaxels in galaxies above 10$^{10}$\msolar, the f-factor is in the range of 0.3-0.7 with a scatter of $\sim$0.1-0.15. 
We have examined the dependence of our SFR79 estimator on the value of E(B$-$V)$_{\rm old}$/E(B$-$V)$_{\rm young}$ at solar metallicity.  Increasing E(B$-$V)$_{\rm old}$/E(B$-$V)$_{\rm young}$ by 0.1, the resulting SFR79 show an overall 0.03 dex offset with respect to the old ones, which is much smaller than the scatter of SFR79 (0.23 dex) we measured in Section \ref{sec:4}. 
We conclude that the dust attenuation corrections of the three parameters are only a secondary effect due to the fact that they are relative values measured at fixed wavelength. The uncertainty of 
the dust attenuation is even much smaller than the value of the applied correction for the three parameters, and is therefore not likely to be a big concern.

\section{Application to MaNGA galaxies}
\label{sec:3}

In this section of the paper, we apply the \RSFR\ estimator constructed in Section \ref{sec:2} to  
spatially-resolved spectroscopic data from the MaNGA survey.  In Section \ref{subsec:3.1}, 
we will give a brief introduction for the sample selection and the measurements of the 
relevant parameters. In Section \ref{subsec:3.2new}, we examine the consistency for the overall change in SFR. 
In Section \ref{subsec:3.2}, we then derive a small additional correction of \RSFR\ 
parameter to correct for an unexpected apparent dependence of \RSFR\ on stellar surface density. 

\subsection{The sample galaxies and the measurement of parameters} \label{subsec:3.1}

MaNGA is one of the largest integral field spectroscopic surveys, 
aiming at obtaining the two-dimensional spectra for
$\sim$10,000 galaxies in the redshift range of $0.01<z<0.15$ \citep{Bundy-15}.  
The wavelength covered by MaNGA is 3600-10300 \AA\ at a spectral resolution R $\sim$2000, 
which is sufficient to accurately measure the 
three diagnostic parameters \citep{Li-15} used in this paper.  

In this work, we utilize the well-defined sample of star-forming galaxy from \citetalias{Wang-19}. Here we therefore only
briefly describe the sample definition, and refer the reader to \citetalias{Wang-19} for further details. 

The galaxy sample is originally selected from SDSS Data Release 14 \citep{Abolfathi-18}, 
 excluding the mergers, irregulars, heavily disturbed galaxies, 
as well as galaxies for which the median S/N of the
5500 \AA\ continuum is less than 3.0 at their effective radii.  The quenched
galaxies are excluded based on the stellar mass and SFR diagram. 
The stellar mass and SFR are measured within the effective radius, i.e. \mstar\ and 
SFR($<$\re)\footnote{Here the SFR is determined by H$\alpha$ luminosity, which represents 
the star formation within the most recent 5 Myr. }, based on the MaNGA spectra.  
Our final sample consists of 976 SF galaxies, and is a representative sample 
of SF main-sequence galaxies in the low-redshift Universe.   

The stellar mass maps of MaNGA galaxies are derived from the public fitting code {\tt STARLIGHT}
\citep{Cid-Fernandes-04}, using SSPs with {\tt Padova} isochrones 
from \cite{Bruzual-03} and the \cite{Chabrier-03} IMF.  The SFR maps are determined 
by the extinction-corrected H$\alpha$ luminosity adopting the conversion formula to SFR 
from \cite{Kennicutt-98}, again using a \cite{Chabrier-03} IMF.  
The uncertainty in determining the SFR via this approach is 15\% (or $\sim$0.06 dex),
due to the variations in the electron temperature in the range $T_{\rm e}=$5000-20000 K \citep{Osterbrock-06}. As above, we also refer to this uncertainty as model uncertainty.

Since in this work our purpose is primarily to investigate the SFR of galaxies on different timescales, 
we denote the SFR directly determined from the H$\alpha$ luminosity
as SFR7, i.e. SFR$_{\rm 5Myr}$ (see Figure \ref{fig:ha_hd}).
%XXX do we still want the footnote??
%\footnote{Of course, as seen in Figure \ref{fig:ha_hd}, the \ewhae\ is not a perfect step-function with  stellar age for SSP models. This suggests that the H$\alpha$-based SFR  is not exactly SFR$_{\rm 5Myr}$.  However, this should not be a concern, because in the second paper of this series, we find the variation of SFHs in a few Myr is negligible with respect to overall  variation in SFHs, with assuming that the intrinsic scatter of SFMS is much less than the scatter of SFR9. }. 
The intrinsic extinction for nebular emission 
is measured based on the Balmer decrement, assuming the CCM dust attenuation
curve \citep{Cardelli-89} and Case B recombination with ab intrinsic flux ratio
of H${\alpha}$/H$\beta$ = 2.86. 
The E(B$-$V) for nebular emission is then a good estimation for the 
E(B$-$V)$_{\rm young}$, i.e. the color excess for the stellar population 
younger than 10 Myr.  
We note that the IMF, isochrones, and
dust attenuation curve that are used to obtain the stellar masses and SFR$_{\rm 5Myr}$ are 
consistent with those used in Section \ref{sec:2}. 

The strengths of the emission lines are measured based on the stellar continuum-subtracted 
spectrum by fitting a Gaussian profile to the lines. 
The \dindex\ and \ewhda\ are directly measured based on the observed 
spectra after subtracting emission lines, rather than from the best-fit continuum spectra. 
This avoids the uncertainties of the measurements due to the possible systematic offset 
(especially at 3800-4200 \AA) between the model spectra and observed ones. 
For many SF galaxies, the bottom of the H$\delta$ absorption is usually accompanied 
by weak H$\delta$ emission, which make the \ewhda\ difficult to measure. In this work,
we take advantage of the $\chi^2$ minimization spectral fitting code developed by 
\cite{Li-05}, which can effectively mask the emission-line regions iteratively
during the fitting. This is critical to accurately model the absorption troughs 
and also characterize the emission lines (see examples in \cite{Li-15}).

Before applying our \RSFR\ estimator on the MaNGA galaxies, we first correct  
the three diagnostic parameters for dust attenuation based on Equation \ref{eq:3}. 
Since both the estimator and the dust attenuation correction depend
on the stellar metallicity of galaxies, we adopt the stellar mass-metallicity 
relation from \cite{Zahid-17} to estimate the stellar metallicity of individual galaxies. 
The relation can be written as: 
\begin{equation}\label{eq:4}
    \log_{10}\frac{Z}{Z_{\odot}} = Z_0 + \log_{10}\left[1-\exp\left(-\left[\frac{M_*}{M_0}\right]^\gamma\right)\right], 
\end{equation}
where $Z_0$ = 0.075, $M_0$ = 10$^{9.79}$\msolar, and $\gamma$ = 0.56. 
This relation is determined by modelling the galaxy spectra with a 
linear combination of sequential single burst model spectra.  
For a given set of three observational parameters at given metallicity, we first calculate
the correction based on Equation \ref{eq:3} and Table \ref{tab:2} at
the two closest metallicities. Then we use linear interpolation 
in $\log_{10}Z/Z_{\odot}$ to obtain the corrections of the observational three diagnostic 
parameters at the required metallicity.  In the similar way, we then obtain the
\RSFR\ using the (dust-corrected) values of the three diagnostic parameters by calculating
the \RSFR\ at the two closest metallicities based on Equation \ref{eq:2}
and Table \ref{tab:1} then obtain the \RSFR\ at the required metallicity
via linear interpolation in $\log_{10}Z/Z_{\odot}$. 

\subsection{Consistency check: the overall change in SFR} \label{subsec:3.2new}

Having calculated \RSFR\ for all spaxels in our MaNGA sample, we can now carry out an important consistency check by calculating the total SFR of all spaxels in the sample, averaged over the last 5 Myr, and the total SFR averaged over the last
800 Myr.  These should be roughly equal.  To be precise, the ratio of these, i.e. $\langle {\rm SFR7}\rangle/\langle {\rm SFR9}\rangle$, should reflect the overall cosmic evolution of the SFR of the SF galaxy population over the last Gyr, i.e. the change in overall SFR of SF galaxies that is implied by integrating the change in the sSFR of the Main Sequence.  The cosmic evolution of 
 $\log_{10}\langle {\rm SFR7}\rangle/\langle {\rm SFR9}\rangle$ that is expected for
the ensemble of main-sequence 
galaxies is calculated to be $-$0.025 dex based on the evolution of the sSFR of the SFMS
from \cite{Lilly-16}, and the stellar mass function for SF galaxies 
from \cite{Peng-10}. 

The ratio of these two total SFR in the MaNGA data (using our estimator of \RSFR\ to calculate individual SFR9) is actually $-0.066$ dex, which is reassuringly close (within 0.04 dex, or 10\%) to the expected value.  This very satisfactory agreement should be taken as a first confirmation that 
our calibration of \RSFR\ is quite accurate and certainly usable.  In fact, given the assumptions that we made, however reasonably, about the effect of reddening, about the form of the IMF and about the choice of stellar evolution isochrones, the very close agreement to within 0.04 dex should probably be seen as fortuitous.  This is explored further in the next subsection.

\subsection{Correction of a dependence of $\Sigma_*$} \label{subsec:3.2}

\begin{figure}
  \begin{center}
    \epsfig{figure=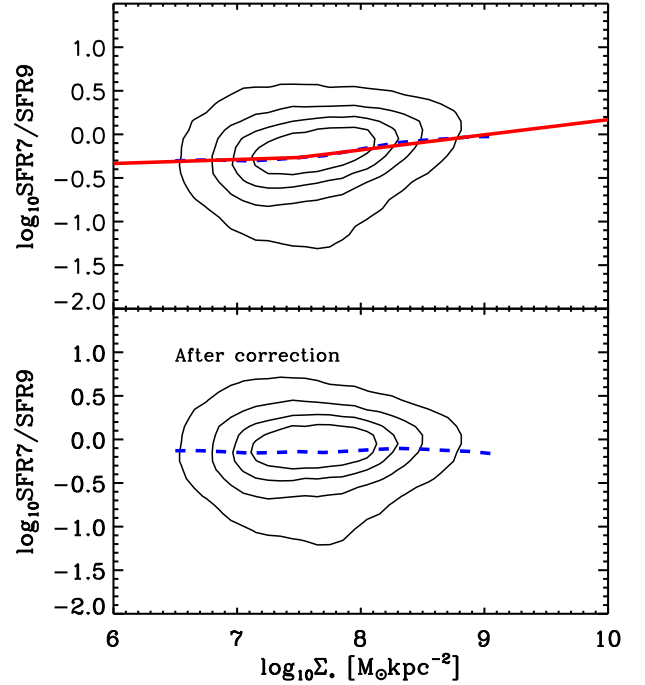,clip=true,width=0.45\textwidth} % size
    \end{center}
  \caption{Correction of the presumed spurious dependence of 
 \RSFR\ on $\Sigma_*$.  The blue dashed
  lines show the median \RSFR\ at given $\Sigma_*$.  The upper panel shows the raw \RSFR\ (upper panel) as a function of
 %and
 % \RSFR\ after correcting the dependence on $\Sigma_*$ (lower panel) 
 % as a function of 
  $\Sigma_*$ for the spaxels selected from MaNGA galaxies. Since, as discussed in the text, the increase with $\Sigma_*$ is thought to be spurious, this median relation is used to correct the \RSFR\ using a simple broken power-law shown with the red line.  
The lower panel shows the distribution of \RSFR\ in the spaxels after this correction. In both panels, the
  contours show the number density distribution of the spaxels in the sample.
 }
  \label{fig:rsfr_cor}
\end{figure}

The top panel of Figure \ref{fig:rsfr_cor} shows the \RSFR\ for all the 
spaxels in our sample galaxies as a function
of de-projected stellar mass surface density ($\Sigma_*$) . 
The contours show the overall number density distribution of spaxels in this diagram. These contours enclose 
30\%, 50\%, 70\% and 90\% of all spaxels, from the inside out. 
The blue dashed line shows the median \RSFR\ at given $\Sigma_*$. 
Based on \RSFR, we can obtain $\log_{10}$SFR9, i.e. \SFRNI, for each individual spaxel of 
the sample galaxies. 
%XXX  Need to put blue dashed on top of red solid line in the figure!

It is clear in this figure that the mean \RSFR\ appears to slightly increase 
with increasing $\Sigma_*$, 
suggesting that SF galaxies have on average a slightly negative \RSFR\ radial 
gradient.  We suspect that this small effect is unlikely to be real,  
 since it would suggest that the SFR was declining more at large radii, leading to a negative gradient in the rate of change of the sSFR of galaxies.  Since galaxies generally have a positive gradient in sSFR at the current epoch, this would imply that this gradient was weakening.  If anything, we would expect the opposite trend in any ``inside-out" scenario of galaxy evolution.
%since 
%a small gradient in \RSFR\ would result in a significant change 
%in the overall radial gradient of \SFRSE\ only assuming that  
%the small gradient of \RSFR\ can keep for a few Gyrs. 
%Therefore a flat overall \RSFR\ profile is necessary for
%keeping the stability of the overall profile of \SFRSE\ for SF 
%galaxy population.  Furthermore, the relative high \RSFR\ at higher 
%$\Sigma_*$ end indicates that the inner regions of massive galaxies 
%are increasing their star formation with respect to the outer regions 
%as a whole. This is inconsistent with the picture of the inside-out growth of galaxies.  
 There are other reasons to question this small gradient in \RSFR.
Not least, although the dust-correction is in principle computed locally, we did not consider any radial variation of the form of the dust-correction, nor of the
stellar metallicity \citep[e.g.][]{Zheng-17, Goddard-17}, nor any (possible) radial variation of
IMF \citep{Gunawardhana-11} in the estimation of \RSFR.  We suspect that some combination of these may be the cause of the trend in the upper panel of Figure \ref{fig:rsfr_cor}.

Accordingly, we therefore perform a small {\it ad hoc} correction to the values of \RSFR\ as a function of $\Sigma_*$ (only).  This is constructed so as to eliminate 
the dependence of \RSFR\ on $\Sigma_*$, while not significantly perturbing the total rates of star-formation, as discussed in the previous sub-section.  We first fit the median 
\RSFR-$\Sigma_*$ relation with a piecewise linear function, shown
as the red line in Figure \ref{fig:rsfr_cor}.  The red line is in the
form of 
\begin{equation} \label{eq:5}
 y =
   \begin{cases}
   0.04581x-0.6089,   &  x<7.5, \\
   0.1737x-1.568,     &  x\geq7.5, \\
   \end{cases}
\end{equation}
where y = $\log_{10}$SFR79, and x=$\log_{10}\Sigma_*$ [$M_{\odot} {\rm kpc}^{-2}$]. For each spaxel,  
we then subtract this median \RSFR\ at the corresponding
$\Sigma_*$.  
In order to preserve the total star-formation rates, as discussed in the previous sub-section,
we then add a uniform $-0.136$ dex to the \RSFR\, computing this value to 
exactly match the $\log_{10}\langle {\rm SFR7}\rangle/\langle {\rm SFR9}\rangle$
with the value of the cosmic evolution for SF main-sequence galaxies (i.e. including the 0.04 dex offset discussed in the previous sub-section).
The bottom panel of Figure \ref{fig:rsfr_cor} shows the \RSFR\ as a function 
of $\Sigma_*$ after this correction. 

This correction to \RSFR\ is of course quite arbitrary. However, we stress that this correction only makes the 
overall \RSFR\ profile flat, and does not change the {\it scatter} of \RSFR\ 
at given $\Sigma_*$.  Since the variability of SFHs is indicated by the 
{\it scatter} of \RSFR\ across the population, rather than by its average or median value, this correction will not significantly
affect any of our conclusions in the following analysis. In the remainder
of this work, the \RSFR\ for both individual spaxels and individual 
galaxies will be corrected according to the above approach.  

\subsection{The \RSFR\ maps and profiles} \label{subsec:3.3}

\begin{figure*}
  \begin{center}
    \epsfig{figure=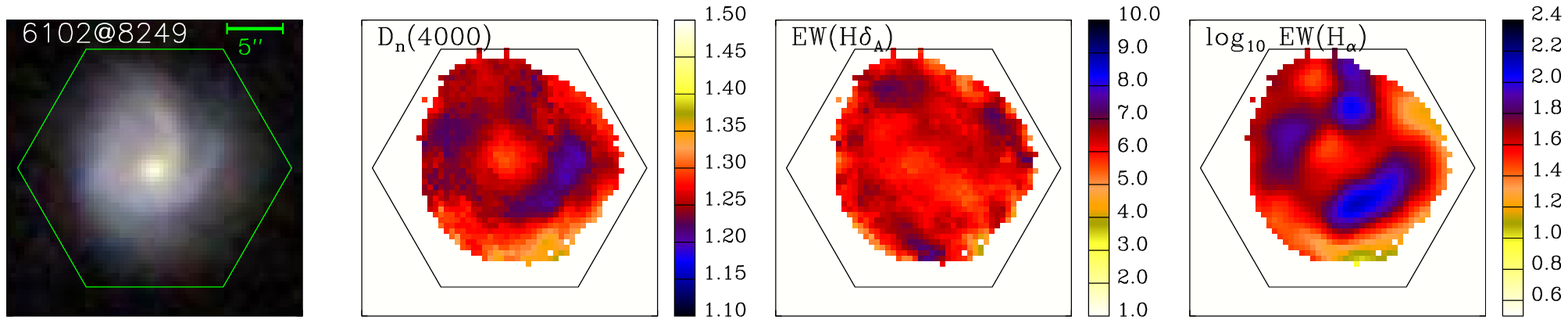,clip=true,width=0.99\textwidth} % size
    \epsfig{figure=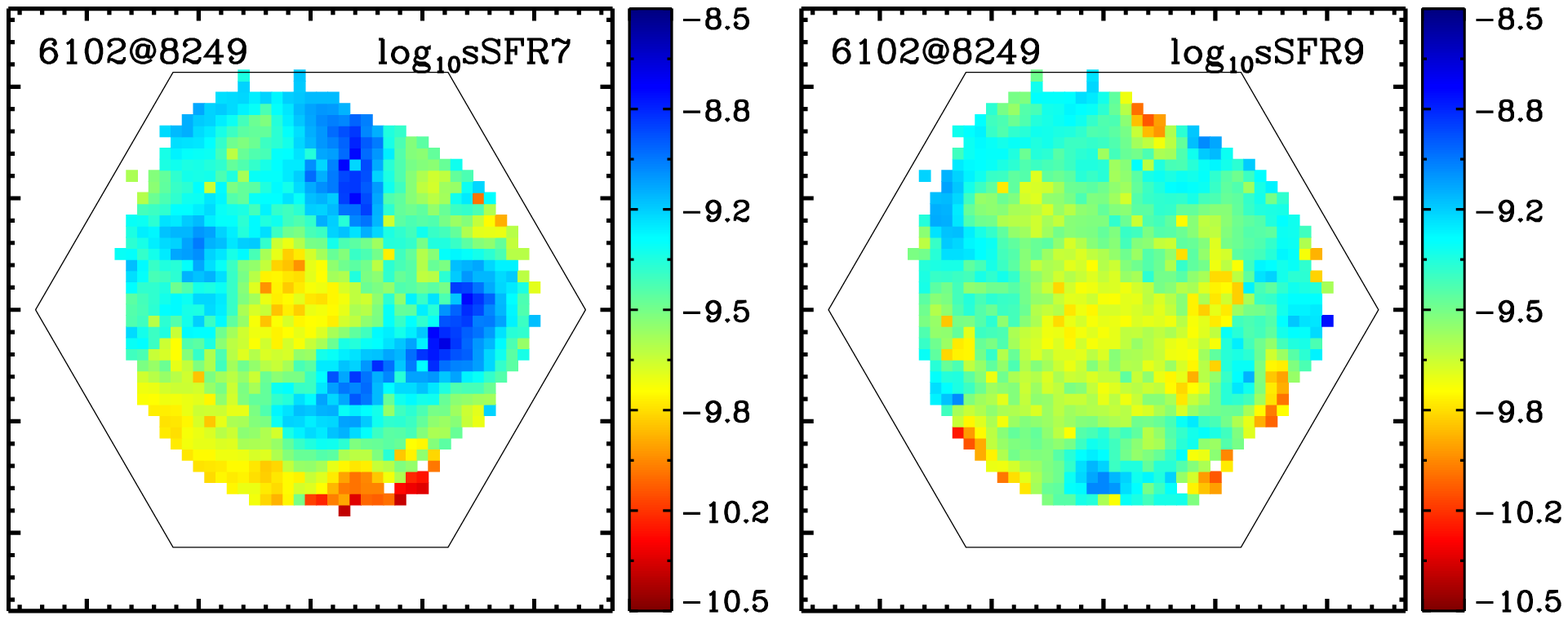,clip=true,width=0.495\textwidth}
    \epsfig{figure=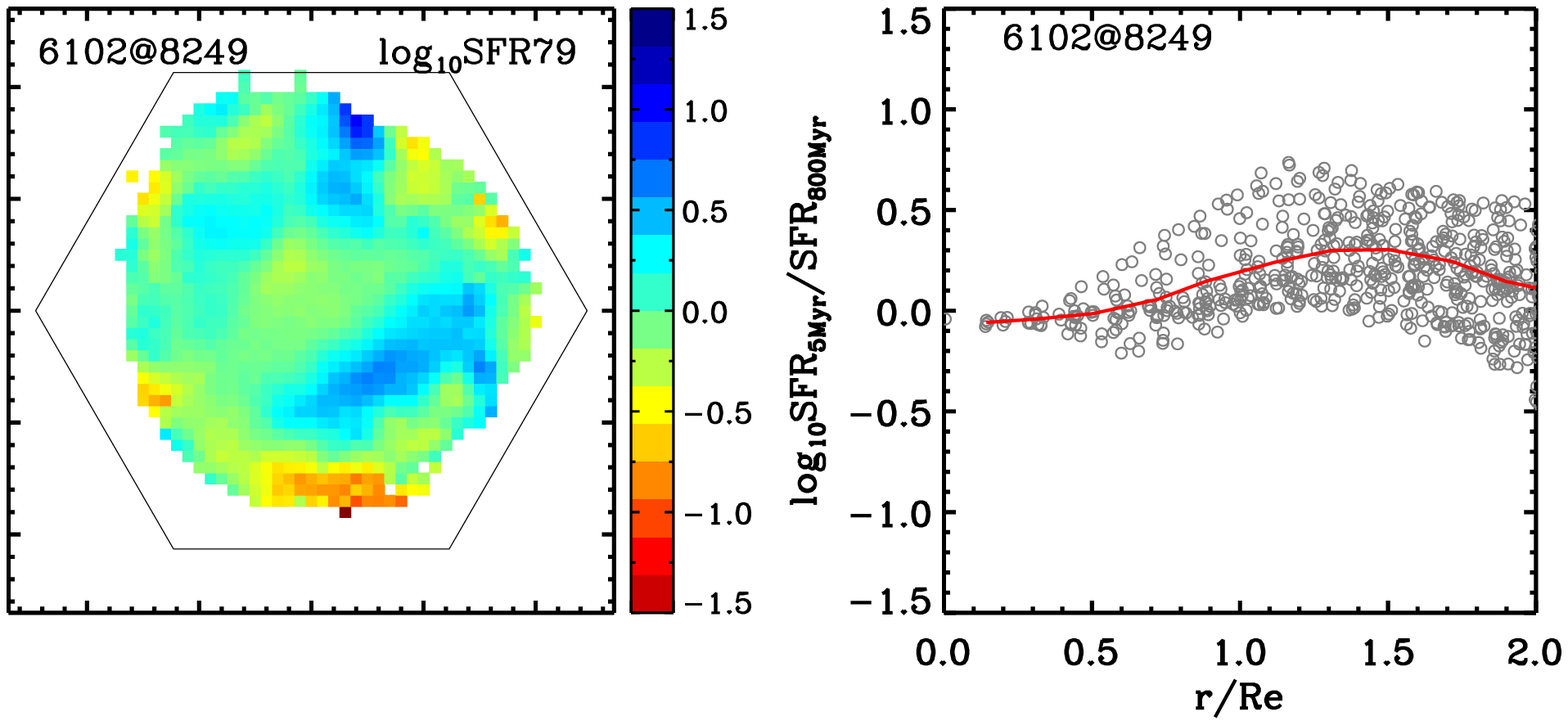,clip=true,width=0.495\textwidth} % size
    \end{center}
  \caption{ An example of the measurements and derived quantities for
  the MaNGA galaxy 6102@8249.
 The top panels show from left to right, the SDSS color image, and maps of the 4000\AA\ break, the
  H$\delta_{\rm A}$ index, and $\log_{10}{\rm EW}({\rm H}\alpha)$. The hexagons represent 
  the area covered by the IFS bundles on this galaxy.  The bottom panels show from left to right, the maps of $\log_{10}$sSFR7, of $\log_{10}$sSFR9, and of $\log_{10}$SFR79, together with  
  the \RSFR\ profile of this galaxy. 
  In the bottom rightmost panel, the gray circles represent the individual spaxels, 
  and the red line show the \RSFR\ profile, which are calculated based on the 
  binned diagnostic parameters within a set of annulus. }
  \label{fig:rsfr_example}
\end{figure*}

Based on the approach in Section \ref{sec:3}, we can now obtain the \RSFR\ maps of each sample galaxy. 
Figure \ref{fig:rsfr_example} shows an example of the measurements for one typical galaxy 
(MaNGA-ID: 8249-6102). 
The top panels of Figure \ref{fig:rsfr_example} shows the SDSS color image, \dindex, \ewhda, 
and \lgewhae\ maps from left to right, respectively. It should be noted that the three diagnostic parameters are shown prior to the correction for dust attenuation. 
The bottom panels show the maps of sSFR7, sSFR9,  and \RSFR, as well as the profile of \RSFR\ for this galaxy, from left to right, respectively.  
%XXXX  WOULD IT BE POSSIBLE TO SHOW THE SFR7 and SFR9 mapes too??  Perhaps normaized to some average radial gracient in SFR9?  Also hy not blow up the colour image to the same size>

% Will rewok this next para when we have new plot?

As shown, the \ewhae\ map shows clumpy features (blue clumps), corresponding to the 
regions with high recent star formation within 5 Myr.
However, these regions do not show high \ewhda, suggesting 
that the star formation in these regions were not unusually active in forming stars during the last
$\sim$1 Gyr.  Thus, the current star formation in these
blue clumps of \lgewhae\ map is triggered recently (much less than 800Myr). 
Consistent with this, these regions are seen as positive (blue) on the \RSFR\ map that 
blue clumps are seen in the same regions. On the other hand, some regions with
high \ewhda\ but relative low \ewhae, are visible on the \RSFR\ map with
yellow clumps. The star formation of these regions is reduced in a 
time scale much less than 800 Myr.  
This simple example indicates that the \RSFR\ we measured is indeed meaningful, 
and provide the quantitative description for the above effect.   

In the bottom rightmost panel of Figure \ref{fig:rsfr_example},  the small gray 
circles indicate the \RSFR\ derived spaxel-by-spaxel in this galaxy, while the red line shows the \RSFR\ 
profile obtained from the binned spectra in annular bins.  
%XXX check: from the spectra or by average SFR79?? The binned ones ... 
It can be clearly seen that, between 0.5-1.0\re\, the distribution
of \RSFR\ is asymmetric: most of the spaxels have a relative low \RSFR, while a 
small fraction of spaxels have increased \RSFR.  This is due no doubt to the 
duty cycle of star formation.   At any given time, star-formation in a given region of galaxy is found in a limited number of active regions, which themselves are active for only a short fraction of the time. 

In the current work, we do not wish to study the variation of SFR (or \RSFR) that is caused by 
this small scale local effect but rather focus on the variations in SFR on larger scales.  Therefore we use a binning scheme to average out these small scale spatial effects.

The diagnostic parameters for a given radial bin 
are calculated in the following way. For instance, for \dindex,
we first calculate the flux density of the blue and red bandpass near 4000 \AA\ 
as described in Section \ref{subsec:2.3} for all the spaxels (with 
S/Ns greater than 3 at 5500 \AA) in a given radial bin. The  \dindex\ of 
this radial bin is then obtained from the ratio of the sum of the flux densities in the red bandpass to the 
sum of them in the blue bandpass, summing over all the spaxels in this radial annular bin.  
In this process, we also estimate the observational error in the flux density
for the blue and red bandpasses. We first assume the flux errors of all the binned spaxels 
have no covariance in obtaining the binned error ($\sigma_{\rm no,cov}$). This 
error is obviously less than the real error due to the covariance of binned spaxels that arises because the spatial-resolution ($\sim$2.5 arcsec) of the MaNGA survey is much 
larger than the size of each individual spaxel (0.5 arcsec).
To correct this, we adopt an empirical function from \cite{Law-16}: 
\begin{equation} \label{eq:6}
\frac{\sigma_{\rm cov}}{\sigma_{\rm no, cov}} = 
      \begin{cases}
          1.0 + 1.6\log_{10} N_{\rm pixel}  & N_{\rm pixel} \leq 100 \\
          4.2  & N_{\rm pixel} > 100  \\
      \end{cases}
\end{equation}
where $\sigma_{\rm cov}$ is the error with considering the covariance of
binned spaxels, and $N_{\rm pixel}$ is the number of binned spaxels.  We then obtain the observational error of \dindex\ by error propagation. 
%% NEED TO TALK ABOUT THIS WITH YOU....
In a similar way, we obtain the binned \ewhda\ and \ewhae, as well as their error for a given bin.  
The E(B$-$V)$_{\rm young}$ in a 
given radial bin is also calculated based on the ratio of total H$\alpha$ flux 
to total H$\beta$ flux within the bin. Finally, the \RSFR\ in the bin is calculated 
with the binned diagnostic observational parameters using Equation \ref{eq:2}, 
and the observational error of \RSFR\ is calculated based on the 
observational error of three diagnostic parameters via error 
propagation\footnote{In this process, we also combine the uncertainties 
invoked by the dust attenuation in Section \ref{subsec:2.5}
for the three diagnostic parameters (see Table \ref{tab:2}). }. 

The advantage of the binning scheme is 1) to improve the accuracy of the measurements 
of \RSFR, 2) to reduce the variation of \RSFR\ caused by the small scale effect discussed above
(including the duty cycle of star formation).
We set 
the radial bin width to be 0.2\re, large enough to eliminate the scale effect, and calculate the \RSFR\ and its observational error for a given radial bin 
via the above approach.  In the current paper,
we focus on the radial profiles of \RSFR\ for the sample galaxies, 
and will not consider the local variation of \RSFR\ within these radial bins. 

\section{The SFR9-based SFMS of the sample galaxies} \label{sec:4}

\begin{figure*}
  \begin{center}
    \epsfig{figure=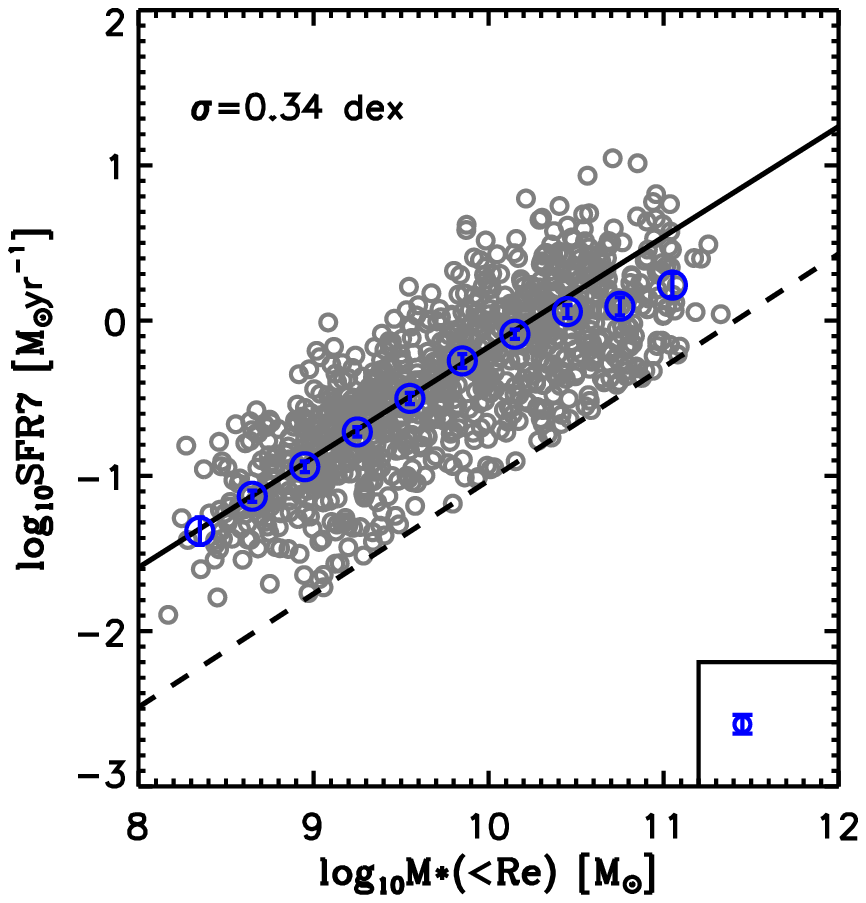,clip=true,width=0.42\textwidth} 
    \epsfig{figure=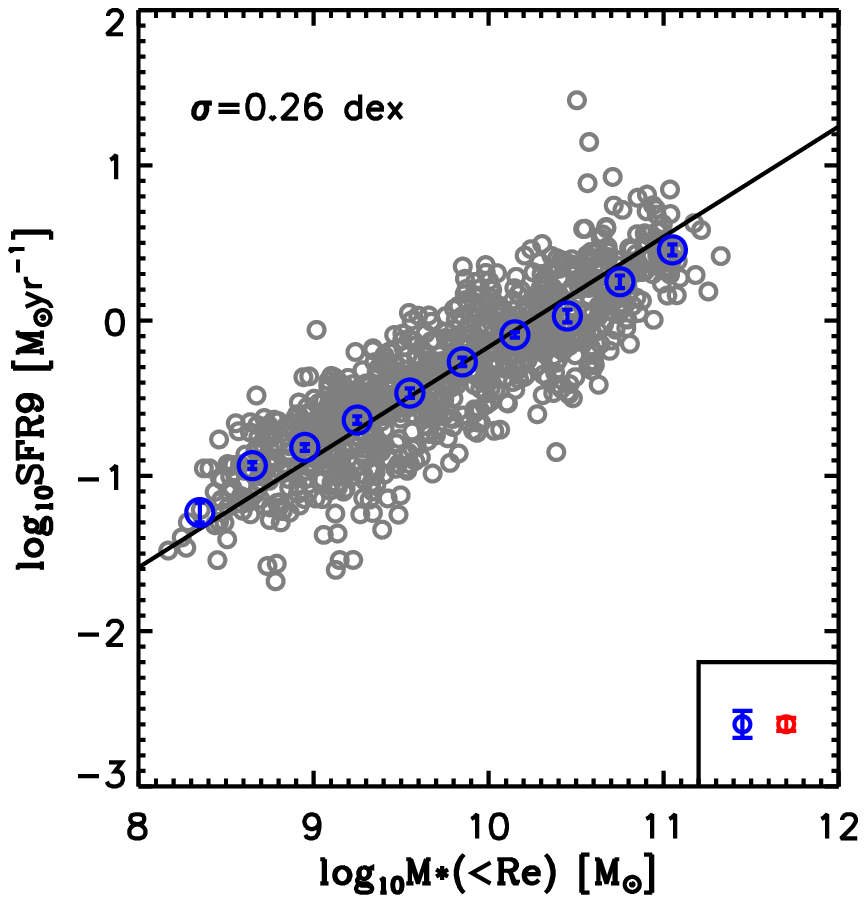,clip=true,width=0.42\textwidth}  
    \epsfig{figure=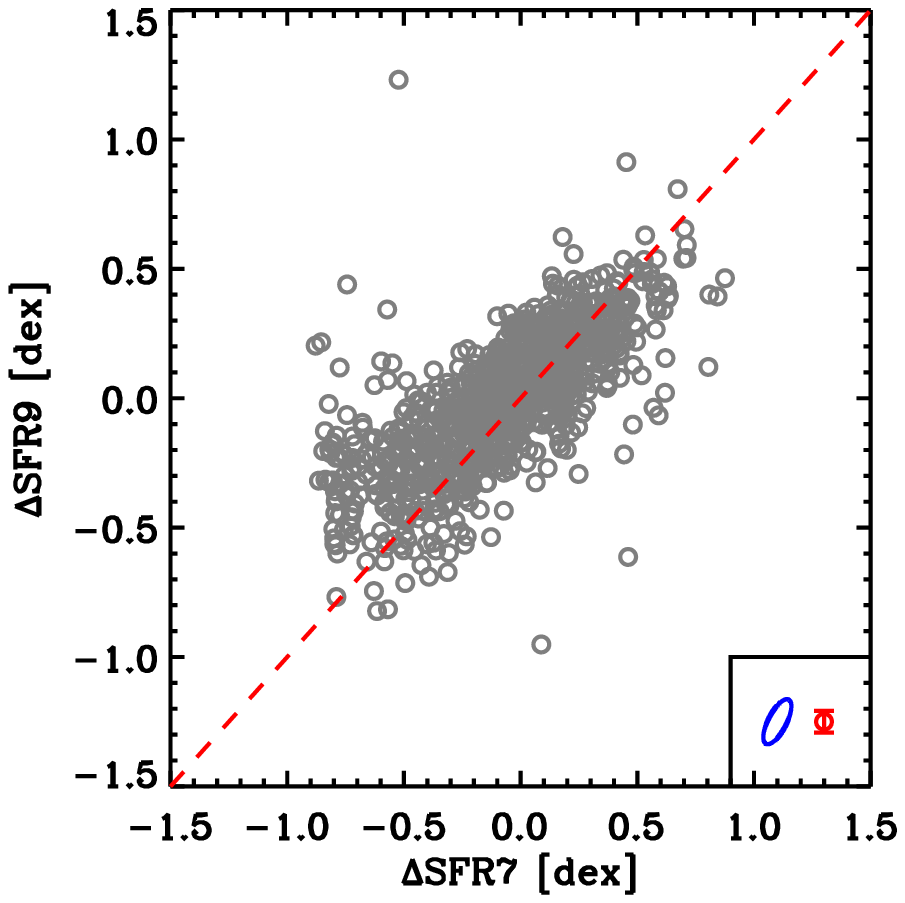,clip=true,width=0.42\textwidth} 
    \epsfig{figure=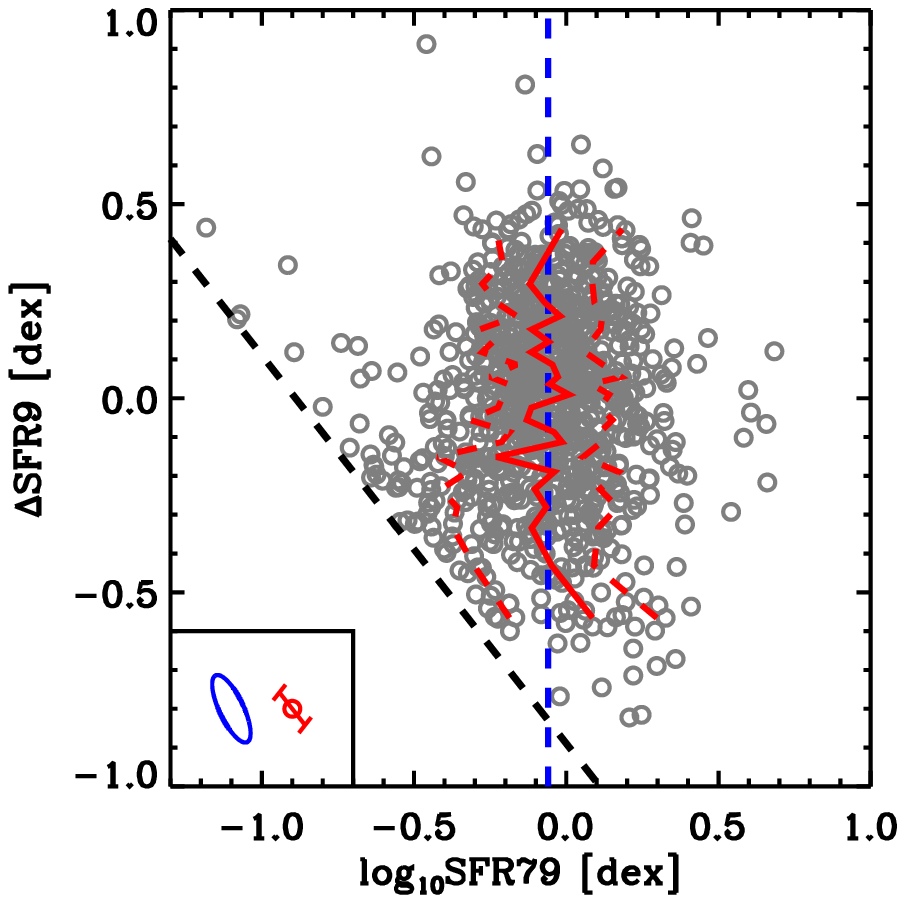,clip=true,width=0.42\textwidth}  
    \end{center}
  \caption{Top left panel: the SFR7 versus \mstar\ for the sample galaxies.
  Top right panel: the SFR9 versus \mstar\ for the sample galaxies. 
  In these two top panels, the blue data points are the median SFR in different 
  stellar mass bins. The black solid line is the best-fit line to the median 
  SFR7-\mstar\ relation of the galaxies with \mstar$<10^{10}$\msolar, 
  which is found to be a good description of the SFMS for SFR9. 
  The dashed line in the left panel indicates the selection boundary of SF galaxies 
  from \citetalias{Wang-19}. 
  Bottom left panel: the $\Delta$SFR7 versus $\Delta$SFR9 for
  the sample galaxies.  The dashed red line shows equality between these two.  Bottom right panel: the $\Delta$SFR9 as a function 
  of \RSFR\ for the sample galaxies. The dashed and solid red lines show
  the 16\%, 50\% and 84\% of \RSFR\ at different $\Delta$SFR9. The
  vertical blue dashed lines show the overall median \RSFR. The black dashed line 
  carries over the selection criteria of SF galaxies from the top left panel. 
  In each panel, we show the median errors (including the direction of the errors) 
  of the data points in the corner box: the blue one is the model uncertainty, and the red one shows the uncertainty coming from the measurements from the observation. In the top left panel, the median measurement uncertainty is negligible with respect to the model uncertainty for SFR7, and therefore we do not show the typical measurement uncertainty of SFR7. 
  }
  \label{fig:global}
\end{figure*}

\begin{figure*}
  \begin{center}
    \epsfig{figure=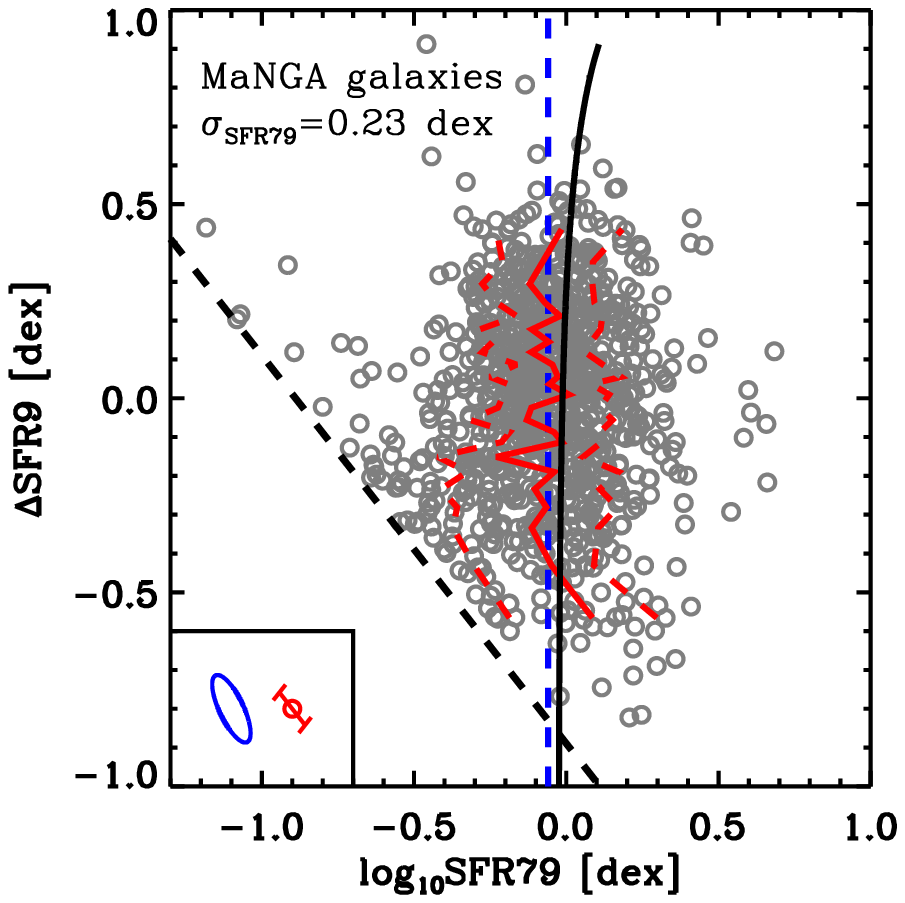,clip=true,width=0.42\textwidth} 
    \epsfig{figure=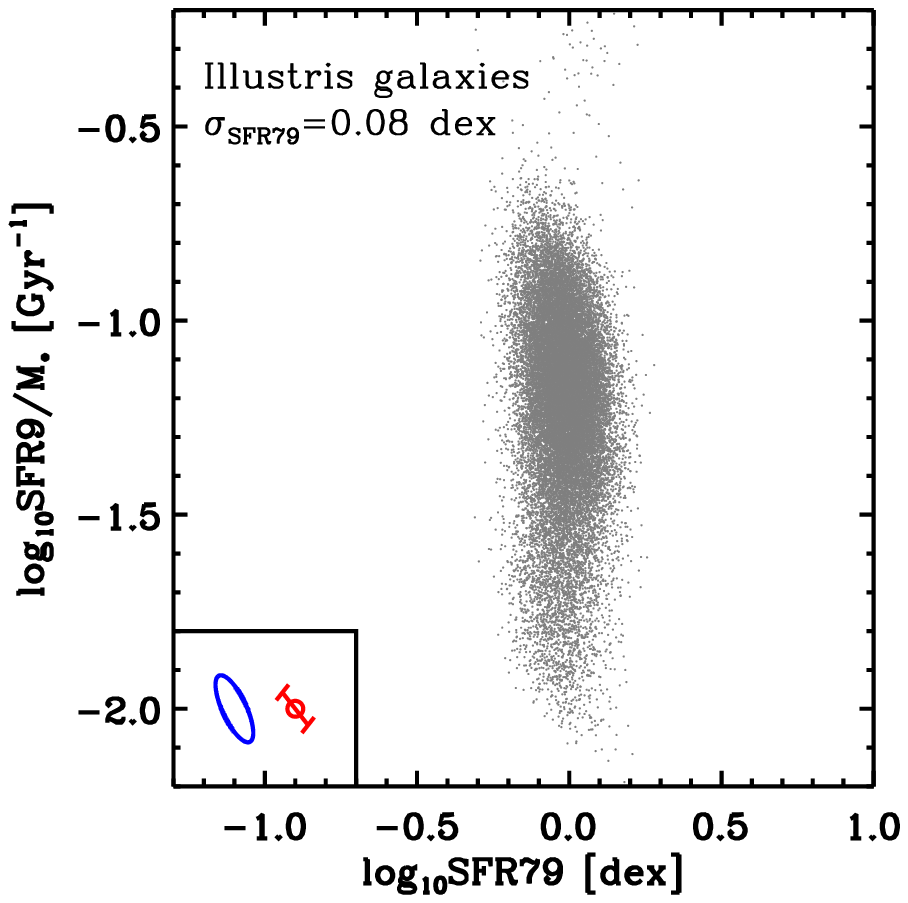,clip=true,width=0.42\textwidth} 
    \end{center}
  \caption{The left-hand panel is the same as the bottom right panel of Figure \ref{fig:global}, 
  but superposes 
  the relation of 
  \RSFR-$\Delta$SFR$_{\rm 800Myr}$ that is obtained for a series of SFHs that are
  parallel to the evolution of the SFMS, i.e. which have constant $\Delta$SFR$_{\rm 800Myr}$, as the black solid line.   The change in SFR, and thus in \RSFR\, due to maintaining a constant $\Delta$SFR$_{\rm 800Myr}$, is negligible.    Right panel: the relation between the SFR79-SFR9/$M_*$ in logarithmic space for the best-fit log-normal SFHs of the SF galaxies 
  (sSFR7 $>10^{-2}$ [Gyr$^{-1}$]) selected from Illustris from \cite{Diemer-17}.  
  Using the best-fit log-normal SFHs smooths out all short-term variations of the SFHs. For fair comparison with the left panel, we broaden the 
  distribution of galaxies in the right panel with the typical uncertainties from both models and observations, as indicated in the bottom left-hand corner of each plot. The resulting dispersions in \RSFR\ are 0.08 dex and 0.23 dex respectively. This comparison emphasizes that the distribution of the observed \RSFR\ in MaNGA is dominated by relatively short term variations of the SFR that are absent in the log-normal SFHs. 
  }
  \label{fig:check_novar}
\end{figure*}

\subsection{The SFR7-based and SFR9-based SFMS} \label{subsec:4.1}

We first examine the global SFR79 and SFR9 for the sample galaxies and use these to examine the SFMS when defined using the measures of star-formation rate on the two timescales of 5 Myr and 800 Myr. 
Consistently with the measurements of global SFR7 (i.e. SFR$_{\rm 5Myr}$)
and stellar mass, 
the global SFR9 is calculated for each individual galaxy by summing the flux in all the spaxels within the effective radius to obtain the integrated 
SFR7, SFR79 and thus SFR9 for the sample galaxies.  

%%% IS THIS A NEW SELECTION OR THE SAME DISCUSSED BEFORE.....
We exclude 16 
galaxies that are located in the Seyfert regions on the Baldwin-Phillips-Telervich (BPT)
diagram \citep{Baldwin-81, Kewley-06}, based on
the emission line flux ratios within the effective radius, since in these galaxies
the H$\alpha$ emission is largely contaminated by the contribution of the AGN.  
We also exclude 4 galaxies which have the three diagnostic parameters 
beyond the valid range of the calibrator (see details in Section \ref{subsec:2.3}).
These leave 956 galaxies.  

In the top two panels of Figure \ref{fig:global}, we present the SFMS based on
SFR7 (left-hand panel) and SFR9 (right-hand panel).  In both panels, 
the blue circles show the median SFR in different stellar mass bins. 
The black solid line is the best-fit straight line to the median SFR7-\mstar\ 
relation for galaxies with \mstar\ less than $10^{10}$\msolar. Following the work of
\citetalias{Wang-19}, we define the solid line as the ``nominal'' SFMS for 
SFR7. Interestingly, the 
line also matches the SFMS with SFR9 very well. This is consistent with 
the fact that the evolution of the SFMS is small within the last 800 Myr 
\citep[e.g.][]{Brinchmann-04, Pannella-09, Stark-13, Schreiber-15} and is a reflection of the fact that our average SFR79 is very close to unity, as discussed above. 
We therefore adopt the solid line as the ``nominal'' SFMS 
for both SFR7 and SFR9. 

The typical errors of SFR7 and SFR9, including the uncertainties 
from the observations and calibrator, are shown in the corner box of each panel in 
Figure \ref{fig:global}.  The uncertainty of SFR7 is 0.06 dex due to 
the conversion formula from \cite{Kennicutt-98}. The measurement uncertainty of the H$\alpha$
luminosity is negligible with respect to the uncertainty in the 
conversion formula to SFR, and therefore we do not show it in the top left panel.  
The typical uncertainty of SFR79 is 0.076 dex, obtained by combining the uncertainty from the calibrator
is 0.063 dex, and the measurement error is 0.042 dex.  Comparing with the intrinsic 
scatter within the galaxy population of SFR79 ($\sim$0.23 dex), the measurement and calibration uncertainty of SFR79 only broadens
the distribution of $\log_{10}$SFR79 by less than 10\%. This indicates that 
the dispersion of $\log_{10}$SFR79 
in the bottom right panel of Figure \ref{fig:global} is real, which provides 
the basic condition to study the variability of the SFHs. 

Comparing the top two panels in Figure \ref{fig:global}, it is noticeable 
that the scatter of the SFMS is much smaller when using SFR9 than when using SFR7.  This is to be expected.
Averaging the SFR over 
800 Myr eliminates the variation of SFH 
on shorter timescales. It should be noted that, as an extreme example of this,
the SFMS would have zero scatter, if the SFR was computed as the average over 
the age of Universe.  By examining galaxies from the EAGLE simulation, \cite{Matthee-19} also found that 
the scatter of SFMS becomes smaller when using the SFR averaged over longer
timescales.   

%This gives further confidence that the \RSFR\ estimator works well.  If the \RSFR\ measurements were meaningless, the scatter in SFR$_{\rm 800Myr}$ would be larger than in SFR$_{\rm 5Myr}$.

\subsection{The evolution of SFMS indicated by the change parameter} \label{subsec:4.2}

Based on the ``nominal'' SFMS, we now define two parameters to quantify the 
deviation in logarithmic SFR space of each galaxy from the main-sequence at its mass, 
$\Delta$SFR7 and $\Delta$SFR9 (in dex) of the galaxy, i.e. the vertical distance from the 
``nominal'' SFMS (see Figure \ref{fig:global}).  
The bottom left panel of Figure \ref{fig:global} shows the correlation between 
 $\Delta$SFR7 and $\Delta$SFR9.
 
This plot contains additional important information about how galaxies move above and 
below the ``nominal'' SFMS.   An extreme scenario in which  all SF galaxies 
evolved parallel to the ``nominal'' SFMS would produce
$\Delta$SFR9 always equal to $\Delta$SFR7, 
and therefore galaxies would exactly follow the one-to-one line with zero scatter 
on the $\Delta$SFR7-$\Delta$SFR9 diagram. 
We could imagine the opposite extreme case, in which the scatter of the 
SFMS is purely due to the variation of SFR on very short timescales ($<<$800 Myr).  In this case, 
we would expect that the $\Delta$SFR9 for all SF galaxies would be close to zero, and therefore galaxies would lie on a flat sequence with almost zero 
scatter on the $\Delta$SFR7-$\Delta$SFR9 diagram. 
For cases in between, galaxies would be located on a sequence with the slope 
between zero and one. The slope and the dispersion of the sequence on 
the $\Delta$SFR7-$\Delta$SFR9 diagram therefore indicates the
relative contributions to the dispersion of the SFMS on long and short timescales. 
In the second paper of this series, we will constrain the PSD of the specific 
SFHs of galaxies based on the location of galaxies on the 
$\Delta$SFR7-$\Delta$SFR9 diagram. We do not discuss this plot 
further here and refer the reader to that second paper. 

The bottom right panel of Figure \ref{fig:global} shows the relation
between $\Delta$SFR9 and \RSFR. The dashed and solid red lines
show the 16\%, 50\% and 84\% percentiles of \RSFR\ for galaxies at different 
$\Delta$SFR9.  The lack of galaxies in the bottom-left
corner is due to the fact that the definition of our SF galaxies was based on the SFR7-based SFMS
(see the dashed line in the top-left panel of Figure \ref{fig:global}). 
Galaxies (if any) below the dashed black line in the bottom right panel of Figure \ref{fig:global} 
would not be included in the sample selection.  

As discussed above, the \RSFR\ parameter determines the position of a galaxy on the  
SFMS (defined using the SFR over the last 5 Myr), i.e. $\Delta$SFR7, relative to its position on the SFMS that is defined using the star-formation averaged over the last 800 Myr, $\Delta$SFR9.  This latter quantity will to first order be the average $\Delta$SFR7 over the last 800 Myr.  Therefore, apart from the small offset due to the overall evolution of the SFR in star-forming galaxies with cosmic time, the sign of SFR79 indicates whether the galaxy is generally moving up or down in sSFR (i.e. its present position relative to its average position over the last 800 Myr).  Galaxies to the right of the vertical dashed line in the lower right panel of Figure \ref{fig:global} are in this sense increasing their sSFR, or ``going up'', while those to the left are ``going down'' relative to the SFMS.  

We have already commented that the value of  $\log_{10}\langle {\rm SFR7}\rangle/\langle {\rm SFR9}\rangle$ is closely matched to the overall evolution of the sSFR of the SFMS.  We furthermore here see in the lower right panel of Figure \ref{fig:global} 
that there is
no apparent correlation between \RSFR\ and $\Delta$SFR9. The absence of a correlation between \RSFR\ and $\Delta$SFR9 is required if the scatter of the SFMS is to remain more or less constant over cosmic time.  A strong 
positive correlation between $\Delta$SFR9 and \RSFR\ means that galaxies 
in the upper (lower) part of the SFMS would tend to move up (down)
with respect to the ``nominal'' SFMS, leading over time to an increased dispersion in the SFMS. Similarly, a strong anti-correlation between $\Delta$SFR9 
and \RSFR, would lead to a reduced dispersion over time.  

A roughly
constant scatter of the SFMS is indeed seen over a wide range of cosmic epochs \citep[e.g.][]{Speagle-14, 
Whitaker-14, Schreiber-15, Barro-17}, requiring that there should be no correlation between \RSFR\ and $\Delta$SFR9.  The fact that this is indeed seen in our \RSFR\ estimates is an important external consistency check that provides further
confirmation that our estimator for \RSFR\ works well.

Furthermore, it is noticeable that the distribution of \RSFR, at given $\Delta$SFR9, is quite {\it symmetrical} about the median value.  This symmetry is in contrast to the pronounced asymmetry in \RSFR\ that is visible in the bottom rightmost panel of Figure \ref{fig:rsfr_example}.  In broad terms, this symmetry implies that, for an individual galaxy as a whole, the timescales of ``above average star-formation" and ``below average star-formation" are broadly similar.   As an example, short periods of highly elevated SFR superposed on longer periods of constant SFR would produce an asymmetric distribution in \RSFR\ with a peak at slightly negative \RSFR\ and a tail to high positive \RSFR. However, we do not see this in the data (the bottom right panel of Figure \ref{fig:global}). 

We return here to a point touched on earlier, namely that a galaxy with a constant sSFR will
have slightly positive \RSFR. This is because the SFR increases as the stellar mass
increases.  We show this in the left-hand panel of  Figure \ref{fig:check_novar} which reproduces the data from 
the bottom right panel of Figure \ref{fig:global}. If all galaxies had a fixed position
$\Delta$sSFR relative to the (slowly evolving) SFMS, and experienced no other
variability in their (s)SFR, then they would lie precisely along the solid black line in the left
panel of Figure \ref{fig:check_novar}, displaced to left and right only by observational scatter in determining \RSFR.
Because the sSFR of the SFMS is low at the current epoch, i.e. ${\rm sSFR}^{-1}>>$ 1 Gyr,
the \RSFR\ produced by constant $\Delta$sSFR is negligible for SFMS galaxies, essentially because the mass change of the galaxy during one
Gyr is negligible. 

This emphasizes that the observed scatter in \RSFR\ within the
population is completely dominated by time-variability in the SFR (and sSFR) of galaxies on
timescales of 1 Gyr or less, and not by a range of (unvarying) sSFR within the population. 
This is further illustrated in the right hand panel of Figure \ref{fig:check_novar} in which
we plot the current-epoch \RSFR\ of the 26,485 log-normal SFH of star-forming
(sSFR7 $> 10^{-2}$ Gyr$^{-1}$) Illustris galaxies from \cite{Diemer-17} that were discussed above.
These log-normal smooth SFH will by construction not have short-term variability.  The \RSFR\ are computed
directly from the SFH, but we add Gaussian observational scatter to the points to simulate
the real data.  
%Except for a handful of outliers (15), representing less than 0.1\% of the
%population, the vast majority of these smooth SFH lie along a very narrow sequence in the diagram.  As an aside, the few outliers are all galaxies undergoing declining star-formation after a recent narrow peak.  Whether such ``late-blooming'' SFH represent real massive galaxies in the Universe is unclear to us. 
The dispersion in \RSFR\ of this simulated population of log-normal galaxies is only 0.08 dex (produced almost entirely by the addition of observational uncertainties), much less than the observed dispersion of 0.23 dex.  This comparison emphasizes that the much broader scatter in
\RSFR\ in the real MaNGA data is caused by real short-term temporal variations in the SFR of
galaxies that are not present in the log-normal SFH fits given by \cite{Diemer-17}.  

%In addition, \RSFR\ is nearly independent of $\Delta$SFR$_{\rm 800Myr}$, indicating 
%that the movement of galaxies up and down on the SFMS at most recent 5 Myr does not 
%depend on the position on the SFR9-based SFMS. Indeed, this is a vital 
%condition to maintain the scatter of the main-sequence. 

\section{The spatially-resolved analysis of \RSFR\ } \label{sec:5}

The \RSFR\ profile for each galaxy is constructed as in \citetalias{Wang-19}.  We divide the spaxels for each galaxy into a set of 
non-overlapping elliptical annular bins with a constant radial interval in deprojected radius of 
$\Delta(r/Re)$=0.2.  For a given galaxy, we compute the deprojected radius 
from the center of the galaxy based on the minor-to-major axis ratio from 
the NSA (NASA-Sloan Atlas) catalog \citep{Blanton-11}.
Then the three diagnostic parameters, as well as the 
E(B-V) of the nebular emission are determined from the spaxels within each of these annuli following the approach 
described in Section \ref{subsec:3.3}.

\begin{figure*}
  \begin{center}
    \epsfig{figure=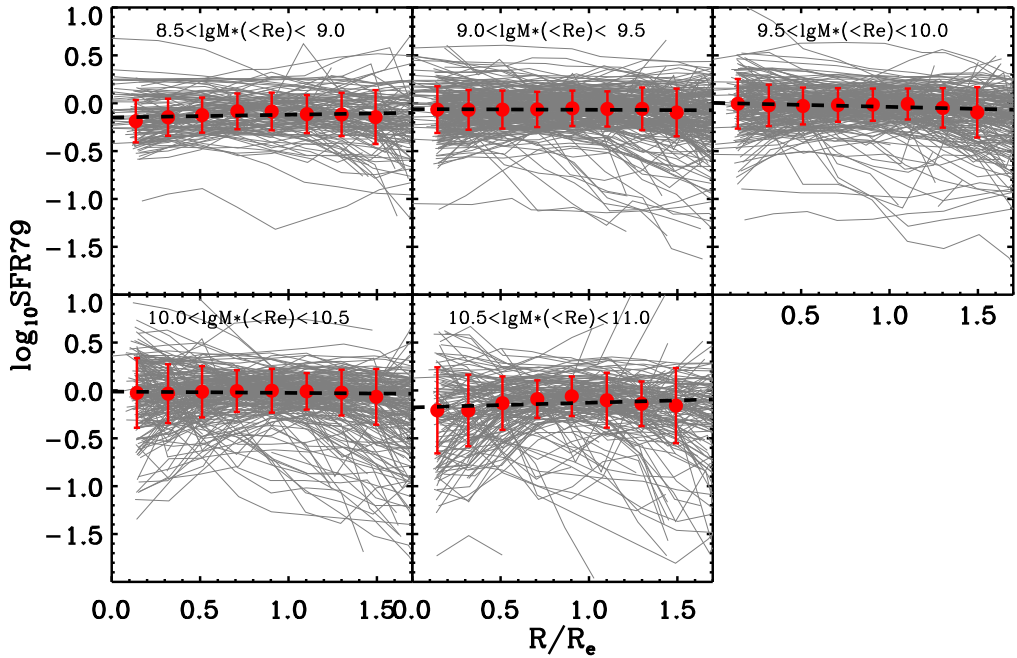,clip=true,width=0.82\textwidth} 
    \end{center}
  \caption{The \RSFR\ profiles for individual MaNGA galaxies.  We
  separate galaxies into five equal stellar mass bins in logarithmic
  space.
  %, 8.5$<$\lgmstar$<$9.0, 9.0$<$\lgmstar$<$9.5, 9.5$<$\lgmstar$<$10.0,
  %10.0$<$\lgmstar$<$10.5, and 10.5$<$\lgmstar$<$11.0, respectively. 
  In each panel, 
  the red dots show the median \RSFR\ profile of the sample galaxies 
  in the corresponding stellar mass bin.  The error bars on these points indicates the 1 $\sigma$ scatter 
  of \RSFR\ at a given radius. }
  \label{fig:rsfr_prof}
\end{figure*}

Figure \ref{fig:rsfr_prof} shows the \RSFR\ profiles for the individual 
galaxies in five stellar mass bins of 0.5 dex, from \lgmstar=8.5 to \lgmstar=11.0. 
Figure \ref{fig:rsfr_prof} shows that the median \RSFR\ profiles (indicated by red dots)
of each of the five mass sub-samples are overall flat. This is not surprising 
and would be expected since we applied an ad hoc $\Sigma_*$ adjustment as described in Section \ref{subsec:3.2}.  Of more interest is the scatter in \RSFR\ within the population, which is 
shown with the red error bars. These show the dispersion $\sigma$(\RSFR) computed using a three-sigma-clip algorithm. We will return to this point below.

In \citetalias{Wang-19}, we studied the star-formation profiles of this same sample of galaxies.  Specifically, we looked at the
radial elevation or suppression of the star-formation surface density (as measured on 5 Myr timescales) as a function of the overall displacement of the galaxy from the mid-line of the SFMS.  We found that galaxies 
that are significantly above the SFMS in their overall SFR show enhanced star formation surface densities at all galactic radii and that, conversely,
galaxies that are significantly below the SFMS in overall SFR show suppressed star formation surface densities at all galactic
radii.  Interestingly, we found that this relative enhancement (or suppression) of star formation is greater in the central regions than in the outer regions for galaxies with \lgmstar$>$9.5.  

We illustrated this further by showing  
(see Figure 9 of \citetalias{Wang-19}) that the dispersion in the radial SFR surface density at a given relative radius (for a given stellar mass bin), which we parameterized as $\sigma(\Delta\Sigma_{\rm SFR})$ in the notation of \citetalias{Wang-19}, strongly depended on the apparent effective gas depletion timescale.  The depletion timescale was estimated using different proxies 
\citep{Shi-11, Krumholz-12} rather than from a direct measure of the gas content, so the observed trend could equally well be viewed as a trend with stellar surface mass density $\Sigma_*^{-0.5}$ or the other proxies used in these relations.

We interpreted the \citetalias{Wang-19} result as possibly reflecting the dynamical response of a gas-regulator system to changes in the gas inflow rate. We constructed a heuristic toy model in which an idealized gas-regulator \citep{Lilly-13} was driven by a 
periodic inflow of different forms, either a sinusoidal function or the inverse
error function.  The amplitude (and to a lesser extent the phase) of the SFR response of the regulator system to this periodic variation of gas inflow varies with the ratio of the driving period and the effective gas depletion timescale, since the latter sets the response time of the regulator (see \cite{Lilly-13} for details).    The observed variation of $\sigma(\Delta\Sigma_{\rm SFR})$ across galaxies and the broad 
features of the SFR7 profiles could therefore be explained through this simple mechanism.   

We can now use the \RSFR\ profiles derived in the current paper to extend this result in two ways.  First, rather straightforwardly, we can use the \RSFR\ profile to construct the $\Delta\Sigma_{\rm SFR9}$ profiles of galaxies, to complement the analysis of the $\Delta\Sigma_{\rm SFR7}$ profiles presented in \citetalias{Wang-19}.  Second, and of more originality, we can use the \RSFR\ profiles to directly examine the {\it temporal} variations of SFR within galaxies.  This is an important point.  In \citetalias{Wang-19}, we interpreted the dispersion in $\Delta\Sigma_{\rm SFR}$ as arising from temporal variations in SFR, but this was precisely that: an interpretation.  One could, at least in principle, imagine that galaxies evolve in such a way as to maintain a constant displacement of their overall sSFR from the mid-line of the SFMS, with a corresponding constant offset in $\Delta\Sigma_{\rm SFR7}$.  In this case, the scatter of the SFMS and the scatter in star-formation surface density $\sigma(\Delta\Sigma_{\rm SFR7})$ could have nothing to do with any temporal variations in the SFR of individual galaxies, but rather reflect ``intrinsic'' (i.e. time-independent) differences between galaxies.

The information on \RSFR\ in the current work allows us to break this interpretational ambiguity decisively.  Significant variations in \RSFR\ can {\it only} arise because of real temporal variations in the SFR within {\it individual} galaxies.  It is trivial that a constant SFR in galaxies will always produce an \RSFR\ that is precisely unity. In the previous Section, we showed that, while a constant sSFR will produce values of \RSFR\ that deviate from unity, these effects are completely negligible for the sSFR of interest (see the bottom right panel of Figure \ref{fig:global}). Unless the SFMS is a transitory phenomenon (which it is not) the scatter in \RSFR\ within the population is therefore completely dominated by real temporal variations of SFR (and sSFR) within individual galaxies.   Therefore, if the interpretation of the $\sigma(\Delta\Sigma_{\rm SFR7})$ relation that we advanced in \citetalias{Wang-19} is correct, then we should definitely expect to see a correlation between the dispersion in \RSFR\ with the gas depletion timescale.  

In the next subsection of the paper, we construct first the $\Sigma_{\rm SFR9}$ profiles and carry out a completely analogous analysis to that of $\Sigma_{\rm SFR7}$ that was presented in \citetalias{Wang-19}.  Then, in the subsequent subsection, we then turn to examine the \RSFR\ profiles for more direct information on temporal variability.

\subsection{The profiles of $\Sigma_{SFR9}$} \label{subsec:5.1}

\begin{figure*}
  \begin{center}
    \epsfig{figure=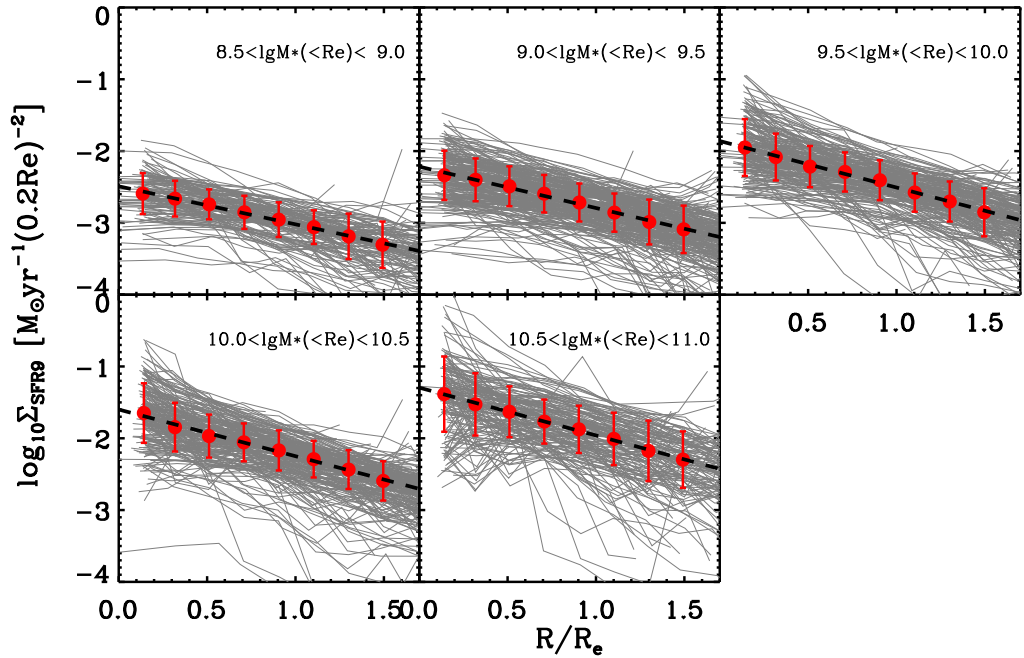,clip=true,width=0.82\textwidth} 
    \end{center}
  \caption{The $\Sigma_{\rm SFR9}$ profiles for individual galaxies. We display the
  profiles in the same five stellar mass bins as in Figure
  \ref{fig:rsfr_prof}. In each panel, the red dots show
  the median $\Sigma_{\rm SFR9}$ profile, and the error bars show the 
  scatter of $\Sigma_{\rm SFR9}$ at given radii.  In each panel, the dashed line 
  is the fit of the median profile by an exponential disk. 
  }
  \label{fig:sfr9_prof}
\end{figure*}

\begin{figure*}
  \begin{center}
    \epsfig{figure=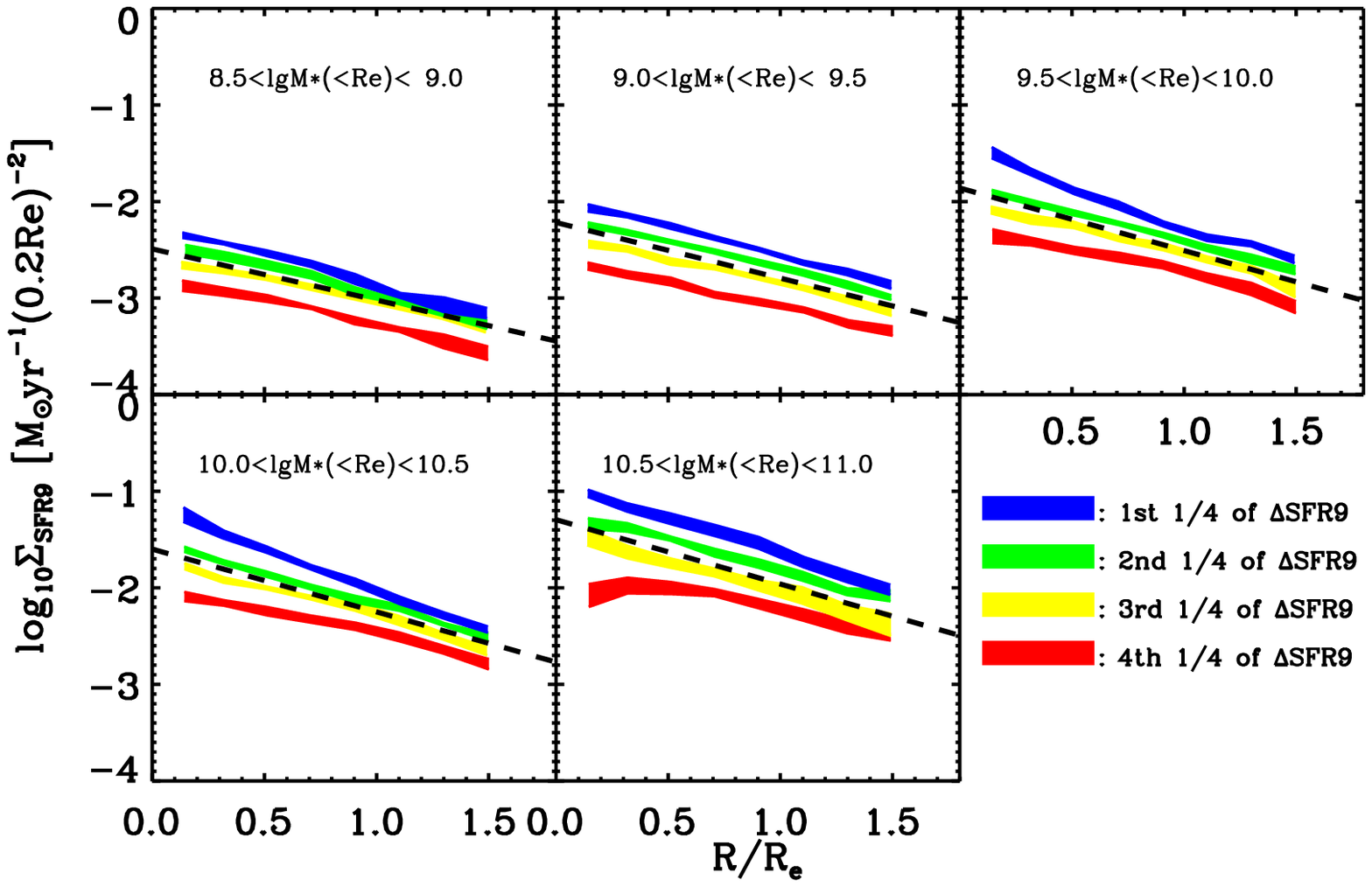,clip=true,width=0.82\textwidth} 
    \epsfig{figure=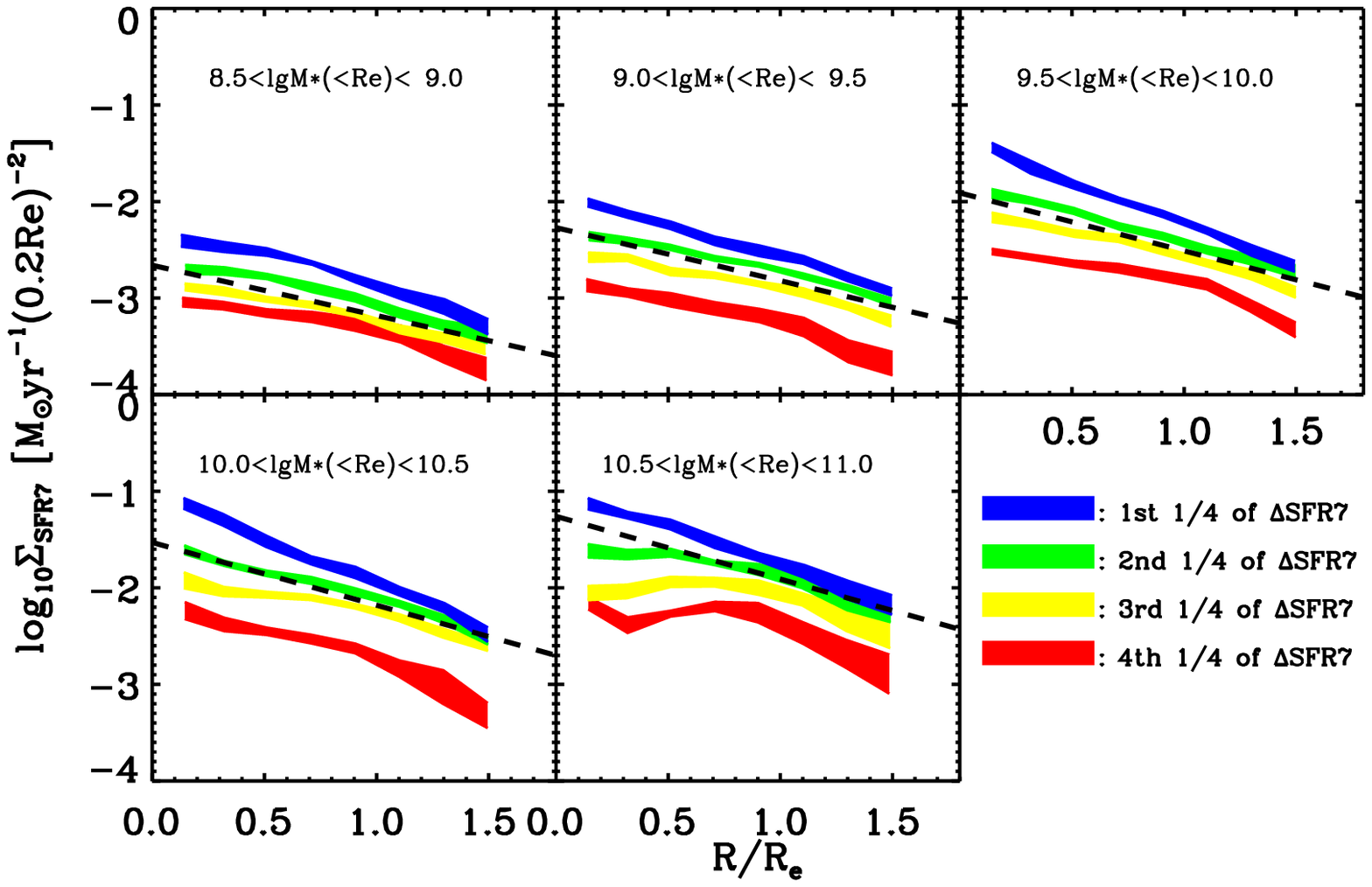,clip=true,width=0.82\textwidth} 
    \end{center}
  \caption{Top five panels: the median $\Sigma_{\rm SFR9}$ profiles of galaxies
  in each quartile of $\Delta$SFR9 bins, in each of the five mass ranges.  Bottom five panels: the median $\Sigma_{\rm SFR7}$
  profiles for galaxies in each quartile of $\Delta$SFR7. 
  In each stellar mass bin, we divide galaxies into 
  four subsamples according to global $\Delta$SFR9 (or $\Delta$SFR7),
  i.e. the deviation from the SFR9-based (or SFR7-based) SFMS. 
  In both sets of panels, the black dashed lines are the median
profiles of all galaxies, taken from Figure \ref{fig:sfr9_prof} (or figure 4 of \citetalias{Wang-19}). In each panel, the thresholds of the quartiles for $\Delta$SFR7 or $\Delta$SFR9 used to separate galaxies are listed in Table \ref{tab:3}, as well as the total number of galaxies in the corresponding stellar mass bin (the last column in Table \ref{tab:3}).
}
  \label{fig:bin_sfr9_prof}
\end{figure*}

\begin{figure*}
  \begin{center}
    \epsfig{figure=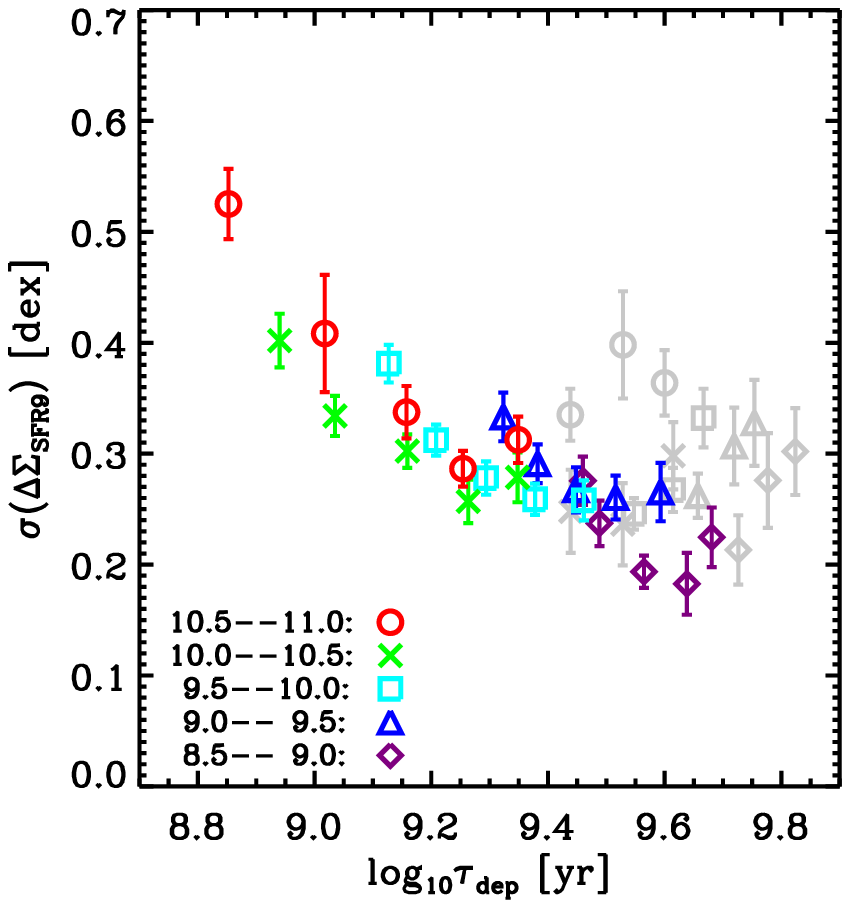,clip=true,width=0.42\textwidth} 
    \epsfig{figure=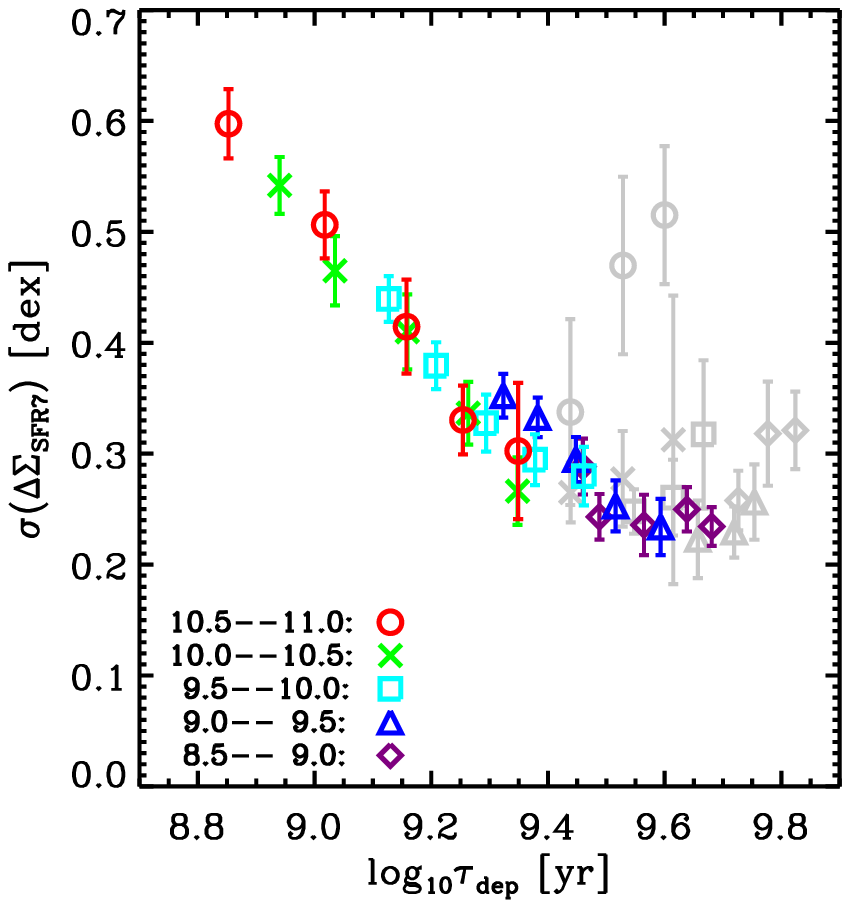,clip=true,width=0.42\textwidth}
  \end{center}
  \caption{Left panel: The scatter of $\Delta{\Sigma}_{\rm SFR9}$ as a function of the indirectly estimated gas depletion time.
  Right panel: The scatter of $\Delta{\Sigma}_{\rm SFR7}$ as a function of gas
  depletion time, taken from \citetalias{Wang-19}. In both panels, the different colors are for galaxies in
  different stellar mass bins, as denoted in the bottom-left corner. 
  Data points with the radius larger than \re, are indicated in gray, as in \citetalias{Wang-19}.}
  \label{fig:scatter_dsfr9}
\end{figure*}

\begin{table*}[ht]
\renewcommand\arraystretch{1.5}
\begin{center}
\caption{The thresholds of the quartiles of $\Delta$SFR7, $\Delta$SFR9 and SFR79 used to classify galaxies in Figure \ref{fig:bin_sfr9_prof} and \ref{fig:bin_rsfr_prof}. } \label{tab:3}
\begin{tabular}{@{}lrrrrrrrrrrrrrr@{}}
\tableline
\tableline

\multirow{2}{*}{$\log M_*(<Re)$} & & 
\multicolumn{3}{c}{$\Delta$SFR7} & & 
\multicolumn{3}{c}{$\Delta$SFR9} & & 
\multicolumn{3}{c}{SFR79} &  & 
N$_{\rm gal}$ \\
 & & 
25\% & 50\% & 75\% & & 
25\% & 50\% & 75\% & & 
25\% & 50\% & 75\% &  & 
  \\

\cline{3-5}  \cline{7-9} \cline{11-13}
% $\log M_*(<Re)$&     &  $\Delta$SFR7  &   &  &  $\Delta$SFR9 &    &  & SFR79 &   &  N$_{\rm gal}$ \\
%\tableline
%\multicolum{test} \\ \cline{1-3}

[8.5,9.0] & & $-$0.134 & $-$0.009  &   0.176 & &  0.009  &    0.131   &   0.287 & &   $-$0.256  &   $-$0.113  &  0.002 &  & 129  \\
 
[9.0,9.5]  &  & $-$0.219 & $-$0.009 &    0.178 & & $-$0.144  &   0.061    &  0.198 & &  $-$0.185   & $-$0.068  &   0.067 &  & 240  \\
 
[9.5,10.0]  &  &  $-$0.246 &  $-$0.024  &   0.207 & & $-$0.188  &   0.014    &  0.194 & & $-$0.140  &  $-$0.024   &  0.087 & & 200 \\

[10.0,10.5]  &  &  $-$0.349  & $-$0.061  &   0.161 & & $-$0.231  &  $-$0.075   &   0.105 & &    $-$0.133   & $-$0.012    &  0.106 & & 198  \\

[10.5,11.0]  &  &  $-$0.502   & $-$0.219 &   0.027 & & $-$0.263 &   $-$0.080   &  0.070 & &   $-$0.300  &   $-$0.147   &  0.027 & & 149  \\

\tableline
\tableline
\end{tabular}
\end{center}
\end{table*}

The profiles of $\Sigma_{\rm SFR9}$ for 
each individual galaxies, shown in Figure \ref{fig:sfr9_prof}.  For each galaxy,  
the $\Sigma_{\rm SFR9}$ profile is generated by combining its \RSFR\ profile (shown in
Figure \ref{fig:rsfr_prof}) and the $\Sigma_{\rm SFR7}$ profile in figure 3 of \citetalias{Wang-19}. 
When presenting the $\Sigma_{\rm SFR7}$ profiles in \citetalias{Wang-19}, 
we normalized the radius for each galaxy by dividing by the effective radius
of the galaxy. This common procedure removes the effect due to the variation of galaxy size \citep[e.g.][]{Gonzalez-Delgado-16, Ellison-18, Medling-18, Guo-19}. 
However, more originally,
we also normalized the $\Sigma_{\rm SFR}$ with (0.2\re)$^2$, i.e. computed the
surface density of the SFR per area of (0.2\re)$^2$.  
This ensures that the integration of a profile on a given surface 
density-radius diagram reflects the actual integrated quantity in physical terms. 
In the current work, we therefore present the $\Sigma_{\rm SFR9}$ profiles in the same way.

As shown in Figure \ref{fig:sfr9_prof}, the $\Sigma_{\rm SFR9}$ profiles vary from 
galaxy to galaxy. Analogous to the result of $\Sigma_{\rm SFR7}$ profiles in \citetalias{Wang-19},  
the scatter in $\Sigma_{\rm SFR9}$ across the galaxy population shows a larger 
scatter in galactic center than the outer 
regions ($<Re$), at least for the three highest stellar mass bins.  
The median profile of $\Sigma_{\rm SFR9}$ (red dots) can be well fitted by
a straight line (the black dashed line), indicating a typical exponential star 
formation disk with no suppression of star formation (or quenching) in the central regions of galaxies.  The slopes of the median 
$\Sigma_{\rm SFR9}$ profile are $-$0.53, $-$0.58, $-$0.65, $-$0.65 and $-$0.67 from 
low to high stellar mass bins, which are almost the same to the slopes 
of median $\Sigma_{\rm SFR7}$ profile computed in \citetalias{Wang-19} at 
corresponding mass bins.  This agrees well with the notion that star 
formation within main-sequence galaxies varies in a quasi-steady 
state within an exponential disk, ignoring other structural properties
such as the presence or absence of a bulge.  

We split the galaxies in each stellar mass bin into four 
quartiles of the global $\Delta$SFR9, i.e. the position relative to the SFMS as defined by the SFR over the last 800 Myr.  The thresholds of the quartiles for $\Delta$SFR9 are listed in Table \ref{tab:3} for each stellar mass bin. 
The top set of panels in Figure \ref{fig:bin_sfr9_prof} show the median profiles of 
$\Sigma_{\rm SFR9}$ for the four quartiles in global $\Delta$SFR9 for each of the five stellar mass bins.
For comparison, we also evenly divide galaxies into four subsamples according to 
the global $\Delta$SFR7, and present their median $\Sigma_{\rm SFR7}$ profiles 
in the bottom panels of Figure \ref{fig:bin_sfr9_prof}.  
The thickness of the median profile indicates its uncertainty computed by the bootstrap method. 

It can be seen from Figure \ref{fig:bin_sfr9_prof} that 
galaxies with higher (or lower) global SFR9 show enhanced  
(or suppressed) star formation surface density (within the last 800 Myr) at all galactic radii.
This elevation (or suppression) in galactic center is more pronounced 
in more massive galaxies. 
This result is entirely consistent with and analogous to the result shown in \citetalias{Wang-19} for $\Sigma_{\rm SFR7}$,
%and also in the bottom panels of Figure \ref{fig:bin_sfr9_prof}, 
that galaxies 
with higher integrated SFR7 show higher SFR7 at all galactic radii. 
%We conclude that the profiles of $\Sigma_{\rm 800Myr}$ strength the result in 
%\citetalias{Wang-19}, which is consistent with the prediction of the gas-regulator model
%driven by an oscillating gas inflow. 

Given the similar results in SFR7 and SFR9 in Figure \ref{fig:bin_sfr9_prof}, one might worry that this is somehow due to SFR9 being a derived quantity based on SFR7 and SFR79, the ratio of SFR7/SFR9.   However, it is easy to see that this in fact makes the measurement of SFR9 largely independent of that of SFR7.  The former is ultimately linked to the strength of H$\delta$ absorption and the latter to the strength of H$\alpha$.  
Clearly, if the H$\alpha$ emission due to star formation is over-estimated, for instance because of 
the contribution of an AGN component, then both the SFR7 and SFR79 would be more or less equally perturbed.  Ultimately the value of SFR9 is based primarily on the \ewhda\ value.  
The result in Figure \ref{fig:bin_sfr9_prof} is therefore largely independent of the analogous 
result in \citetalias{Wang-19}. 

\citetalias{Wang-19} defined a parameter, $\Delta \Sigma_{\rm SFR7}$, to quantify the 
deviation from the median profile for a given stellar mass. In the same way, 
we now define the deviation from the median profile of $\Sigma_{\rm SFR9}$ 
to be $\Delta \Sigma_{\rm SFR9}$.  Since the median profiles of $\Sigma_{\rm SFR7}$
have almost the same slopes of $\Sigma_{\rm SFR9}$ (and also almost the same 
intercepts) at all the five stellar mass bins, we use the 
same set of median profiles as in \citetalias{Wang-19} to calculate these deviations.  We note that this does not affect the measurement of the {\it scatter} 
of $\Delta \Sigma_{\rm SFR9}$ at a given galactic radius. 

We show in Figure \ref{fig:scatter_dsfr9} the scatter of $\Delta \Sigma_{\rm SFR9}$ and $\Delta \Sigma_{\rm SFR7}$ 
(the second is taken from the right panel of figure 9 in \citetalias{Wang-19} for comparison)
as a function of gas depletion time.  As in \citetalias{Wang-19}, the gas depletion time (i.e. the inverse of the star formation efficiency SFE) is calculated based on an empirical 
formula from \cite{Shi-11}, the so-called extended-Schmidt law: 
\begin{equation} \label{eq:8}
  \frac{\rm SFE}{yr^{-1}} = 10^{-10.28} \left[\frac{\Sigma_*}{M_{\odot} {\rm pc}^{-2}}\right]^{0.48}
\end{equation}
This formula is derived from integrated observations of individual galaxies 
over five orders of magnitude in stellar mass density, but was found to also
be valid for spiral galaxies at sub-kiloparsec resolutions and 
low-surface-brightness regions. In \citetalias{Wang-19}, we also presented 
another empirical formula for the gas depletion time by using 
the orbital timescale \citep{Krumholz-12}. Since the two give similar result, in this work
we only present the one computed with the extended-Schmidt law.  
%The colors and symbols 
%in Figure \ref{fig:scatter_dsfr9} are the same as those in Figure \ref{fig:scatter_rsfr}, and
Each data point represents one galactic radial bin within a given galactic stellar mass bin,
as in \citetalias{Wang-19}. 
The colored data 
points are for the radii less than \re, while gray radii are for the radii greater than \re. 
Here, we will only focus on those regions within the effective radius as done also in \citetalias{Wang-19}, 
because the outer regions are more likely affected by environmental effects, 
such as ram-pressure stripping, and etc.. 

Consistent with the result of \citetalias{Wang-19}, the scatter in $\Delta \Sigma_{\rm SFR9}$ 
is also evidently a decreasing function of the gas depletion time.  
The results on the SFR surface density on 800 Myr timescales, $\Sigma_{\rm SFR9}$, presented in this subsection extend and confirm the results on the SFR surface density on 5 Myr timescales, $\Sigma_{\rm SFR7}$, discussed in \citetalias{Wang-19}.  The basic issue of whether variations in $\Sigma_{\rm SFR}$ reflect temporal variations or intrinsic (time-independent) differences from galaxy to galaxy however remains.  To resolve this issue, we now turn to examine the \RSFR\ profiles more directly.

\subsection{The radial profiles of \RSFR\ } \label{subsec:5.2}

\begin{figure*}
  \begin{center}
    \epsfig{figure=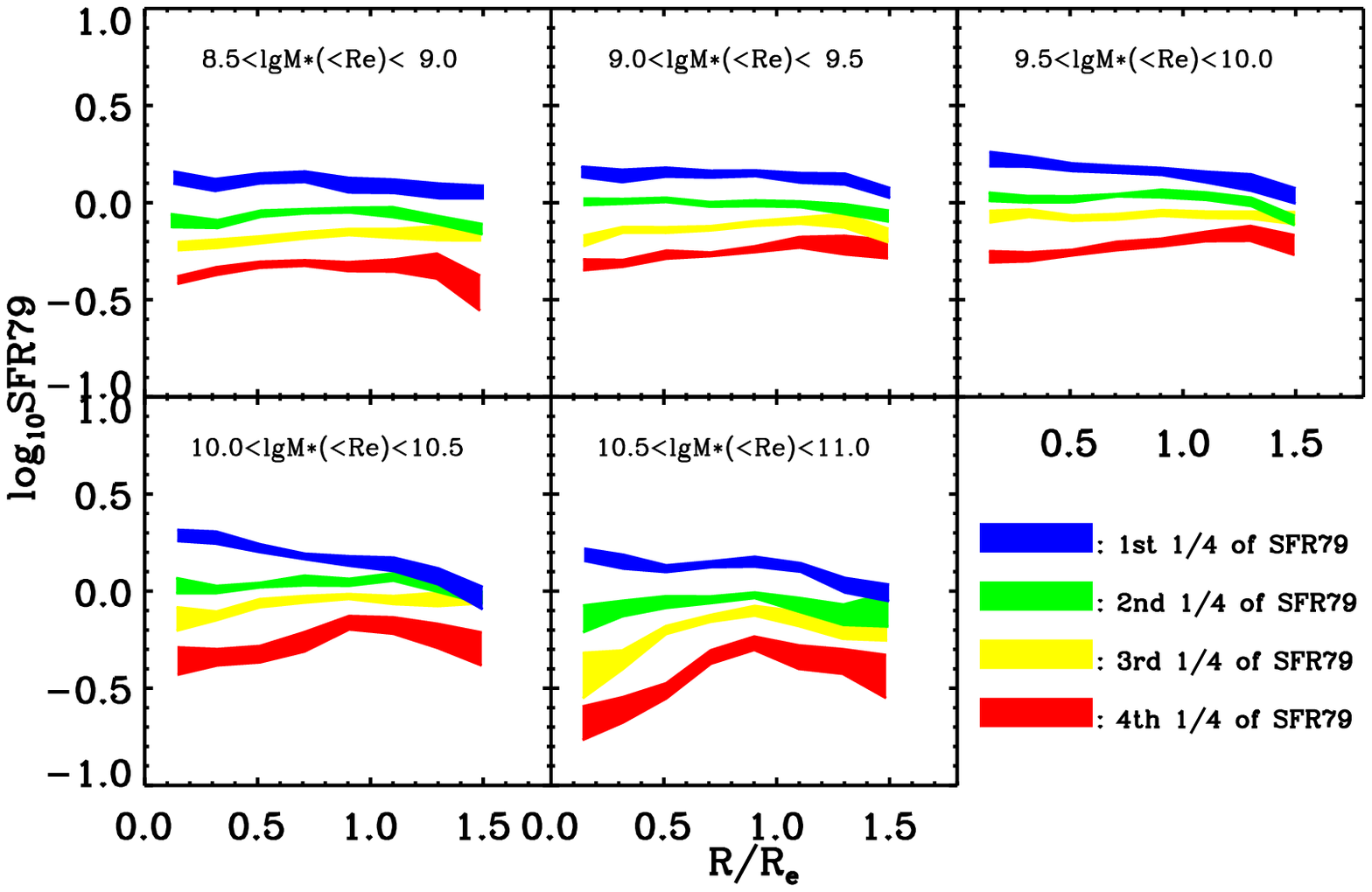,clip=true,width=0.82\textwidth} 
    \epsfig{figure=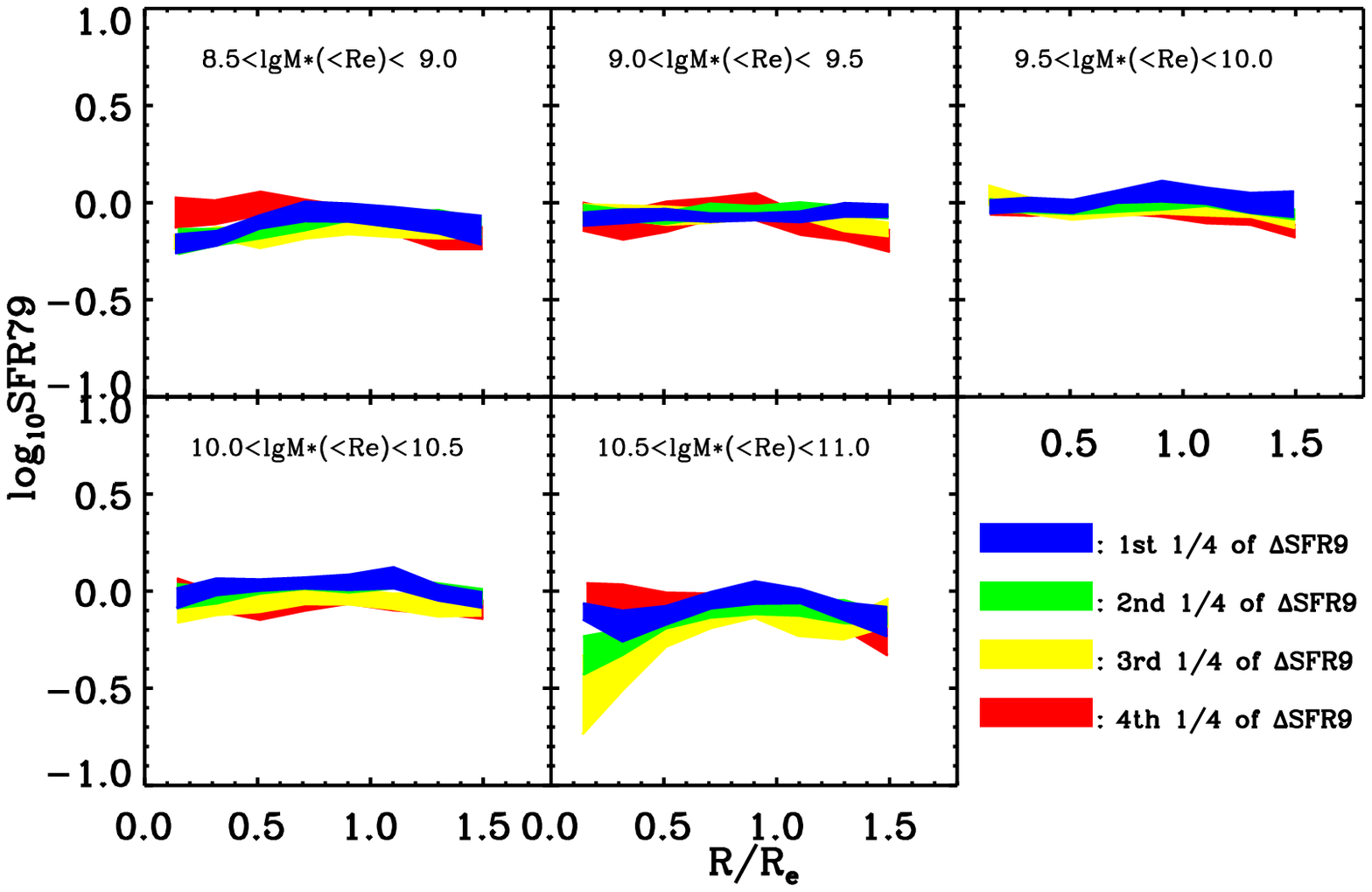,clip=true,width=0.82\textwidth} 
    \end{center}
  \caption{Top five panels: the median \RSFR\ profiles for galaxies in quartiles of the overall \RSFR. 
  Bottom five panels: the median \RSFR\ profiles for galaxies in quartiles of the overall $\Delta$SFR9.
In each stellar mass bin, the thresholds of the quartiles for SFR79 or $\Delta$SFR9 used to separate galaxies are listed in 
 Table \ref{tab:3}.
  %For each stellar mass bin, we separate galaxies into four subsamples according to the 
  %global \RSFR\ (or $\Delta$SFR9). 
  %We first sort the sample galaxies by the global \RSFR\ (or $\Delta$SFR9)from high %to low. 
  %The blue, green, yellow and red lines shows the median  \RSFR\ (or $\Delta$SFR9) %profile 
  %of galaxies in the first quarter, the second quarter, the third quarter and
  %the bottom quarter, respectively.   
  }
  \label{fig:bin_rsfr_prof}
\end{figure*}

The \RSFR\ profiles of the galaxies in the sample were shown in Figure \ref{fig:rsfr_prof}.
We now divide the galaxies in each stellar mass bin into four quartiles according 
to their global \RSFR\ as measured within their effective radii in Section \ref{sec:4}.  
The top group of panels of
Figure \ref{fig:bin_rsfr_prof} shows the median \RSFR\ profiles of the 
four quartiles in overall \RSFR\ for each stellar mass bin.   The thickness of each line reflects the
uncertainty of the median \RSFR\ profile as computed via a bootstrap approach. 

It can be seen that galaxies with high global \RSFR\ appear to have larger \RSFR\ 
at all galactic radii, and galaxies with low global \RSFR\ appear to have lower
\RSFR\ at all galactic radii.  This means that a galaxy with a certain global change in SFR, i.e.
an elevation (or suppression) of the SFR
with respect to the average over the last 800 Myr, experiences a corresponding elevation (or suppression)
at all galactic radii.  
Furthermore, it can be seen that, at least for the three more massive bins of galaxy mass, this change 
(either elevation or suppression) is more pronounced at small galactic radii. 

In the bottom panels of Figure \ref{fig:bin_rsfr_prof}, we present the analogous \RSFR\ profiles
for the same five stellar mass bins but now dividing galaxies into four quartiles of $\Delta$SFR9, i.e. the position of the galaxies relative to the SFMS defined in terms of the SFR averaged on 800 Myr timescales. 
For all the mass bins, the \RSFR\ profiles appear
to be overlapped together, independent of the $\Delta$SFR9.  This means that 
the \RSFR\ profile does not depend on the location of galaxies relative to the SFR9-based SFMS. This is completely consistent with the result for the {\it integrated} \RSFR\ that was presented above in Section \ref{subsec:4.2}. 

These results in Figures \ref{fig:rsfr_prof} and \ref{fig:bin_rsfr_prof} are consistent but quite 
distinct from the result in \citetalias{Wang-19}.  \citetalias{Wang-19} found that 
galaxies with higher SFR with respect to the SFMS show higher SFR at all galactic radii. However, there was no direct information on the temporal changes in the SFR in \citetalias{Wang-19}, only an inference of spatially coherent temporal variations in the SFR. 
The new result in the top panels of Figure \ref{fig:bin_rsfr_prof} allows us to look directly at these temporal changes. Galaxies with a larger (temporal) {\it change} in their overall SFR, show larger (temporal) changes 
in their SFR at all galactic radii. 
However, as shown in the bottom panels of Figure \ref{fig:bin_rsfr_prof}, we do not find larger (temporal) changes in the SFR in galaxies that have different overall levels of star-formation. 
This emphasizes that the \RSFR\ parameter, characterizing the {\it change} of SFR, gives a different perspective on galaxies than the SFR itself.

We now return to examine the dispersion in \RSFR\ from galaxy to galaxy at a given relative radius within a given stellar mass bin. This dispersion was shown in Figure \ref{fig:rsfr_prof} as the error bars on each point and was computed with a sigma-clipping 
algorithm by iteratively rejecting points beyond 3$\sigma$. 
For the two lowest stellar mass bins, 
it is evident that the scatter of \RSFR\ across the galaxy population does not
change significantly with radius. However, for the more massive bins in
stellar mass, the dispersion in \RSFR\ clearly increases towards the centers of galaxies relative to the scatter in the outer regions. 

Since it is improbable that widely separated galaxies are varying coherently in their SFR, apart from the slow overall ``cosmic'' evolution of the SFMS discussed previously, 
the scatter in \RSFR\ across a population of galaxies (at a given epoch) is, as discussed earlier, 
unambiguously a measure of the {\it variability} of the SFR within individual objects.  
This is quite different from the scatter in $\Sigma_{\rm SFR7}$, which could reflect temporal changes in individual galaxies, but also, conceivably, could reflect intrinsic (time-independent) differences from galaxy to galaxy.  
Whereas temporal variability could only be postulated in \citetalias{Wang-19} as an explanation for the correlation of $\sigma(\Delta \Sigma_{\rm SFR7})$ with the depletion time, in this paper we are able to measure temporal variability directly.
 
A clear predication of the gas-regulator model of \cite{Lilly-13} is that 
the variability of the SFR in response to variations in the inflow should strongly depend on the gas depletion time.  Specifically, \citetalias{Wang-19}
drove the gas-regulator system with a sinusoidal inflow rate, and showed that 
the amplitude of the variation in the resulting SFR ($\sigma_{\rm SFR}$) is the amplitude of the variation of the inflow
($\sigma_{\rm \Phi}$) multiplied by a frequency dependent response curve.  
The response curve can be written as 
\begin{equation} \label{eq:7}
    f = \frac{1}{[1+(2\pi\tau_{\rm eff,dep}/\tau_{\rm P})^2]^{1/2}},  
\end{equation}
where $\tau_{\rm P}$ is the period of the cold gas inflow, and 
$\tau_{\rm eff,dep}$ is the ``effective'' gas depletion timescale, defined as the 
gas depletion timescale 
$M_{\rm gas}/{\rm SFR}$ divided by the mass-loading factor of any outflow.
As shown in Equation \ref{eq:7}, for a given inflow, a shorter $\tau_{\rm eff,dep}$
leads larger variations of SFR, and vice versa. 
Although Equation \ref{eq:7} was derived for an idealized sinusoidal input of inflow rate, similar shapes of the response curve are seen for a range of more complicated forms for the inflow rate (see details in \citetalias{Wang-19}). 

\begin{figure}
  \begin{center}
    \epsfig{figure=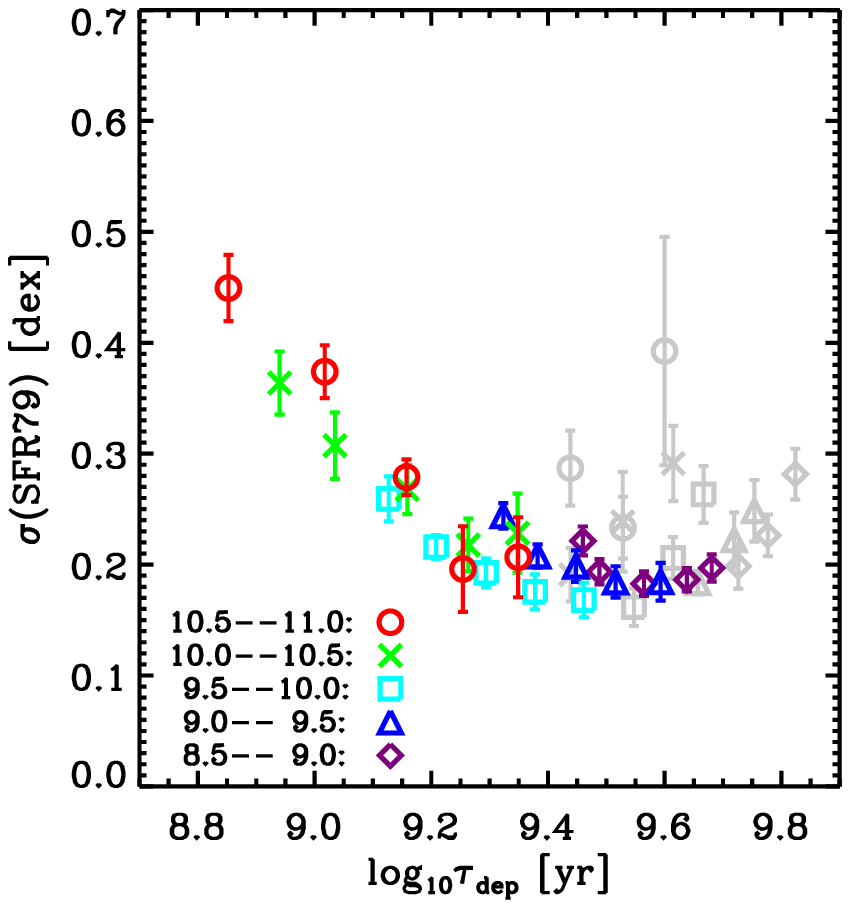,clip=true,width=0.42\textwidth} 
  \end{center}
  \caption{The scatter of \RSFR\ as a function of the indirectly derived gas depletion time.
  The different colors are for galaxies at
  different stellar mass bins, as denoted in the bottom-left corner. 
  Data points with the radius larger than \re, are indicated in gray.}
  \label{fig:scatter_rsfr}
\end{figure}

Figure \ref{fig:scatter_rsfr} shows the scatter of \RSFR\ at different galactic radii as 
a function of the inferred gas depletion time for different radii in the five stellar mass bins, computed as described in the previous subsection. 
In Figure \ref{fig:scatter_rsfr}, each data point
represents one galactic radial bin at a given galactic stellar mass bin. According to Equation \ref{eq:8},
the inner regions of galaxies correspond to the shorter gas depletion timescale. 

As shown, we find that there is 
a tight correlation between the observed dispersion of \RSFR\ and 
the inferred gas depletion timescale, 
with the shorter $\tau_{\rm dep}$ associated with the larger range of \RSFR.  The scatter in \RSFR\ from galaxy to galaxy may be taken as a direct measure of the variability of the SFR in galaxies on timescales between $10^7$ and $10^9$ years.  

This new result is completely independent from that presented in \citetalias{Wang-19}, but is entirely consistent
with the heuristic model presented therein.  Regions with 
higher SFE, i.e. with shorter depletion times, do indeed appear to show a larger response in their SFR to changes in the inflow (see figure 12 in \citetalias{Wang-19}). 
This supports the idea that 
the dynamical response of gas regulator model to a time-varying inflow is the origin of the 
variation of SFR within and across galaxies. This further supports the idea that the
narrow SFMS is indeed the result of the quasi-steady state between the inflow, outflow
and star formation \citep[e.g.][]{Schaye-10, Bouche-10, Lilly-13, Tacchella-16, Wang-19}. 

\subsection{Discussion}

Throughout this work, we have excluded the regions that are located on the Seyfert regions on the BPT diagram in the analysis, because the contribution of H$\alpha$ emission by AGN would likely lead to an over-estimate of SFR79 (see Section \ref{subsec:4.1}). However, some inner regions of galaxies may still be contaminated by a low-luminosity AGN, although this effect must be weak for SF regions.

The connection between AGN and the instantaneous star formation has been investigated by many authors \citep[e.g.][]{Stanley-15, Stanley-17, Harrison-17, Bernhard-18, Dai-18, Scholtz-18, Schulze-19}. Although both positive and negative AGN feedback have been 
proposed in the literature \citep[e.g.][]{Silk-13, Bieri-16, Kalfountzou-17, Shin-19}, convincing observational evidence for the impact of star formation
by AGN is still lacking \citep{Stanley-15, Bernhard-18, Ramasawmy-19, Scholtz-20}. Therefore, it is not clear to us what the SFR79 profile should be for an AGN-host galaxy.

In addition to AGN, other physical processes have been proposed to play roles in changing the instantaneous SFR via different mechanisms, such as the existence of bar \citep[e.g.][]{Wang-12, Lin-17}, and  tidal/ram-pressure stripping \citep{Gunn-72, Moore-96, Abadi-99, Poggianti-17}. For instance, the existence of the bar can effectively transfer the cold gas to the galactic center, and lead to central enhanced star formation \citep{Lin-17}. 

We emphasize that our star formation change parameter, SFR79, allows examination for the change of SFR in response to such processes. In contrast with previous studies, the SFR79 characterizes the change of star formation with respect to the level in the past, rather than to an assumed ``control'' population. 

Many studies suggest that massive galaxies are assembled and quenched from inside outwards \citep[e.g.][]{Perez-13, Tacchella-15, Abdurrouf-18}. Under the ``inside-out'' scenario, the profile of SFR79 would show a drop in the center of ``quenching'' or ``newly quenched'' galaxies (but see \cite{Lilly-16} for how sSFR gradients can arise without differential radial quenching). This seems at first sight consistent with the result in Figure \ref{fig:bin_rsfr_prof}, that galaxies with an overall lower SFR79 show a significant drop of SFR79 at the galactic center, at least for two highest stellar mass bins. However, we note that our galaxies are all SF galaxies as defined with the SFR7-based SFMS, which means that we may not expect to see the SFR79 profiles for newly quenched galaxies in our analysis. Instead, the SFR79 characterizes the movement of SF galaxies on the SFMS. Furthermore, analogously to our analysis in \citetalias{Wang-19}, galaxies with overall higher SFR79 are also seen to have elevated SFR79 in their centers.

Under the dynamical gas regulator model \citep{Lilly-13, Wang-19}, the amplitude of the change of SFR is larger for regions of short gas depletion time, and the change of SFR can be both suppression and elevation. This is also in good agreement with our result. Therefore, the drop of SFR79 in galactic center for massive galaxies with overall lower SFR79, may have little to do with quenching {\it per se}, and may be purely due to the drop of overall inflow rate. 

Finally, as discussed in  \citetalias{Wang-19},  there is no need for any specific physical processes that operate in the center to quench a galaxy ``inside-out'', such as AGN feedback. Instead, the cut-off of the global inflow would naturally lead to an inside-out suppression of star-formation, due to the short gas depletion time in galaxy centers. 

\section{Summary and Conclusion}
\label{sec:6}

In this work, we have introduced and calibrated a new indicator of the change in SFR in individual galaxies, \RSFR.  Observationally, this is based on the equivalent widths of H$\alpha$ 
emission and H$\delta$ absorption, and the amplitude of the 4000 \AA\ break. These three parameters are good indicators for 
recent SFHs \citep{Worthey-97, Kauffmann-03, Li-15, Wang-18} on different timescales. 
Specifically, the H$\alpha$ luminosity traces the SFH within 
the most recent 5 Myr, while \ewhda\ traces the SFHs within the last $\sim$1 Gyr, and
\dindex\ is sensitive to the light-weighted stellar age within 2 Gyr. 

The parameter SFR79 is equal to SFR7/SFR9, i.e. SFR$_{\rm 5Myr}$/SFR$_{\rm 800Myr}$, where the subscripts refer to the preceding time interval over which the SFR is averaged.  Similarly parameters have been called  
burstiness in the literature \citep{Weisz-12, Guo-16, Broussard-19}, but we find this misleading in the present case.  
Our SFR79
spans over two orders of magnitude in timescales, which is much larger than that of 
UV-to-H$\alpha$ flux ratio. As distinct from the ratios of fluxes at different wavelengths, 
the three diagnostic spectral parameters used in the estimator of \RSFR\
are not in principle sensitive to dust attenuation, even if some effects can enter because of spatial variations in the extinction within galaxies. The
change parameter SFR79 characterizes the changes in the SFR, in the sense that a positive
(or negative) value means an elevation (or suppression) of the SFR in the last 5 Myr with respect 
to the average value over the last 800 Myr.  

We calibrate the \RSFR\ by constructing millions of mock SFHs spanning a huge range of 
different SFR behaviours, then generating mock spectra using the SSP models for six different metallicities.  
%We
%measure the \RSFR\ of the mock SFHs, and the three diagnostic parameters. Third, we %explore the 
%relation between \RSFR\ and the combination of three diagnostic parameters, and examine %how large
%is the scatter of the established relation.  

We find that SFR79 can be determined with an uncertainty of 0.06-0.09 dex for different metallicities
(see Equation \ref{eq:2}, Figure \ref{fig:rsfr_fit} and Table \ref{tab:1}). We also examined 
the stability of this calibration by using a different IMF and a different stellar evolution model 
(i.e. isochrones). These only cause an overall 
shift of \RSFR\, and do not change the distribution 
(i.e. also the scatter) of the \RSFR. 
%This does not affect the investigation of the stochasticity 
%of SFHs for galaxy population, because the width of the burstiness characterizes the %stochasticity 
%of recent star formation, rather than the average value of burstiness %\citep{Broussard-19}. 

We compute a small second-order correction for dust 
attenuation that arise from a stellar age dependent E(B$-$V) model 
from \cite{Charlot-00}.  We assume the E(B$-$V) for stellar population older than 10 Myr 
is the one-third of that for stellar population younger than 10 Myr (Section \ref{subsec:2.5}).  

We then apply the calibrator to a well-defined sample (taken from \citetalias{Wang-19}) of SF main-sequence galaxies selected from the MaNGA survey,
and obtain the maps and profiles of \RSFR\ and SFR9 for each individual galaxies, using a metallicity taken from the overall mass-metallicity relation.

Based on the new information, we investigate the variation of \RSFR\ and $\Sigma_{\rm SFR9}$ within 
and across the galaxy population. Our main results are summarized as follows. 

\begin{itemize}

\item  We have first measured the SFR79
%$\log_{10}\langle {\rm SFR7}\rangle/\langle {\rm SFR9}\rangle$  
of the whole galaxy population, i.e. the integrated (over all galaxies) SFR7 divided by the integrated SFR9.  This ratio was $-$0.066 dex, reflecting a small decline in the SFR.  This is remarkably close to that expected value ($-$0.025 dex) from the cosmic evolution of the characteristic sSFR of the SFMS.  This reassuring agreement attests to the empirical validity of our calibration.   

\item  We have applied a small {\it ad hoc} zero-point correction to \RSFR\ as a function of the stellar surface mass density, to remove a small radial dependence of \RSFR\ that we suspect is not real and may instead reflect an uncorrected radial dependence of the metallicity, or possibly the IMF.  

%Considering the slowly evolved SFMS in the last 1 Gyr, the value of \RSFR tells 
%whether each individual galaxies move up or down on the SFMS in last 5 Myr.  This is 
%new information in galaxy evolution derived from the observations. 

\item The global \RSFR\ (measured within \re) is nearly independent of  $\Delta$SFR9, i.e. the deviation 
from the SFR9-based SFMS.  This means that the movement of galaxies (up or down) on the 
SFMS does not depend on their average positions on the SFMS during the last 800 Myr.  
This is required if the scatter of the SFMS is to be preserved over time, as observed, since a strong positive (or negative) relation between \RSFR\ and $\Delta$SFR9
would broaden (or compress) the SFMS over time.  This provides a further empirical support for the validity of our \RSFR\ estimator.

\item  The scatter in the SFR9-based SFMS is 0.26 dex, noticeably smaller than the SFR7-based SFMS (0.34 dex).  

\item Galaxies with higher (or lower) global 
SFR9 with respect to the ``nominal'' SFMS, show enhanced (or suppressed) star formation 
at all galactic radii with respect to median $\Sigma_{\rm SFR9}$ profile at given stellar mass. 
In addition, the star formation in central regions of galaxies are more enhanced 
(or suppressed) than outer regions (with respect to the median profile), 
at least for galaxies with \mstar$>$9.5.  
These results are the equivalent for 800 Myr timescales to those already 
shown for the shorter 5 Myr timescales in \citetalias{Wang-19}. 

\item  Of greater novelty, we show that galaxies with higher (or lower) global \RSFR\ 
appear to have higher (or lower) \RSFR\ at all galactic radii with, again, noticeably larger range of \RSFR\ in the central regions of more massive galaxies. 
This means that galaxies with a recent enhancement in SFR (i.e. within 5 Myr with respect 
to the SFR averaged over the last 800 Myr) have enhanced star formation at all galactic radii
(again, with respect to the star formation in the past). 
%This strongly supports the picture that the star formation in galaxies is primarily driven 
%by the overall inflow of cold gas, which is the basis of the gas regulator model. 

\item  Finally, in the most important result of the paper, we show that the dispersion in the \RSFR\ across the galaxy population, at a given galactic 
radius and stellar mass, is strongly anti-correlated with the inferred
gas depletion time for these locations.   The scatter in \RSFR\ across the population is a direct measure of the {\it temporal variability} of the SFR within individual objects, since it would be completely unreasonable to suppose that the temporal variations of widely separated galaxies are coherent (apart from a slow cosmic evolution of the SFMS). 

\end{itemize}

In \citetalias{Wang-19}, we interpreted the observed dependence of the dispersion of $\Sigma_{\rm SFR7}$ across the population with the gas depletion timescale to be reflecting the dynamical {\it temporal} response of the gas regulator system to variations of the inflow rate. This was because, in a gas-regulator system, the amplitude of changes in the SFR to changes in the inflow rate is predicted to be strongly anti-correlated with the gas depletion time. 

In the current work, we are able to measure directly these temporal changes by measuring the dispersion of the change parameter SFR79 across the population of galaxies.  The fact that this dispersion is seen, once again, to anti-correlate with the inferred gas depletion timescale, further strengthens the case for the simple gas-regulation picture of galaxies and for interpreting variations in the star-formation rate of galaxies in terms of variations in the inflow rate.

%(and also in $\Delta \Sigma_{\rm SFR9}$) 

%\citetalias{Wang-19} investigated the variation of \SFRSE\  within and across galaxies, 
%and find the dynamical response of gas regulator model to a periodic inflow can 
%rather well explain the $\Sigma_{\rm SFR7}$ profiles and the correlation between the 
%variation of $\Sigma_{\rm SFR7}$ and gas depletion time.  However, \citetalias{Wang-19} 
%only presented the SFR measured with H$\alpha$ luminosity, and do not give 
%any evolving information for SFHs of individual galaxies.  
%This means that \citetalias{Wang-19} use a time evolving model to explain a static 
%observational result.  In the current work, by using an independent 
%approach from \citetalias{Wang-19}, we for the first time quantify the star formation
%change parameter, \RSFR, 
%based on the three diagnostic parameters, and confirm the result of 
%\citetalias{Wang-19} from the perspective of time evolution. This strongly strengths 
%the scenario proposed in \citetalias{Wang-19}. 

The star formation change parameter proposed in this work is a new parameter, which contains 
valuable information of the time-varying SFH for the galaxy population. 
It provides a new approach to study the physical processes that
govern the SFR within galaxies on different timescales. The SFR79 also opens a new window for testing models of galaxy formation and evolution in future hydrodynamical simulations and semi-analytic models.   
In the second paper of this series, we will explore how the power-spectrum of SFR variability in galaxies, 
i.e. the contribution of the variation in SFR at different timescales, 
based on the result in the present work.   

\acknowledgments

Funding for the Sloan Digital Sky Survey IV has been provided by
the Alfred P. Sloan Foundation, the U.S. Department of Energy Office of
Science, and the Participating Institutions. SDSS-IV acknowledges
support and resources from the Center for High-Performance Computing at
the University of Utah. The SDSS web site is www.sdss.org.

SDSS-IV is managed by the Astrophysical Research Consortium for the
Participating Institutions of the SDSS Collaboration including the
Brazilian Participation Group, the Carnegie Institution for Science,
Carnegie Mellon University, the Chilean Participation Group, the French Participation Group, 
Harvard-Smithsonian Center for Astrophysics,
Instituto de Astrof\'isica de Canarias, The Johns Hopkins University,
Kavli Institute for the Physics and Mathematics of the Universe (IPMU) /
University of Tokyo, Lawrence Berkeley National Laboratory,
Leibniz Institut f\"ur Astrophysik Potsdam (AIP),
Max-Planck-Institut f\"ur Astronomie (MPIA Heidelberg),
Max-Planck-Institut f\"ur Astrophysik (MPA Garching),
Max-Planck-Institut f\"ur Extraterrestrische Physik (MPE),
National Astronomical Observatory of China, New Mexico State University,
New York University, University of Notre Dame,
Observat\'ario Nacional / MCTI, The Ohio State University,
Pennsylvania State University, Shanghai Astronomical Observatory,
United Kingdom Participation Group,
Universidad Nacional Aut\'onoma de M\'exico, University of Arizona,
University of Colorado Boulder, University of Oxford, University of Portsmouth,
University of Utah, University of Virginia, University of Washington, University of Wisconsin,
Vanderbilt University, and Yale University.

This research has been supported by the Swiss National Science Foundation. 
The authors thank Benedikt Diemer for providing the SFHs of Illustris galaxies 
by private communication. 

\bibliography{rewritebib.bib}

\clearpage
%\appendix

%%%%%%%%%%%%The End%%%%%%%%%%%%%%%%%%%%%%%%%%%%%%%%%%%%%%%%%
\label{lastpage}
\end{document}